\documentclass[aps,floatfix,twocolumn,showpacs,pra]{revtex4}

\usepackage{color}
\usepackage{latexsym}
\usepackage{amstext}
\usepackage{amsxtra}
\usepackage{epsfig}
\usepackage{graphicx}

\def\be{\begin{equation}}
\def\ee{\end{equation}}
\def\bea{\begin{eqnarray}}
\def\eea{\end{eqnarray}}
\def\betab{\mbox{\boldmath$\beta$}}
\def\gammab{\mbox{\boldmath$\gamma$}}
\def\chib{\mbox{\boldmath$\chi$}}
\def\Psib{\mbox{\boldmath$\Psi$}}

\begin{document}
\title{Three fully polarized fermions
close to a $p$-wave Feshbach resonance}

\author{M. Jona-Lasinio$^{1,2}$, L. Pricoupenko$^{3}$ and Y. Castin$^{1}$}
\affiliation
{
$^{1}$Laboratoire Kastler Brossel, Ecole normale sup\'erieure, UPMC, CNRS,
24 rue Lhomond, 75231 Paris Cedex 05, France. \\
$^{2}$ 
LENS - European Laboratory for Non-Linear Spectroscopy,
Via Nello Carrara 1,
I-50019 Firenze, Italy
\\
$^{3}$Laboratoire de Physique Th\'{e}orique de la Mati\`{e}re Condens\'{e}e,
Universit\'{e} Pierre et Marie Curie, case courier 121,
4 place Jussieu,
75252 Paris Cedex 05, France. \\
}

\begin{abstract}
We study the three-body problem for three atomic fermions, in the same
spin state, experiencing a resonant interaction in the $p$-wave
channel {\it via} a Feshbach resonance represented by a two-channel
model.
The rate of inelastic processes 
due to recombination to deeply bound dimers
is then estimated from the three-body solution using a simple prescription.
We obtain numerical and analytical predictions for most of the experimentally
relevant quantities that can be extracted from the three-body solution:
the existence of weakly bound trimers and their lifetime, 
the low-energy
elastic and inelastic scattering properties of an atom
on a weakly bound dimer (including the atom-dimer scattering
length and scattering volume), and the recombination rates for three colliding
atoms towards weakly bound and deeply bound dimers.
The effect of ``background" non-resonant interactions 
in the open channel of the two-channel model is also calculated and
allows to determine which three-body quantities are `universal'
and which on the contrary depend on the details of the model.
\end{abstract}

\maketitle

\section{Introduction}

Fermionic superfluidity with $p$-wave pairing is related to a large 
class of subjects in very different areas of physics including condensed 
matter, astrophysics and particle physics \cite{Volovik0,Volovik1}. As already 
observed in $^3$He experiments, the phase diagram in these systems can 
be very rich \cite{Lee}. Moreover, the possible observation of quantum 
phase transitions together with the existence of exotic topological 
defects in $p$-wave superfluids bring a lot of interest in their study. 

Presently, there is some hope that $p$-wave superfluidity and its 
intriguing properties can be observed with ultra cold atoms \cite{Volovik2}.
%p-wave superfluidity with ultra-cold atoms ?: experimental context
Indeed, thanks to the concept of Feshbach resonance \cite{Feshbach}, 
it is possible to tune the inter-atomic interaction and to achieve 
strongly correlated regimes in ultra cold dilute atomic gases. First 
realized in the $s$-wave channel with bosonic species \cite{Inouye,Cornish}, 
the Feshbach resonance is currently 
used for achieving BEC-BCS crossover experiments for the two-component 
Fermi gas in a regime of temperatures where the system can be superfluid 
\cite{Thomas1,Thomas2,Jin,Salomon1,Jochim,Greiner,Zwierlein,
Bourdel,Bartenstein,Grimm_gap,Hulet_xi,Ketterle_unba,Stewart,
Altmeyer,Luo,Zwierlein2}. 
In one spin component Fermi gases, as a 
consequence of the Pauli exclusion principle, two-body scattering processes 
are forbidden in the $s$-wave channel and at low temperatures are dominant 
in the $p$-wave channel. The two-body cross-section which is usually 
negligible in this channel can be greatly enhanced using a $p$-wave Feshbach 
resonance. This resonant regime is now obtained for $^{40}$K 
\cite{Regal2,Ticknor,Gunter,Gaebler} and $^{6}$Li atoms \cite{Zhang,Schunck,Chevy}. 
The production of $p$-wave shallow dimers in ultra cold 
$^{6}$Li \cite{Zhang} and $^{40}$K gases \cite{Gaebler} opens very interesting 
perspectives for the realization of a superfluid $p$-wave phase. 

In these 
experiments, an external magnetic field tunes the energy of a two-body 
$p$-wave bound state in a closed channel and for a small detuning with respect 
to the open channel, a resonance occurs in a two-body $p$-wave scattering 
process. Moreover, due to the presence of a magnetic field, the interaction 
strength depends on the orbital channels considered --a major difference 
with respect to what happens in superfluid $^3$He \cite {Ticknor}.
%ETUDES DU CROSSOVER BEC-BCS POUR UN PAIRING P-WAVE
As a consequence, the question of the symmetry of the low temperature ground 
state in one component fermionic species is non-trivial. Studies of this many-body 
problem are essentially mean-field and depending on the experimental 
realizations, they predict the occurrence of $p_x+ip_y$ (axial), $p_x$ (polar) 
or intermediate phases \cite{Gurarie1,Ohashi,Jason,Cheng,pseudopot,Gurarie2}. 
These predictions lead to possible studies of quantum phase transitions in such systems.
%Experiments: three-body losses
However, the main issue in the achievement of a $p$-wave superfluid concerns
atom losses which are large in present experiments 
\cite{Regal2,Ticknor,Gunter,Zhang,Schunck,Chevy,Gaebler}. The question 
whether or not it is possible that the system thermalizes is then a crucial point.

%Interest of the few body problem in the crossover 
Concerning $s$-wave resonant Fermi systems, few-body studies have proven to 
be very successful in understanding properties of the superfluid gas in the 
BEC-BCS cross-over 
\cite{TerM,Petrov3F,Petrov4F,Leyronas34,Leyronasmany}. These studies 
explain the large lifetime of the system observed at resonance and also predict
the dimer-dimer scattering length which is involved in the equation of state
for the dilute BEC phase. Surprisingly, although general many-body  properties 
are rather well known in $p$-wave superfluids --thanks to contributions 
from the condensed matter community, few body properties in these systems have 
been less studied \cite{Greene,Macek}. 
However, following the example of the works done in the 
$s$-wave channel, few-body problems for $p$-wave pairwise potential are 
valuable 
for a determination of properties in the strongly interacting dilute gas beyond 
a mean-field analysis. As an example, we note that consequences of the 
existence of trimers first found in the present work 
have already been taken into 
account for an estimation of the lifetime of $p$-wave shallow dimers 
\cite{Gurarie3}.

%Items that are addressed in this paper
In this paper, we consider three identical fermions close to a 
$p$-wave Feshbach resonance. We determine their low energy 
scattering properties together with the possible existence of 
trimers. Our study is also a first step toward an understanding 
of the atom losses observed in present experiments 
\cite{Regal2,Gunter,Zhang,Schunck,Gaebler}.
%Structure and results of the paper
The paper is organized as follows. In section \ref{sec:basics}, we recall
basic properties, for an isotropic short range interaction, of resonant 
two-body $p$-wave scattering processes \cite{Landau}. In the resonant 
regime, two parameters are needed for a description of the low 
energy two-body properties: the scattering  volume ${\mathcal V}_s$ 
and also the $p$-wave equivalent of the effective range parameter 
hereafter denoted by $\alpha$. 
For large and positive values of ${\mathcal V}_s$ there exists a
shallow $p$-wave dimer of internal angular momentum one,
that is with three-fold degeneracy. 
For a potential with a compact 
support of radius $b$, we show that at resonance (${\mathcal V}_s=\infty$) 
the effective range parameter cannot reach arbitrarily small values and 
$\alpha b \geq 1$. Consequently, unlike what happens in $s$-wave 
resonances, there is no scale invariance at low energy
and a unitary regime cannot 
be obtained via a $p$-wave resonance \cite{Yvanscaling}. In 
section \ref{sec:modeling}, we introduce the main model Hamiltonian 
that we use in this
work. It is a two channel model of the $p$-wave Feshbach resonance 
\cite{Chevy}
where free atoms in the open channel interact with a molecular $p$-wave 
state in the closed channel, of threefold degeneracy provided
that one neglects the effect of the dipole-dipole interaction 
in presence of the Feshbach magnetic field. 
The inter-channel coupling 
amplitude, as a function of the relative distance of the two atoms,
 is a Gaussian of range $b$ which mimics the van der Waals range 
of a more realistic two-body potential. We first briefly
determine the two-body collisional 
properties of this model. At large coupling the resonance is broad  
$\alpha b \sim 1$ and for ${\mathcal V_s}$ large and positive the shallow dimer 
is essentially in the open channel. In the opposite regime for a weak coupling, 
the resonance is narrow, $\alpha b \gg 1$ and the shallow dimer is almost
entirely  in 
the closed channel. In section \ref{sec:solution}, 
we derive an integral equation for 
the three body problem. We consider solutions of total angular momentum 
$J=1$ and by using the rotational symmetry of the Hamiltonian, we reduce 
the problem in each involved symmetry sector (odd or even) to a one 
dimensional integral equation. In both sectors, we predict the existence of one 
trimer for sufficiently broad resonances. These trimers can exist in 
a regime where there is no shallow dimer (for large and negative 
values of the scattering volume) and are interesting examples of Borromean 
states \cite{Zhukov}, since we find that 
they are not linked to an Efimov effect, contrarily to \cite{Macek}.
We determine also the atom-dimer scattering length $a_{ad}$ as a function 
of the effective range parameter $\alpha$ and the potential
range $b$, for different values of the 
scattering volume. At resonance (${\mathcal V}_s=\infty$), $a_{ad}$ takes large 
values (that is significantly larger than the potential range $b$)
only in the vicinity of the threshold of existence of a trimer. 
The recombination rate of three incoming atoms into a shallow dimer 
and one outgoing atom is computed; it is shown analytically to 
vary as ${\mathcal V}_s^{5/2}$ for large values of the scattering volume,
away from the trimer formation threshold; this differs from
the ${\mathcal V}_s^{8/3}$ law put forward in \cite{Greene} on
the basis of a dimensional analysis ignoring a possible contribution
of the effective range parameter $\alpha$, but is still compatible with
the numerics of \cite{Greene}; finally, the recombination rate
is shown analytically
to present a Fano profile
as a function of $\alpha$ close to this trimer threshold.
In section \ref{sec:loss}, we calculate the losses due to the recombination 
into deeply bound dimers. Since these losses are not present 
in our model Hamiltonian,
we estimate them from the probability that three atoms are 
within a volume of the order of $b^3$ and we obtain
the lifetime of trimer states, the losses due to atom-dimer inelastic 
scattering and the three body recombination rate toward deep molecular states 
from asymptotically free atoms.
Finally we make the model more realistic by including an attractive
interaction
in the open channel, in addition to the coupling with the
closed channel, in section \ref{sec:enri}: we recalculate the
trimer energies, the atom-dimer scattering length and the recombination
rate to weakly bound dimers, and we physically explain the
impact on these quantities of a non-resonant interaction in the open channel.
We conclude in section \ref{sec:conclusion}.

\section{Basic properties of the two-body $p$-wave scattering}
\label{sec:basics}
\subsection{The scattering amplitude}

We consider in this section two particles of mass $m$
in the same spin state and in the center of mass
frame, scattering in free space
{\sl via} a rotationally invariant short range interaction potential.
We assume for simplicity that the interaction potential scatters only
in the $p$-wave channel, so that at large distances, where the effect of the
potential is negligible, the scattering 
wavefunction of energy $E=\hbar^2 k^2/m$, $k>0$, takes the form
\be
\psi_{\mathbf{k}}(\mathbf{r}) \simeq e^{i\mathbf{k}\cdot\mathbf{r}} + 3 f(k) \hat{\mathbf{k}}\cdot\hat{\mathbf{r}}
\frac{d}{dr}\left(\frac{e^{ikr}}{ikr}\right)
\label{eq:scat_state}
\ee
where $\mathbf{r}$ is the relative position of the two particles,
$\pm \mathbf{k}$ are their incoming wave-vectors,
and we have introduced the unit vectors $\hat{\mathbf{r}}=\mathbf{r}/r$
and $\hat{\mathbf{k}}=\mathbf{k}/k$.
The function
 $f(k)$ is the so-called reduced scattering amplitude 
since the angular dependence
of the scattered wave has been pulled out.
We note that Eq.(\ref{eq:scat_state}) becomes exact (that is one can replace
$\simeq$ by $=$) for a compact support interaction potential, when $\mathbf{r}$ is
out of the support of the potential.

In this subsection, we briefly review some basic properties of this $p$-wave
scattering amplitude $f(k)$.
As a consequence of the unitarity of the $S$-matrix of scattering theory,
it obeys the optical theorem,
\be
\mbox{Im}\, f(k) = k |f(k)|^2
\ee
which implies
\be
f(k) = -\frac{1}{u(k)+ik}
\ee
where $u(k)$ is a real function. For cold atoms, the low energy
scattering properties are crucial and we assume that 
$u(k)$ has the following low-$k$ series expansion,
\be
u(k) = \frac{1}{k^2 \mathcal{V}_s} + \alpha + O(k^2).
\label{eq:lkeu}
\ee
The so-called scattering volume $\mathcal{V}_s$ plays a role similar
to the scattering length in the $s$-wave channel: 
the resonant situation corresponds to the limit $|\mathcal{V}_s|\to\infty$.

Another crucial property of the reduced scattering amplitude 
is that its analytic continuation to negative energies, that is to imaginary values
of $k$, gives information on possible bound states in the two-body problem,
in the form of poles of $f(k)$.
More precisely, setting 
$k=i q$, where $q>0$, the solutions $q_{\rm dim}>0$ of the equation
\be
\frac{1}{f(i q)}=0,
\ee
correspond to bound states of the scattering potential, that is here
to dimers of rotational quantum number $S=1$,
with a binding energy 
\begin{equation}
E_{\mathrm{dim}}=\hbar^2 q_{\mathrm{dim}}^2/m.
\end{equation}
The wavefunction of such a dimer, ``out" of the potential (again,
this has an exact meaning
for a compact support potential), is
a solution of the free Schr\"odinger's equation in the $p$-wave channel, so that we may take
it of the form
\be
\phi(\mathbf{r}) = \mathcal{N}
\left(\frac{3}{4\pi}\right)^{1/2}
\frac{r_\gamma}{r} \frac{d}{dr}\left(\frac{e^{-q_{\mathrm{dim}}r}}{r}\right)
\label{eq:phidim}
\ee
where $\mathcal{N}$ is a normalization factor and $r_\gamma$ is
the component of $\mathbf{r}$
along direction $\gamma=x,y$ or $z$ \cite{wdam}. 

The knowledge of the dimer wavefunction ``inside" the potential requires a full solution
of Schr\"odinger's equation. However it is possible to access the normalization factor
$\mathcal{N}$ directly from the knowledge of the scattering amplitude.
Using the closure relation
\be
\int \frac{d^3 k}{(2\pi)^3} \psi_{\mathbf{k}}(\mathbf{r}) \psi_{\mathbf{k}}^*(\mathbf{r}')=
\delta(\mathbf{r}-\mathbf{r}') - \sum_i \phi_i(\mathbf{r}) \phi_i^*(\mathbf{r}'),
\label{eq:clos}
\ee
in the limit of large $r$ and $r'$, we obtain
\be
|\mathcal{N}|^2 = -\frac{2}{q_{\mathrm{dim}}\left[1-iu'(iq_{\mathrm{dim}})\right]},
\label{eq:norm}
\ee
assuming that $u(k)$ has a series expansion with even powers of $k$ only and using
contour integration in the complex plane to single out the contribution
of the poles of $f(k)$.
As we shall see, this relation (\ref{eq:norm}) may be used to put constraints
on the parameter $\alpha$.

\subsection{Constraint on the parameter $\alpha$ close to resonance}

Whereas the scattering volume can be adjusted at will by a Feshbach
resonance driven by a magnetic field, 
the value of $\alpha$ on resonance cannot be adjusted the same way
so it is important to determine what are
its possible values on resonance.

We assume that $\alpha_{\rm res}\neq 0$, where $\alpha_{\rm res}$
is the value of $\alpha$ on resonance \cite{apz}.
In the resonant limit, we see from the low-$k$ expansion of $u(k)$
that there exists a weakly bound dimer on the side 
$\alpha_{\rm res} \mathcal{V}_s >0$ of the resonance \cite{quasimodo}: 
\be
q_{\mathrm{dim}} \sim \frac{1}{\sqrt{\alpha_{\rm res} \mathcal{V}_s}}.
\label{eq:qdim_approx}
\ee
From Eq.(\ref{eq:lkeu}) and Eq.(\ref{eq:norm}) we obtain
\be
|\mathcal{N}|^2 \sim \frac{1}{\alpha_{\rm res}}.
\ee
This imposes $\alpha_{\rm res}>0$. This is in sharp contrast
with the case of $s$-wave scattering, where the effective range
$r_e$ can take any sign on resonance.

For a compact support potential, vanishing outside a sphere of radius $b$, that
is for $r>b$, the normalization of the
dimer wavefunction to unity imposes
\be
\int_{r>b} d^3r\, |\phi(\mathbf{r})|^2 \leq 1.
\ee
Calculating the resulting integral with the expression Eq.(\ref{eq:phidim}) 
leads to \cite{Lpseudo}
\be
|\mathcal{N}|^2 q_{\mathrm{dim}} \left[\frac{1}{2}+\frac{1}{q_{\mathrm{dim}}b}\right]
e^{-2q_{\mathrm{dim}}b} \leq 1.
\ee
In the limit $|\mathcal{V}_s|\to\infty$ this leads to \cite{MurPopov}
\be
\alpha_{\rm res} \geq \frac{1}{b}
\label{eq:ineg}
\ee
where, again,  
$\alpha_{\rm res}$ is the value of $\alpha$ on resonance $|\mathcal{V}_s|=\infty$.
In the zero range limit $b\to 0$, we see that
$\alpha_{\rm res}$ cannot tend to zero, but on the contrary has to diverge!
This is in sharp contrast with the $s$-wave case, where one can find models for the interaction
potential where $r_e\to 0$ in the zero range limit $b\to 0$.

\begin{figure}
\includegraphics[width=8cm,clip]{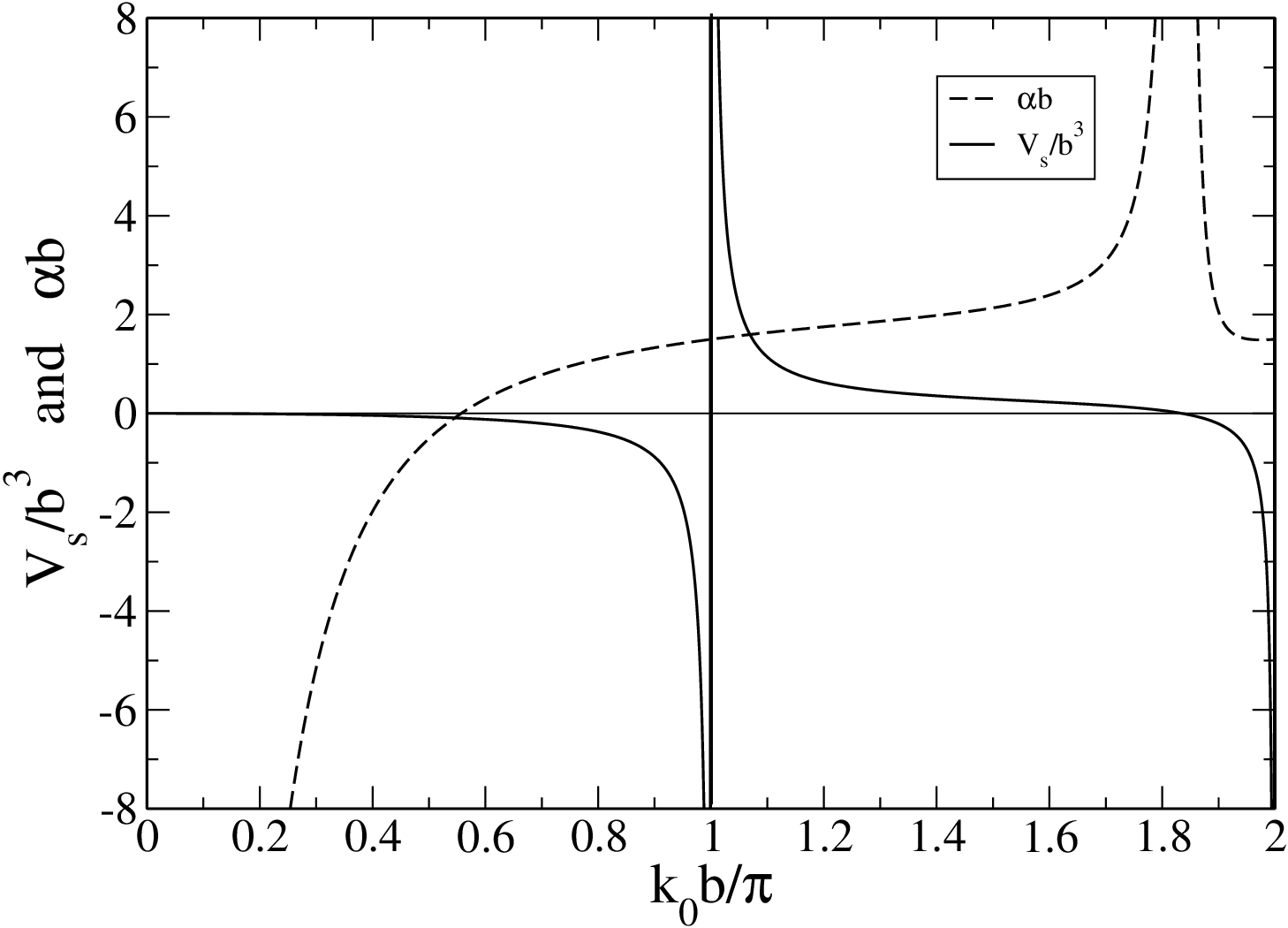}
\caption{For a square well interaction potential $V(r)=
-\frac{\hbar^2 k_0^2}{m}\, 
\theta(b-r)$, where $\theta$ is the Heaviside function,  
values of $\alpha$ in units of $1/b$ (dashed line) and $\mathcal{V}_s$ in units of $b^3$ (solid line)
as functions of $k_0 b$. Note the divergence of $\alpha$ when $\mathcal{V}_s=0$.
\label{fig:sw}}
\end{figure}

To illustrate these properties on a simple example, we give in
Fig.\ref{fig:sw} the values of $\alpha$ and $\mathcal{V}_s$ for a square well
interaction potential, as functions of the well depth.
We see on the figure that (\ref{eq:ineg}) is satisfied at resonance, and that
$\alpha$ is no longer constrained by this condition away from resonance, and
may even vanish and become negative.

\section{Modeling of the resonant $p$-wave interaction}
\label{sec:modeling}

In this section, we introduce the main model used in this paper
to describe the $p$-wave
interaction between same spin state fermions close to a resonance.
It is simply a two-channel model of a Feshbach resonance, 
that is a direct generalization of the $s$-wave two-channel model
\cite{2channel-model} to the $p$-wave case, in the spirit of \cite{Chevy}.
It is extended in section \ref{sec:enri} to include direct interactions
among atoms in the open channel.

\subsection{Model Hamiltonian}

As is standard in a two-channel model, the atoms may populate either the open channel,
where they are treated explicitly as fermionic particles, or the closed
channel, where they exist only under the form of specific tight two-body bound states,
here referred to as molecules; these molecules are treated as bosons, and have
an internal rotational state of spin $S_{\rm mol}=1$ 
since they are $p$-wave two-body bound states.
We assume that the three rotational sublevels of a molecule are degenerate: even
if this is not exactly true in practice because of the effect of the
dipole-dipole
interaction in presence of the magnetic field used
to produce the Feshbach resonance \cite{Ticknor}, this will make our model rotationally
invariant and greatly simplify the algebra for the three-body problem.
For simplicity, we also assume that there is no direct interaction among the fermionic
particles, the resonant $p$-wave atomic interaction being taken into account through the coupling
between fermions and molecules. As already mentioned, this simplifying
assumption is removed in section \ref{sec:enri}.

\begin{figure}
\includegraphics[width=8cm,clip]{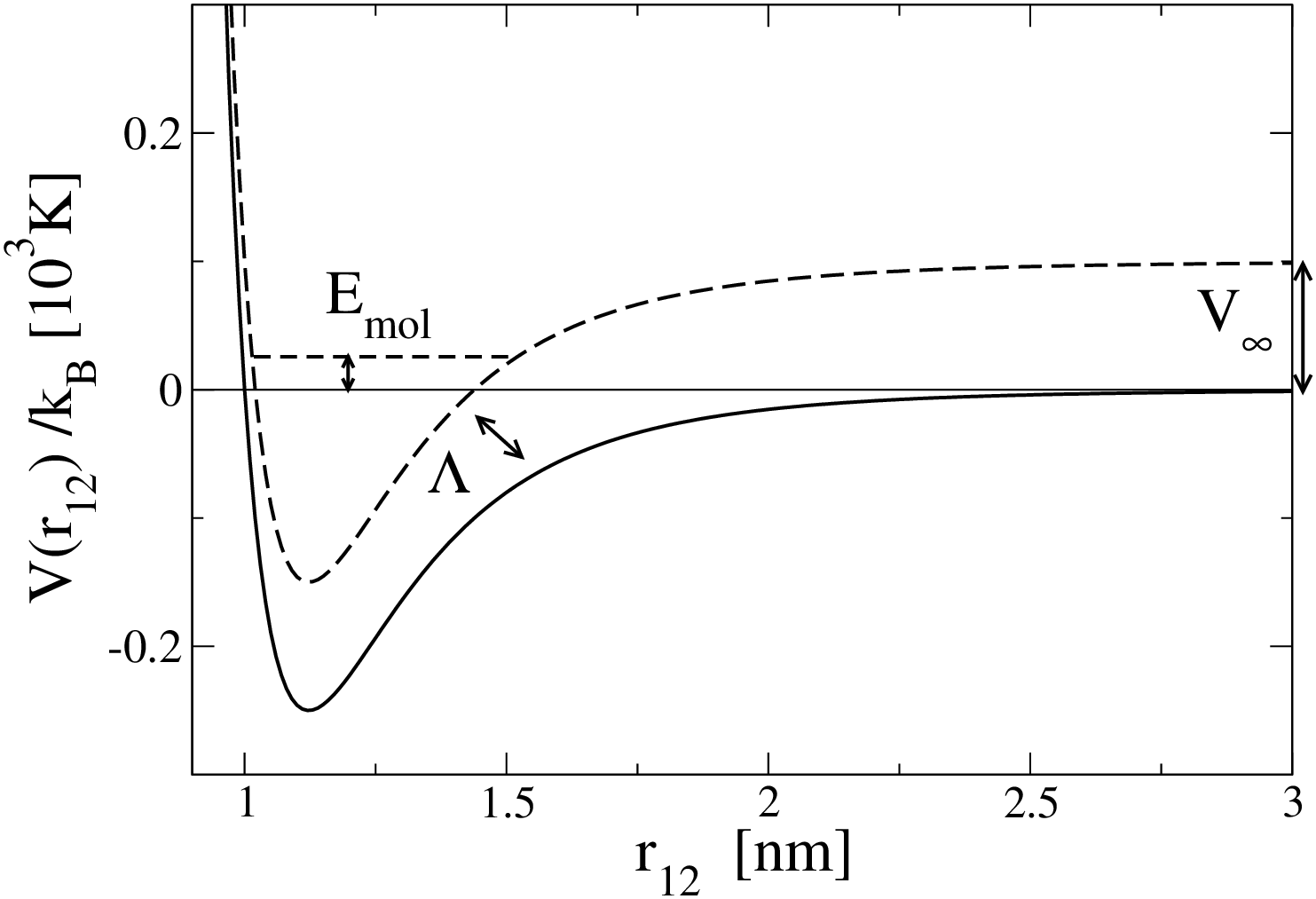}
\caption{Schematic view of a Feshbach resonance configuration: the atoms interact
via two potential curves, plotted as a function of the interatomic distance.
Solid line: open channel potential curve. Dashed line: closed channel potential curve.
When one neglects the coupling $\Lambda$ between the two curves, the closed channel
has a molecular state of energy $E_{\rm mol}$ with respect to the dissociation
limit of the open channel. Note that the energy dependence of the two curves
is purely indicative, and the spacing between the solid curve
and the dashed curve was greatly exaggerated for clarity.
}
\label{fig:feshbach}
\end{figure}

The situation is represented schematically in Fig.\ref{fig:feshbach}.
Mathematically, it corresponds to the following free space Hamiltonian
written in second quantized form:
\bea
&H& = \int \frac{d^3k}{(2\pi)^3} 
\left[\frac{\hbar^2k^2}{2m} a_\mathbf{k}^\dagger a_\mathbf{k}
+ \left(E_{\mathrm{mol}} + \frac{\hbar^2 k^2}{4m}\right) 
\sum_\gamma b_{\gamma,\mathbf{k}}^\dagger b_{\gamma,\mathbf{k}} \nonumber \right] \\
&+& \Lambda \int \frac{d^3k d^3k'}{(2\pi)^6}
\left[\sum_\gamma \chi_\gamma^*\left(\frac{\mathbf{k}-\mathbf{k}'}{2}\right)
b^\dagger_{\gamma,\mathbf{k}+\mathbf{k}'} a_{\mathbf{k}} a_{\mathbf{k}'}
+\mbox{h.c.}\right]. 
\label{eq:hamil}
\eea
The annihilation and creation operators for fermions (that is for the atoms in the
open channel) in plane waves of wave vectors $\mathbf{k}$ and $\mathbf{k}'$
obey the anticommutation relation
\be
\{a_\mathbf{k},a_{\mathbf{k}'}^\dagger\} = (2\pi)^3\, \delta(\mathbf{k}-\mathbf{k}')
\ee
which corresponds to the convention 
$\langle \mathbf{r}|\mathbf{k}\rangle = e^{i\mathbf{k}\cdot\mathbf{r}}$
for the plane-wave.
The operator $b_{\gamma,\mathbf{k}}$ annihilates a molecule (in the closed channel),
with a center of mass momentum $\hbar\mathbf{k}$,
in one of the three degenerate
internal states $\gamma$ in the $S_{\rm mol}=1$ molecular rotational manifold;
we take here for $\gamma$ one of the directions $x$, $y$ or $z$,
which amounts to using the chemistry basis $\{|\gamma\rangle\}$,
where $|\gamma\rangle$ is an eigenstate of zero angular momentum along
direction $\gamma$, rather than the standard basis $\{|m=0,\pm 1\rangle\}$. 
As we mentioned, molecules are treated as bosons so that the $b$'s obey commutation
relations
\be
[b_{\gamma,\mathbf{k}},b_{\gamma',\mathbf{k}'}^\dagger] =
\delta_{\gamma\gamma'} (2\pi)^3\delta(\mathbf{k}-\mathbf{k}').
\ee
Also, the $a,a^\dagger$ fermionic operators commute with the bosonic ones $b$ and $b^\dagger$.
In addition to its center of mass kinetic energy, each molecule has an internal 
energy
$E_{\mathrm{mol}}$, defined 
in the absence of coupling between the open and the closed channels, and counted
with respect to the dissociation energy of the open channel.

Whereas the first contribution in the right-hand side of (\ref{eq:hamil})
simply corresponds to non-interacting gases of atoms and molecules,
the second contribution describes the coupling between the two species, that is 
between the open and closed channels,
responsible for the $p$-wave resonance.
This inter-channel coupling
depends on the relative momentum between two atoms through the functions $\chi_\gamma$;
here, we are in the case of a $p$-wave coupling so we take
\be
\chib(\mathbf{k}) = \mathbf{k}\, e^{-k^2 b^2/2}
\ee
where $b$ is the range in real space of the inter-channel coupling, of the order of
the radius of the closed channel molecule.
The overall amplitude of the inter-channel coupling is measured by the coupling
constant $\Lambda$, taken here to be real; it has not the dimension 
of an energy, but rather has the same dimension as $\hbar^2 b^{1/2}/m$. 
As we already mentioned,
the model is summarized in Fig.\ref{fig:feshbach}. It holds at low kinetic energies,
below the dissociation limit $V_\infty$ of the closed channel.

\subsection{Two-body aspects}
\label{subsec:two-body-aspects}

Before solving the three-body problem, it is important to understand the two-body
aspects of the model, in the form of the reduced scattering amplitude $f(k)$
and the related properties of possible dimers, according to the general
discussion of section \ref{sec:basics}.

We thus calculate the scattering state of two atoms in the center of mass frame,
that is for a zero total momentum. The most general state vector
is thus a coherent superposition of two atoms 
(in the open channel) and
one molecule (in the closed channel): 
\be
|\Psi\rangle = \sum_\gamma \beta_\gamma b_{\gamma,\mathbf{0}}^\dagger |0\rangle
+\int \frac{d^3k}{(2\pi)^3} A(\mathbf{k}) a_\mathbf{k}^\dagger a_{-\mathbf{k}}^\dagger
|0\rangle.
\label{eq:mgsv2b}
\ee
Since the molecule has a zero total momentum, its state is characterized by
the three complex amplitudes $\beta_\gamma$ in each of the internal rotational
states $\gamma=x,y$ and $z$. On the contrary, the two atoms can have opposite
but arbitrary momenta $\mathbf{k}$ and $-\mathbf{k}$, hence the {\sl a priori}
unknown function $A(\mathbf{k})$.

Injecting this ansatz in Schr\"odinger's equation $(E-H)|\Psi\rangle =0$,
and projecting onto the molecular subspace and the atomic subspace
respectively, one finds that Schr\"odinger's equation is satisfied
when $A$ and $\betab$ satisfy
\bea
(E-E_{\mathrm{mol}})\betab  + 2 \Lambda \int \frac{d^3k}{(2\pi)^3}
A(\mathbf{k})\chib^*(\mathbf{k}) &=& 0 \\
\left(E-\frac{\hbar^2 k^2}{m}\right) A(\mathbf{k}) + \Lambda \betab\cdot
\chib(\mathbf{k}) &=& 0
\label{eq:2b2}
\eea

Equation (\ref{eq:2b2}) does not specify $A(\mathbf{k})$ in a unique way,
for a positive energy $E$,
since $E-\hbar^2 k^2/m$ vanishes for some values of $\mathbf{k}$.
To obtain the scattering state of two atoms, one takes
a more specific form of the ansatz,
corresponding to the superposition in the open channel
of an incoming wave of wavevector $\mathbf{k}_0$ and a purely outgoing scattered wave,
\be
A(\mathbf{k}) = (2\pi)^3 \delta(\mathbf{k}-\mathbf{k}_0) 
-\Lambda \frac{\betab\cdot\chib(\mathbf{k})}{E+i0^+ -\frac{\hbar^2 k^2}{m}}.
\label{eq:ansatz2c}
\ee
Here $E=\hbar^2 k_0^2/m \geq 0$ is the total energy of the scattering state.

The general scattering theory \cite{Taylor} relates the scattering state to the incoming
state $|\Psi_0\rangle$ by $|\Psi\rangle = (1+G_0 T)|\Psi_0\rangle$
where $T$ is the $T$-matrix and $G_0$ the resolvent of the non-interacting Hamiltonian.
From this identity it is then apparent that
the matrix element of the $T$-matrix in Fourier space is related to
the numerator of the last term of Eq.(\ref{eq:ansatz2c}):
\be
\langle\mathbf{k}|T(E+i0^+)|\mathbf{k}_0\rangle = -\Lambda \betab\cdot\chib(\mathbf{k}).
\ee
From the known relation between the scattering amplitude and the $T$-matrix \cite{Taylor},
we get the reduced scattering amplitude
\be
f(k_0)=  \frac{-m k_0^2 e^{-k_0^2 b^2}/(4\pi\hbar^2)} 
{\frac{3(E-E_{\mathrm{mol}})}{2\Lambda^2}
-\int \frac{d^3k}{(2\pi)^3}\, \frac{k^2 e^{-k^2 b^2}}{E+i0^+ -\frac{\hbar^2 k^2}{m}}}.
\label{eq:rsa}
\ee
The choice of the Gaussian envelope in $\chib(\mathbf{k})$
allows an explicit expression for the scattering amplitude.
After complexification of $k_0$ by analytic continuation,
setting $k_0=i q_0$, $q_0>0$, we obtain
\bea
\frac{1}{f(i q_0)} &= &
\frac{4\pi}{q_0^2 e^{q_0^2 b^2}}
\left[-\frac{3\hbar^4}{2m^2\Lambda^2} (q_0^2+m E_{\rm mol}/\hbar^2) \right.
\nonumber\\
&&\left. +\int \frac{d^3k}{(2\pi)^3} \frac{k^2 e^{-k^2 b^2}}{q_0^2+k^2}\right] 
\label{eq:fjolie}
\\
&=& 
e^{-q_0^2 b^2} \left[\frac{1+q_0^2 b^2}{q_0^2 \mathcal{V}_s}-\alpha\right]
+q_0 \mathrm{erfc}\,(q_0 b)
\label{eq:f_expl}
\eea
where $\mathrm{erfc}$ is the complementary error function that tends to unity in zero.
With this complexification technique, it is straightforward to identify the
parameters $\mathcal{V}_s$ and $\alpha$ appearing in the low-$k$ expansion
(\ref{eq:lkeu}) and to get the explicit expressions:
\bea
\label{eq:Vse}
\frac{1}{\mathcal{V}_s} &=& \frac{1}{2\pi^{1/2} b^3} -\frac{6\pi\hbar^2}{m\Lambda^2} E_{\rm mol}
\\
\alpha &=& \frac{b^2}{\mathcal{V}_s} + \alpha_{\rm res}
\label{eq:alphae}
\\
\alpha_{\rm res} &=& \frac{1}{\pi^{1/2} b} + \frac{6\pi\hbar^4}{m^2\Lambda^2}.
\label{eq:alpha_res}
\eea
This illustrates the fact that one can tune $\mathcal{V}_s$
to $-\infty$ or $+\infty$ by shifting the molecular energy $E_{\rm mol}$ 
(in practice with a magnetic field $B$) 
around the value $E_{\rm mol}^0$ such
that the right hand side of (\ref{eq:Vse}) vanishes,
$E_{\rm mol}-E_{\rm mol}^0 \simeq \mu (B-B_0)$.

We have introduced
the convenient quantity $\alpha_{\rm res}$, which is the value of $\alpha$
exactly on the Feshbach resonance.
We see that $\alpha_{\rm res}$
depends on the inter-channel coupling $\Lambda$,
and is bounded from below by the inverse of the potential range,
within a numerical factor depending on the details of the model,
here $1/\pi^{1/2}$.
In principle, $\alpha_{\rm res}$ can take any possible value above this
limit, depending on the value of the interchannel coupling $\Lambda$; in
practice, of course, $\Lambda$ is not easily tunable so $\alpha_{\rm res}$
is fixed for a given experimental configuration.

By a direct generalization of a well established $s$-wave terminology,
we may classify the $p$-wave Feshbach resonances as 
\begin{itemize}
\item a broad resonance ($\Lambda \gg \hbar^2 b^{1/2}/m$):  $\alpha_{\rm res}\sim 1/b$
\item a narrow resonance ($\Lambda < \hbar^2 b^{1/2}/m$): $\alpha_{\rm res} \gg 1/b$.
\end{itemize}
We recall that this terminology can be motivated as follows:
If one assumes that $E_{\mathrm{mol}}$ is an affine
function of the magnetic field $B$ with a slope $\mu$, and that $\mathcal{V}_s=\mathcal{V}_s^{\mathrm bg}
[1-\Delta B/(B-B_0)]$ in a more complete theory including the fact that $\mathcal{V}_s$
takes a finite value $\mathcal{V}_s^{\mathrm bg}$ far from the resonance (due to the direct
interaction in the open channel, neglected here) and presumably of the
order of $b^3$, one finds a resonance width
\be
\mu \Delta B = \frac{m\Lambda^2}{6\pi\hbar^2 \mathcal{V}_s^{\mathrm bg}}.
\label{eq:mudelta}
\ee
It remains to compare this resonance width to the `natural' energy scale
$\hbar^2/m b^2$ to obtain the above mentioned terminology.

The last point to discuss for the two-body problem is the existence
or not of a two-body bound state in the open channel. We shall refer to
such a bound state as a dimer, in order not to confuse
it with the molecular state in the closed channel.
Mathematically, such  a dimer is a zero of $1/f(i q_0)$ with $q_0 >0$. 
The expression in between square brackets
in the right hand side Eq.(\ref{eq:fjolie}) 
is a decreasing function of $q_0$ that tends to $-\infty$ 
for $q_0\to +\infty$. Hence there exists at most one dimer in our
model Hamiltonian. There exists one if and only
if the expression between square brackets is positive in $q_0=0$, that is if and only if
$\mathcal{V}_s>0$ \cite{general}.

When a dimer is present, one can express analytically its wavefunction $\phi(\mathbf{r})$
in the open channel, in terms of exponential and erfc functions,
and one can calculate the occupation probability of the closed channel,
$p_{\rm closed}=|\betab|^2$ after
proper normalization of $|\Psi\rangle$ in the center of mass
frame \cite{aide}:
\be
|\betab|^2+ 2\int \frac{d^3k}{(2\pi)^3} |A(\mathbf{k})|^2 =1.
\ee
An equivalent way to obtain $|\betab|^2$ is to calculate the large $r$ behavior of 
$\phi(\mathbf{r})$, which is proportional to $\beta$ and which is related to the normalization
factor $\mathcal{N}$ in Eq.(\ref{eq:phidim}), and then to use the general relation
Eq.(\ref{eq:norm}) \cite{check_ok}. Both ways lead to the expression
\be
\label{eq:pclosed}
\frac{1}{p_{\rm closed}} = 
\frac{m^2\Lambda^2}{6\pi\hbar^4}\frac{e^{q_{\rm dim}^2 b^2}}{|\mathcal{N}|^2}.
\ee
The value of $p_{\rm closed}$ for an infinite scattering volume can be cast
in the very simple forms
\bea
\label{eq:p_closed_infini}
p_{\rm closed}^{\rm res} &=& \frac{6\pi\hbar^4}{m^2\Lambda^2}
\alpha_{\rm res}^{-1} \\
&=& 1-\frac{1}{\pi^{1/2} \alpha_{\rm res}b}.
\label{eq:p_closed_infini_pas_univ}
\eea
The expression (\ref{eq:p_closed_infini}) is quite remarkable since it is 
`universal': It does not involve the interaction range $b$ and, 
as we have checked,
it is not specific to the choice of a Gaussian cut-off function in
$\chib(\mathbf{k})$. It was already derived in \cite{Chevy}, see the unnumbered
equation following equation (9) of that reference.
In the vicinity of the resonance, we see on the expression 
(\ref{eq:p_closed_infini_pas_univ})
that, in the dimer wavefunction, the closed channel 
is strongly occupied for a narrow resonance
and is weakly occupied for a broad resonance; $p_{\rm closed}$  tends to zero
in the broad resonance limit.

To conclude this review of the two-body aspects,
we point out a striking property of the dimer, very different from the usual
$s$-wave case:
In the limit $\mathcal{V}_s/b^3\to +\infty$,
we find that the dimer wavefunction $\phi(\mathbf{r})$
has a well defined, non-zero limit, tending to zero as $O(1/r^2)$ at large $r$.
This can be directly seen in momentum space: for $q_{\rm dim}=0^+$, the function
$A(\mathbf{k})$ is $O(1/k)$ at low $k$, which is indeed square integrable around the
origin $\mathbf{k}=0$.
In other words, at the threshold for the formation of the dimer, the dimer
wavefunction is a well defined non-zero and square integrable function.

\section{Solution of the three-body problem}
\label{sec:solution}

This is the central section, where we solve the three-body problem
within the two-channel model close to a $p$-wave resonance. 
The mathematical structure of the model, with a single molecular state occupied
in the closed channel and no interaction potential in the open channel, 
is such that the three-body problem is amenable to an integral equation
for a one-body `wavefunction'. This integral equation becomes easily solvable numerically
if one further uses the rotational symmetry of the Hamiltonian.
We then obtain predictions for three physical situations,
(i) the existence of three-body bound states, that is of trimers,
(ii) the scattering of an atom
on a dimer, and (iii) the scattering of three atoms, leading to recombination
processes, that is to the formation of a weakly bound dimer and a free atom.

\subsection{Derivation of an integral equation}

We start with the most general ansatz for the three-body problem in the 
center of mass
frame, that is for a zero total momentum.
Because of the conversion of pairs of atoms into molecules and {\sl vice-versa},
the three-body ansatz is a coherent superposition of three fermions
(all three atoms in the open channel) and of one molecule plus one fermion 
(one atom in the open channel and two atoms tightly bound in a molecule in
the closed channel):
\bea
|\Psi\rangle &=& \int\frac{d^3K}{(2\pi)^3} \sum_\gamma \beta_\gamma(\mathbf{K})
b_{\gamma,\mathbf{K}}^\dagger a_{-\mathbf{K}}^\dagger |0\rangle \nonumber \\
&+&\int \frac{d^3k d^3K}{(2\pi)^6}
A(\mathbf{K},\mathbf{k})
a^\dagger_{\frac{1}{2}\mathbf{K}+\mathbf{k}} 
a^\dagger_{\frac{1}{2}\mathbf{K}-\mathbf{k}} 
a^\dagger_{-\mathbf{K}} |0\rangle.
\label{eq:ansatz}
\eea
The one molecule plus one fermion part is parameterized by three one-body `wavefunctions'
$\beta_\gamma$, here in Fourier space; we shall derive an integral equation
for them.
The three fermion part $A$ can be parameterized by two 
Jacobi-like coordinates in momentum space since the total momentum is zero.
For pure convenience, we impose that $A(\mathbf{K},\mathbf{k})$ is an odd function of
$\mathbf{k}$, to reduce the number of terms involving $A$ in the integral
equation for $\betab$.

We inject the general ansatz for $|\Psi\rangle$ in Schr\"odinger's
equation 
$(E-H)|\Psi\rangle=0$, where the total energy $E$ is at this stage of
arbitrary sign.
Projecting Schr\"odinger's equation on the subspace with one molecule and one fermion gives
an equation for $\betab$ with the function $A$ appearing in a source term:
\bea
\left[E-E_{\mathrm{mol}} - \frac{3\hbar^2 K^2}{4m} \right] \betab(\mathbf{K})
+2\Lambda \int \frac{d^3k}{(2\pi)^3} \chib^*(\mathbf{k}) \nonumber \\
\times\left[A(\mathbf{K},\mathbf{k}) 
+ 2 A\left(-\frac{1}{2} \mathbf{K}+\mathbf{k},
-\frac{3}{4} \mathbf{K}-\frac{1}{2} \mathbf{k}\right)\right] 
=0.
\label{eq:mol_atom}
\eea
Projecting Schr\"odinger's equation on the subspace with three fermions leads to:
\bea
\int \frac{d^3K d^3k}{(2\pi)^6}
\left\{\left[E-\frac{\hbar^2 }{m}\left(\frac{3}{4}K^2 + k^2\right) \right]
A(\mathbf{K},\mathbf{k}) \right.\nonumber \\
+\Lambda\, \betab(\mathbf{K})\cdot \chib (\mathbf{k})\Big\}
a^\dagger_{\frac{1}{2}\mathbf{K}+\mathbf{k}}
a^\dagger_{\frac{1}{2}\mathbf{K}-\mathbf{k}}
a^\dagger_{-\mathbf{K}} |0\rangle =0.
\label{eq:brute}
\eea
This equation is satisfied for the choice
\be
A(\mathbf{K},\mathbf{k}) =  A_0(\mathbf{K},\mathbf{k})
-\frac{\Lambda \betab(\mathbf{K})\cdot \chib (\mathbf{k})}
{E+i0^+-\frac{\hbar^2 }{m}\left(\frac{3}{4}K^2 + k^2\right)}.
\label{eq:A}
\ee
For a positive total energy $E>0$:
$A_0$ represents a possible incoming wave of three free atoms, 
and it is an eigenstate of the kinetic energy operator
in the center of mass frame with energy $E$; in presence of such an incoming
free wave, the second term
in $A$ represents the scattered wave in the open channel, which is guaranteed to be outgoing by 
the standard substitution $E\to E+i 0^+$. As we have imposed the convention
that $A(\mathbf{K},\mathbf{k})$ should be an odd function of $\mathbf{k}$,
one has to apply the same convention to $A_0(\mathbf{K},\mathbf{k})$;
note that the last term of Eq.(\ref{eq:A}) is automatically 
an odd function of $\mathbf{k}$, since $\chi(\mathbf{k})$ is.
For a negative total energy $E<0$, the expression between
square brackets in (\ref{eq:brute}) can not vanish, $A_0\equiv 0$ and the $+i0^+$ in
the denominator can be omitted.

Injecting Eq.(\ref{eq:A}) in Eq.(\ref{eq:mol_atom}), we obtain an integral equation
for $\betab$. The term $A(\mathbf{K},\mathbf{k})$ of Eq.(\ref{eq:mol_atom})
gives a contribution
simply proportional to $\betab(\mathbf{K})$, with a $K$-dependent factor; 
collecting it with the factor
in between square brackets in the first term of Eq.(\ref{eq:mol_atom}) 
gives a $K$ dependent expression that can be recognized
as being proportional (with a $K$ dependent factor)
to the inverse of the scattering amplitude of two atoms at the energy
\be
E_{\rm rel} = E - \frac{3 \hbar^2 K^2}{4m}\equiv \frac{\hbar^2 k_{\rm rel}^2}{m},
\label{eq:Erel}
\ee
with the determination $k_{\rm rel}\geq 0$ for $E_{\rm rel}\geq 0$
and $k_{\rm rel}/i > 0$ for $E_{\rm rel} <0$. This relation can be seen as a consequence
of the Jacobi-like parameterization of the momenta of the three fermions that we
have used in (\ref{eq:ansatz}): if three free fermions of total energy $E$
have momenta $\pm\mathbf{k}_{\rm rel}+\mathbf{K}/2$ and $-\mathbf{K}$, then the
modulus $k_{\rm rel}$ will obey (\ref{eq:Erel}).

We finally obtain the general integral equation for the $\beta_\gamma(\mathbf{K})$:
\begin{widetext}
\bea
\frac{k_{\rm rel}^2 e^{-k_{\rm rel}^2 b^2}}{3 f(k_{\rm rel})} 
\, \betab(\mathbf{K}) + 
8\pi \int \frac{d^3k}{(2\pi)^3} \chib^*\left(\frac{1}{2}\mathbf{K}+\mathbf{k}\right)
\frac{\betab(\mathbf{k})\cdot \chib\left(\mathbf{K}+\frac{1}{2}\mathbf{k}\right)}
{K^2+k^2+\mathbf{K}\cdot\mathbf{k}-m(E+i0^+)/\hbar^2}
= \nonumber \\
\frac{4\pi\hbar^2}{m\Lambda} \int \frac{d^3k}{(2\pi)^3} \chib^*(\mathbf{k}) 
\left[A_0(\mathbf{K},\mathbf{k}) + 2 A_0\left(-\frac{1}{2} \mathbf{K}+\mathbf{k},
-\frac{3}{4} \mathbf{K}-\frac{1}{2} \mathbf{k}\right)\right]
\label{eq:integrale}
\eea
\end{widetext}
In what follows we shall solve this integral equation for various physical situations.
(i) In the search for trimers, one assumes an energy $E$ below zero and below the dimer energy
(if there exits a dimer); then $A_0\equiv 0$ and $\betab(\mathbf{K})$ 
is not subjected
to any specific boundary condition.
(ii) In the low energy scattering of an atom on a dimer, the energy is above the dimer
energy but still negative; then $A_0\equiv 0$ 
and one has to introduce a specific
ansatz for $\betab(\mathbf{K})$ to enforce the boundary conditions corresponding
to such a scattering experiment. (iii) In the scattering of three incoming atoms,
the total energy is now non-negative so that $A_0\neq 0$; we shall assume
that this scattering experiment is performed for $\mathcal{V}_s>0$
so that there exists a dimer in the two-body problem, 
which can be formed by a recombination event
in the three-body scattering; then one introduces an
ansatz for $\betab(\mathbf{K})$ describing the presence of a purely outgoing
wave of such a dimer (with an opposite momentum atom).

\subsection{Symmetry sectors from rotational and parity invariance}
\label{subsec:ssfrapi}

Formally
Eq.(\ref{eq:integrale}) is an equation for a spinor
$\betab(\mathbf{K})$, with an internal spin $S_{\rm mol}=1$; here this internal
spin corresponds to the rotational degrees of freedom of the molecule (in the closed
channel); the orbital variable $\mathbf{K}$ here corresponds to the 
relative atom-molecule momentum.
The homogeneous part of Eq.(\ref{eq:integrale}) is invariant by a simultaneous rotation
of the spin and orbital variables of the spinor. The total momentum $J$, obtained
by addition of the spin $S_{\rm mol}$ and the orbital angular momentum $L$, is therefore a
good quantum number.
In this paper, we shall restrict to the manifold $J=1$, which can be obtained from 
$L=0$ plus $S_{\rm mol}=1$, or $L=1$ plus $S_{\rm mol}=1$, or $L=2$ plus $S_{\rm mol}=1$.
In addition, the homogeneous part of Eq.(\ref{eq:integrale}) is invariant by parity
(combining the parity on the spin variables and on the orbital variables).
This decouples the $J=1$ manifold in two sectors, 
\begin{itemize}
\item even sector: $L=1$ plus $S_{\rm mol}=1$
\item odd sector: $L=0$ plus $S_{\rm mol}=1$ and $L=2$ plus $S_{\rm mol}=1$
\end{itemize}

Applying the standard algebra of addition of angular momenta, we obtain the following
ansatz in the odd sector,
\be
\betab(\mathbf{K}) = B_{L=0}(K) \mathbf{e}_z - B_{L=2}(K) 
\frac{\mathbf{K}\cdot\mathbf{e}_z} {K^2} \mathbf{K},
\label{eq:odd}
\ee
where $\mathbf{e}_z$ is the unit vector along $z$ axis.
This ansatz corresponds to a total angular momentum $J=1$ with vanishing angular
momentum component along $z$, $m_J=0$. Considering the other components
$m_J=\pm 1$, or equivalently the states with vanishing angular momentum
component along $x$ and along $y$ respectively, would lead to equivalent 
results, as guaranteed by the rotational invariance of the Hamiltonian.

Similarly, we take as ansatz in the even sector 
\bea
\betab(\mathbf{K}) & =&  
\frac{B_{L=1}(K)}{K}\,  \mathbf{K}\wedge\mathbf{e}_x \nonumber \\
&=& \frac{B_{L=1}(K)}{K} 
\left[(\mathbf{K}\cdot\mathbf{e}_z) \mathbf{e}_y -(\mathbf{K}\cdot\mathbf{e}_y) \mathbf{e}_z\right]
\label{eq:even}
\eea
which corresponds to the even state with $J=1$ and vanishing angular momentum component
along $x$ axis.
After some calculations Eq.(\ref{eq:integrale}) can be turned into an integral equation
for $B_{L=1}$ (in the even sector) or into coupled integral equations for 
$B_{L=0}$ and $B_{L=2}$ (in the odd sector), as detailed in the Appendix
\ref{appen:ad}. The remaining unknown functions depend
on a single real variable $K$ so that a numerical solution is reasonable.

\subsection{Existence of weakly bound trimers}
\label{subsec:ewbt}

We investigate here the existence of three-body bound states, that is of trimers,
in our model Hamiltonian.
These trimers have of course a negative total energy $E$. If one
is on the $\mathcal{V}_s>0$ side of the resonance, where a dimer of energy $-E_{\rm dim}$
exists, one further has $E<-E_{\rm dim}$ to have stability of the trimers
with respect to dissociation into an atom and a dimer; if this condition was not satisfied,
the trimers would not exist as true stationary states but would 
rather be resonances
in the atom-dimer scattering process.

These constraints on the energy have the following mathematical consequences.
Since $E<0$, the source term $A_0$ in (\ref{eq:A}) is identically zero, so that
Eq.(\ref{eq:integrale}) becomes homogeneous.
Since $E<-E_{\rm dim}$, on the side $\mathcal{V}_s>0$ of the resonance, the scattering
amplitude $f(k_{\rm rel})$ in the denominator of the first term
of Eq.(\ref{eq:integrale}) is non zero for all $\mathbf{K}$ and
the linear operator $L(E)$ representing the integral equation has a smooth action
over the class of regular $\betab(\mathbf{K})$ functions.
Numerically, one can then discretize the variable $K$ with no particular care,
and approximate $L(E)$ by a matrix.
The existence of a trimer  corresponds to a non-zero-dimension kernel of the operator $L(E)$;
in practice, we look for the values of $E$ such that the approximating 
matrix has a vanishing eigenvalue.
The explicit form of $L(E)$
for the ansatz in the even and odd sector can be deduced from
the appendix \ref{appen:ad}. In the same appendix, it is also explained how to normalize
the state vector of the trimer.

\begin{figure}[htb]
\includegraphics[width=8cm,clip]{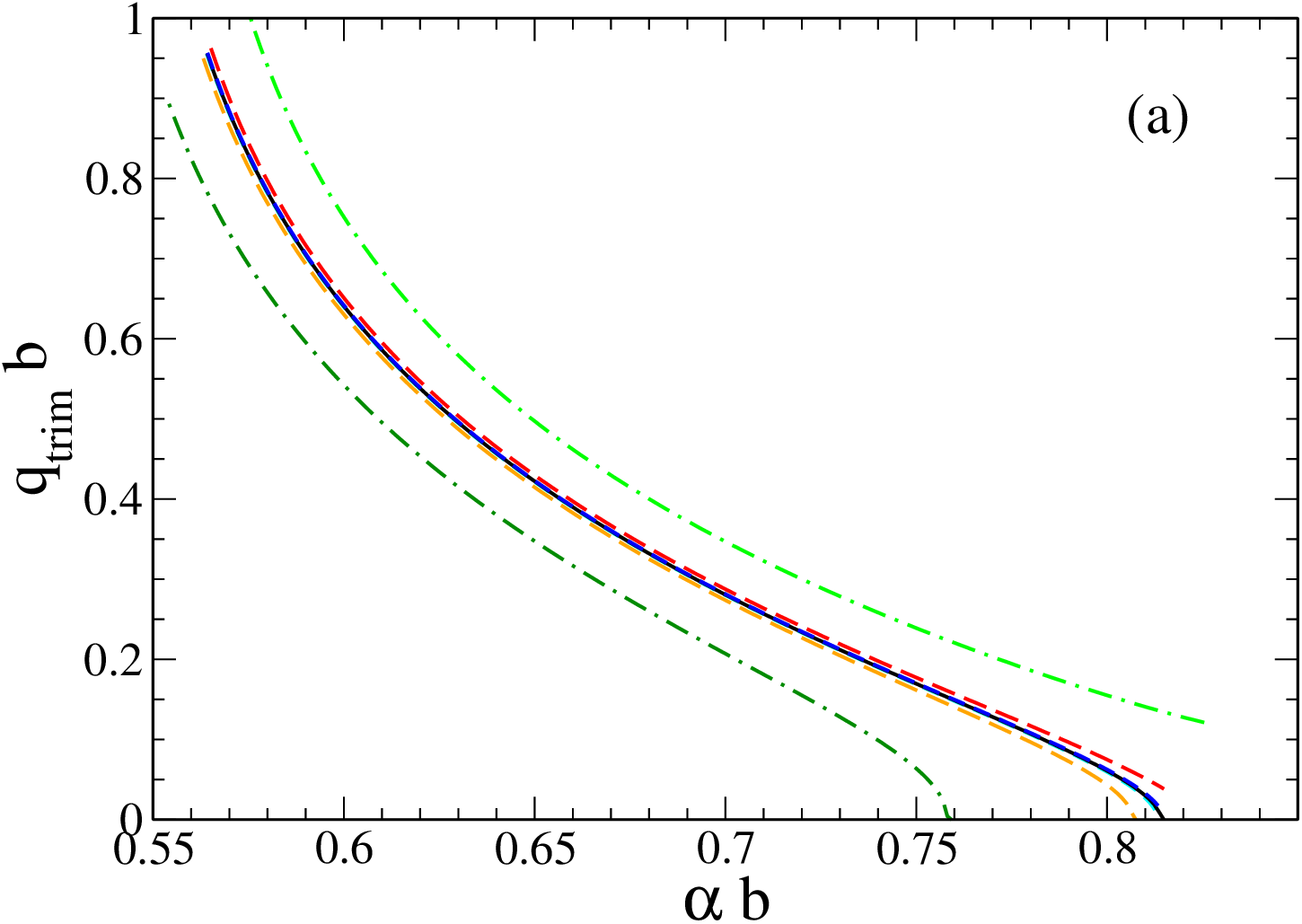}
\includegraphics[width=8cm,clip]{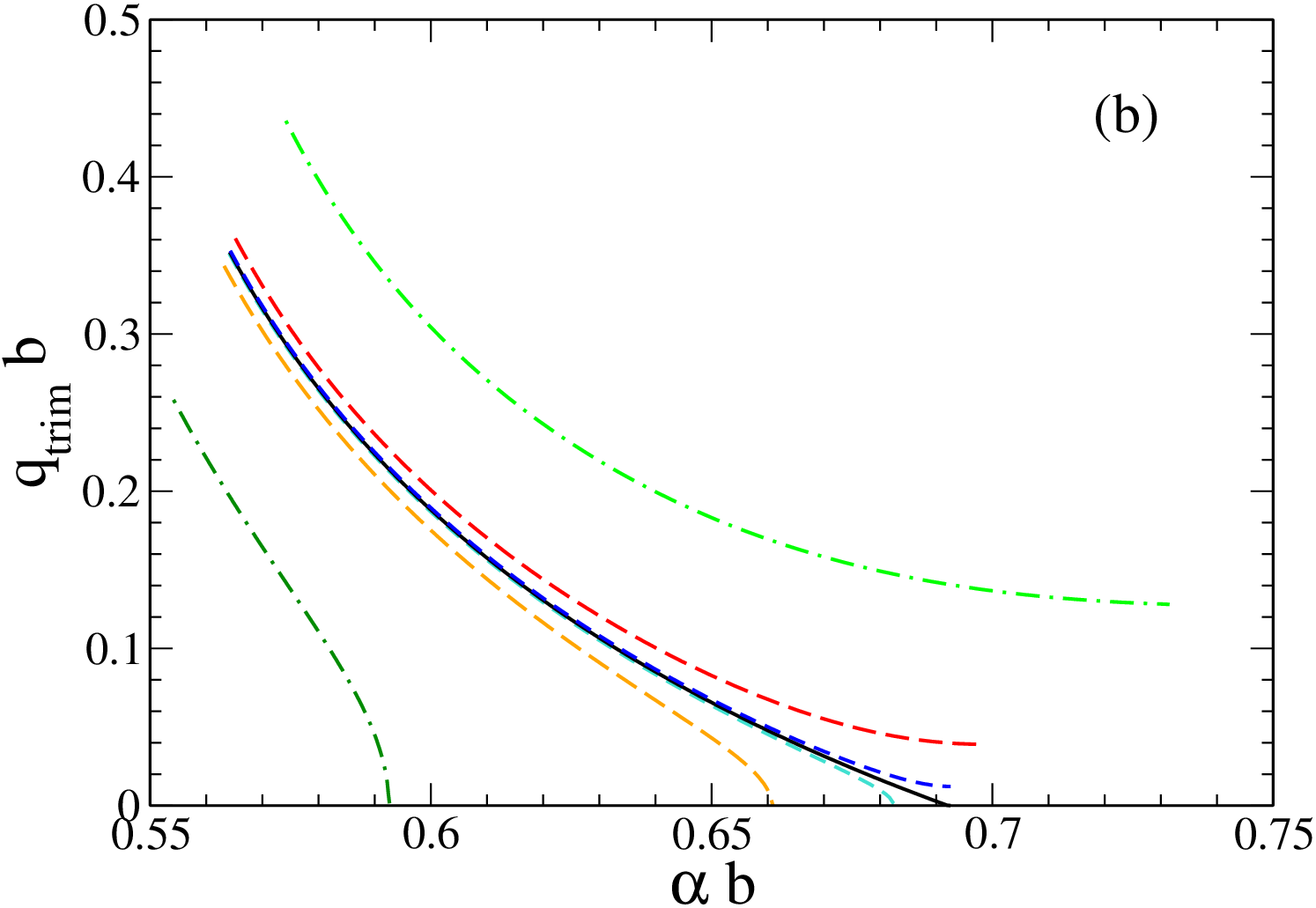}
\caption{(Color online)
For fixed values of the scattering volume $\mathcal{V}_s$, parameter 
$q_{\rm trim}$  of the trimer
(when it exists) as a function of $\alpha b$; $q_{\rm trim}$ is related
to the negative energy $-E_{\rm trim}$ of the trimer by
$E_{\rm trim}=\hbar^2 q_{\rm trim}^2/m$.
(a) Even sector, (b) odd sector, as defined in subsection
\ref{subsec:ssfrapi}.
Solid line (black): $|\mathcal{V}_s|/b^3=\infty$. Above the solid line, positive
values of $\mathcal{V}_s$:
short dashed line (blue): $\mathcal{V}_s=10^4 b^3$;
dashed line (red): $\mathcal{V}_s=10^3 b^3$; 
dashed-dotted (green): $\mathcal{V}_s=100 b^3$.
Below the solid line, negative values of $\mathcal{V}_s$: 
short dashed line (light blue): $\mathcal{V}_s=-10^4 b^3$;
dashed line (orange): $\mathcal{V}_s=-10^3 b^3$; 
dashed-dotted (dark green): $\mathcal{V}_s=-100 b^3$.
At the threshold for the existence of the trimer as a true bound state, on the
$\mathcal{V}_s>0$ side of the resonance, where a dimer exists, the trimer binding
energy vanishes, so that the energy of the trimer
coincides with the one of the dimer, and
$q_{\rm trim}=q_{\rm dim}$ (see text).
\label{fig:trimer}}
\end{figure}

\begin{figure}[htb]
\vspace{2mm}
\includegraphics[width=8cm,clip]{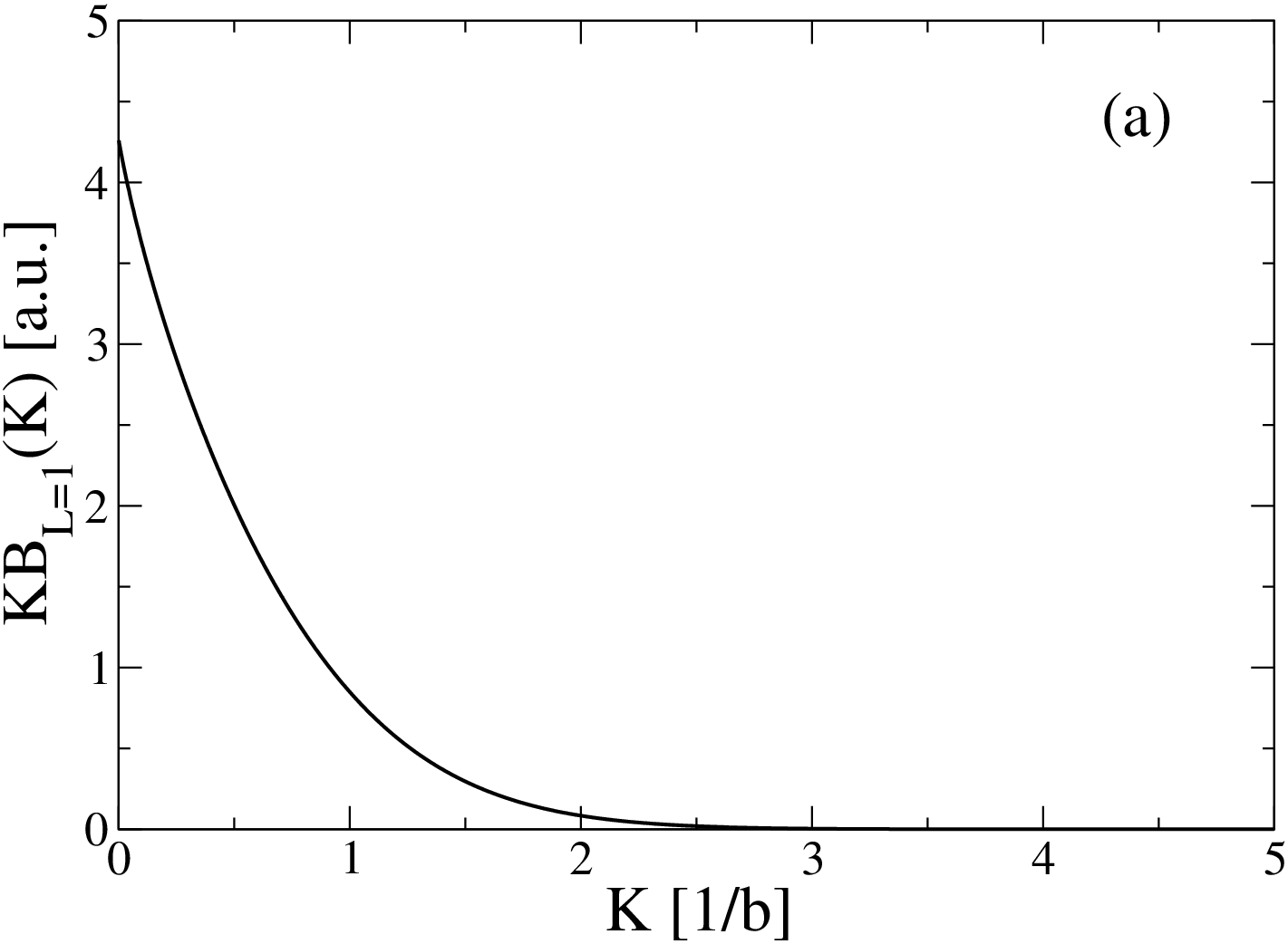}
\includegraphics[width=8cm,clip]{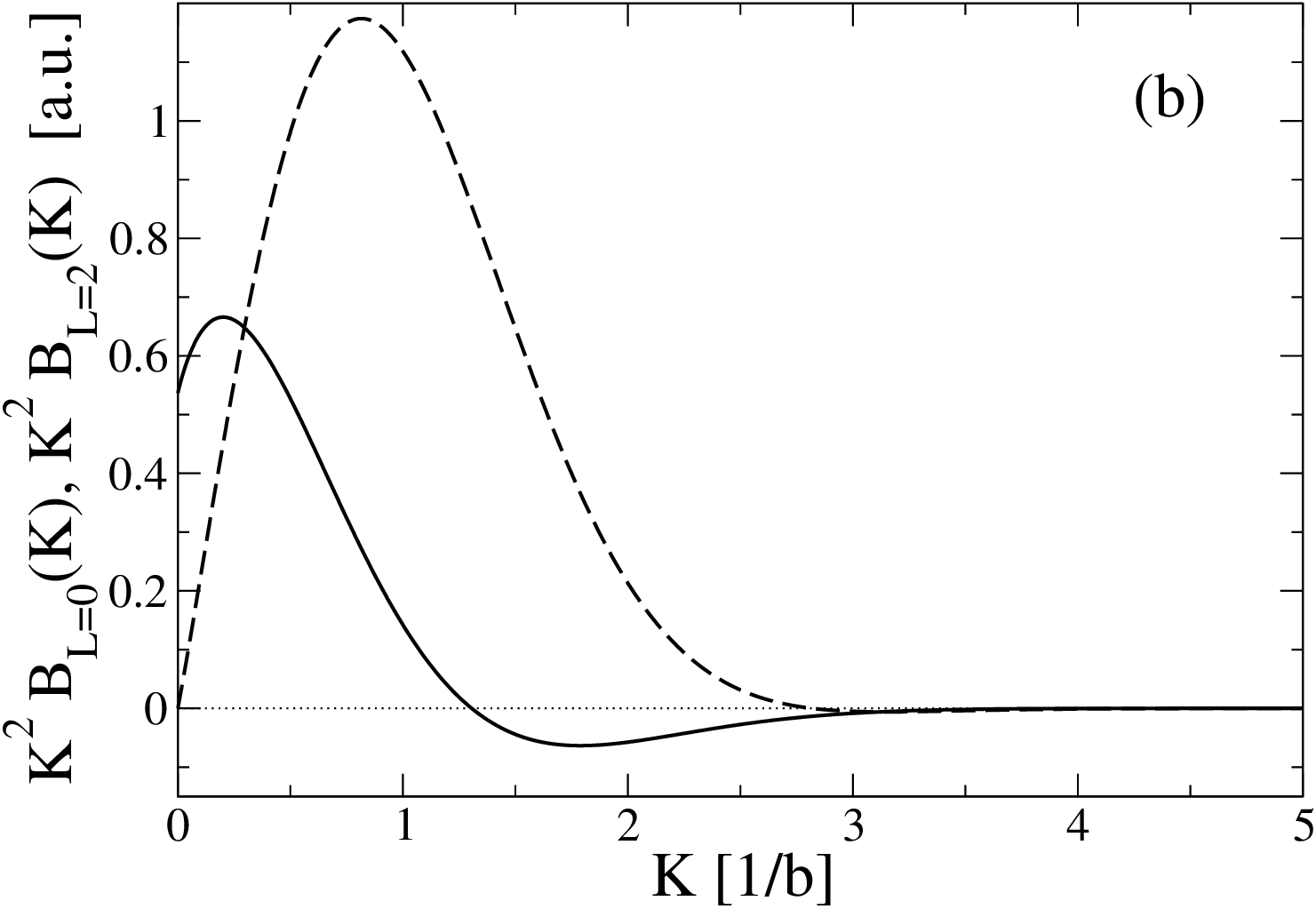}
\caption{For $\mathcal{V}_s=\infty$, and $\alpha_{\rm res} =\alpha_{\rm th}$ 
[right on the thresholds for the
existence of a trimer, see Eqs.(\ref{eq:ath_odd},\ref{eq:ath_even})], 
$K$-dependence of the functions 
(a) $B_{L=1}$ (even sector), 
(b) $B_{L=0}$ (solid line), $B_{L=2}$
(dashed line) (odd sector), for the trimers.
To avoid diverging functions, these functions were multiplied by $K$ in (a)
and by $K^2$ in (b). The normalization is arbitrary.
\label{fig:fot}}
\end{figure}

For values of $|\mathcal{V}_s|\gg b^3$, we have found either zero or one trimer
in each symmetry sector (with threefold rotational degeneracy when the trimer exists).
The energy of the trimer is written as $-\hbar^2q_{\rm trim}^2/m$.
Then $q_{\rm trim}$ 
as a function of $\alpha b$ is given in Fig.\ref{fig:trimer}, for the
even and the odd sectors.
We found no evidence of Efimov effect: in each symmetry sector,
we found at most one trimer, and there is no
oscillation of the $\betab(\mathbf{K})$ with $K$ as a function of $K$,
see Fig.\ref{fig:fot}.
 
We note that, in real experiments with atoms, these trimers may acquire a finite lifetime, 
due to the formation of deeply bound dimers
by three-body collisions. This process is not contained in our Hamiltonian,
since $H$ does not support deeply bound dimers for $|\mathcal{V}_s|\gg b^3$;
its rate is estimated by a simple recipe in subsection \ref{subsec:loss_trim}.

\subsection{Atom-dimer scattering}
\label{subsec:ads}

We consider here the scattering problem of an atom on a dimer, which corresponds to
the positive $\mathcal{V}_s$ side of the resonance and to a total energy
$E\geq -E_{\rm dim}$.
For simplicity, we restrict to the low energy limit of this scattering,
with a relative kinetic energy of the incoming atom and the dimer
much smaller than the binding energy of the dimer:
\be
E+E_{\rm dim} \ll E_{\rm dim}.
\ee
As a consequence, the total energy is negative,
so that energy conservation prevents the dimer from being dissociated by 
the interaction
with the incoming atom and the scattering is elastic.
Furthermore, a multipolar expansion can be performed in terms of
the atom-dimer relative orbital momentum.
In the mathematical limit of a vanishing kinetic energy,
the atom-dimer incoming wave is a $s$-wave and 
the scattering is characterized by
the atom-dimer scattering length $a_{\rm ad}$ that we shall calculate.
To next order of the multipolar expansion the atom-dimer incoming wave 
is a $p$-wave and we shall calculate a corresponding atom-dimer scattering
volume $\mathcal{V}_s^{\rm ad}$.

The property of elastic scattering at $E<0$ rigorously holds for
the model Hamiltonian Eq.(\ref{eq:hamil}), since we have shown that it admits at
most one dimer state (with rotational degeneracy).
Reality with atoms goes beyond this model Hamiltonian: there exist in general
deeply bound dimers, which can make the atom-dimer scattering inelastic even at arbitrarily
low relative kinetic energy.
The corresponding three-body loss rate is estimated in subsection \ref{subsec:loss_ad}.

\noindent{\sl $S$-wave atom-dimer scattering:}
Since we have assumed a negative total energy $E<0$,
there cannot be a free incoming three-atom state so that $A_0\equiv 0$ in Eq.(\ref{eq:A}).
In the center of mass frame, the incoming state is an atom 
impinging on a dimer with vanishing kinetic energy; in the subspace with one atom
and one molecule in the closed channel, this corresponds to a relative orbital
angular momentum $L=0$, that is to a total momentum $J=1$ since the molecule
is of spin unity. According to subsection \ref{subsec:ssfrapi}, the incoming state is in the odd
sector. Mathematically, this scattering experiment corresponds to the
following splitting for $\betab$, into the sum of an incoming
wave of zero momentum (a delta distribution in $\mathbf{K}$ space)
and a scattered wave which is a regular function of $K$:
\be
\betab(\mathbf{K}) = (2\pi)^3\delta(\mathbf{K}) \, p_{\rm closed}^{1/2} \mathbf{e}_z
 + \betab^{\rm out}(\mathbf{K}).
\label{eq:with_delta}
\ee
This is of the form Eq.(\ref{eq:odd}), the delta being in the $L=0$ sector.
Note that the incoming dimer has a probability $p_{\rm closed}$
to be in the closed channel, so that the amplitude of the 
incoming wave for $\betab$, that is in the subspace of one atom and one molecule,
includes a factor $p_{\rm closed}^{1/2}$.

In practice, one injects the form Eq.(\ref{eq:with_delta}) into Eq.(\ref{eq:integrale}).
The $\delta(\mathbf{K})$ term gives a zero contribution
in the first term of the left hand side, 
since $\mathbf{K}=\mathbf{0}$
and $E=-E_{\mathrm{dim}}$ leads to $k_{\rm rel}=i q_{\rm dim}$ and $1/f(k_{\rm rel})=0$.
The $\delta(\mathbf{k})$ inserted in the second term of the left hand side 
of Eq.(\ref{eq:integrale}),
that is the integral term, produces a smooth source term in the left hand side,
\be
\mathbf{T}(\mathbf{K}) = 4\pi p_{\rm closed}^{1/2} (\mathbf{e}_z\cdot\mathbf{K})\mathbf{K}
\frac{e^{-5 b^2 K^2/8}}{K^2+mE_{\rm dim}/\hbar^2}.
\ee
One is left with a linear and 
inhomogeneous system for the vectorial function $\betab^{\rm out}(\mathbf{K})$,
which is then taken of the form Eq.(\ref{eq:odd}), with
coefficients $B_{L=0}^{\rm out}(K)$ and $B_{L=2}^{\rm out}(K)$.
The explicit form of the resulting system is derived in the 
appendix \ref{appen:ad}, and we obtain
\begin{widetext}
\be
D(K) \left(\begin{array}{c} B_{L=0}^{\rm out}(K) \\ B_{L=2}^{\rm out}(K)\end{array}\right)
+\frac{4}{\pi} \int_0^{+\infty} dk\, k^2
e^{-5(K^2+k^2)b^2/8}
M(K,k) \left(\begin{array}{c} B_{L=0}^{\rm out}(k) \\ B_{L=2}^{\rm out}(k)\end{array}\right)=
4\pi p_{\rm closed}^{1/2} \frac{K^2e^{-5b^2 K^2/8}}{K^2+q_{\rm dim}^2}
\left(\begin{array}{c} 0 \\ 1 \end{array}\right)
\label{eq:ieodd}
\ee
\end{widetext}
where we have introduced the diagonal part
\be
D(K) = \frac{k_{\rm rel}^2 e^{-k_{\rm rel}^2 b^2}}{3 f(k_{\rm rel})} 
\label{eq:defD}
\ee
and the two by two matrix $M(K,k)$ is given in the appendix.

Let us start with an intuitive presentation of the results.
We expect that, at low $K$, 
the scattered wave in the $L=0$ channel diverges as $1/K^2$, so that we set
\be
B_{L=0}^{\rm out}(K) \sim -p_{\rm closed}^{1/2} \frac{4\pi}{K^2}  a_{\rm ad}.
\label{eq:we_set}
\ee
In position space this indeed corresponds to the large $r$ behavior $1-a_{\rm ad}/r$, where
$r$ is the distance between the molecule and the atom, so that 
$a_{\rm ad}$ is indeed the atom-dimer scattering length.
In the channel $L=2$, the outgoing wave is expected to scale as $1/r^3$ at large $r$,
because of the centrifugal barrier; this corresponds to $B_{L=2}(K)$ having a finite limit
in $K=0$. 

\begin{figure}
\includegraphics[width=8cm,clip]{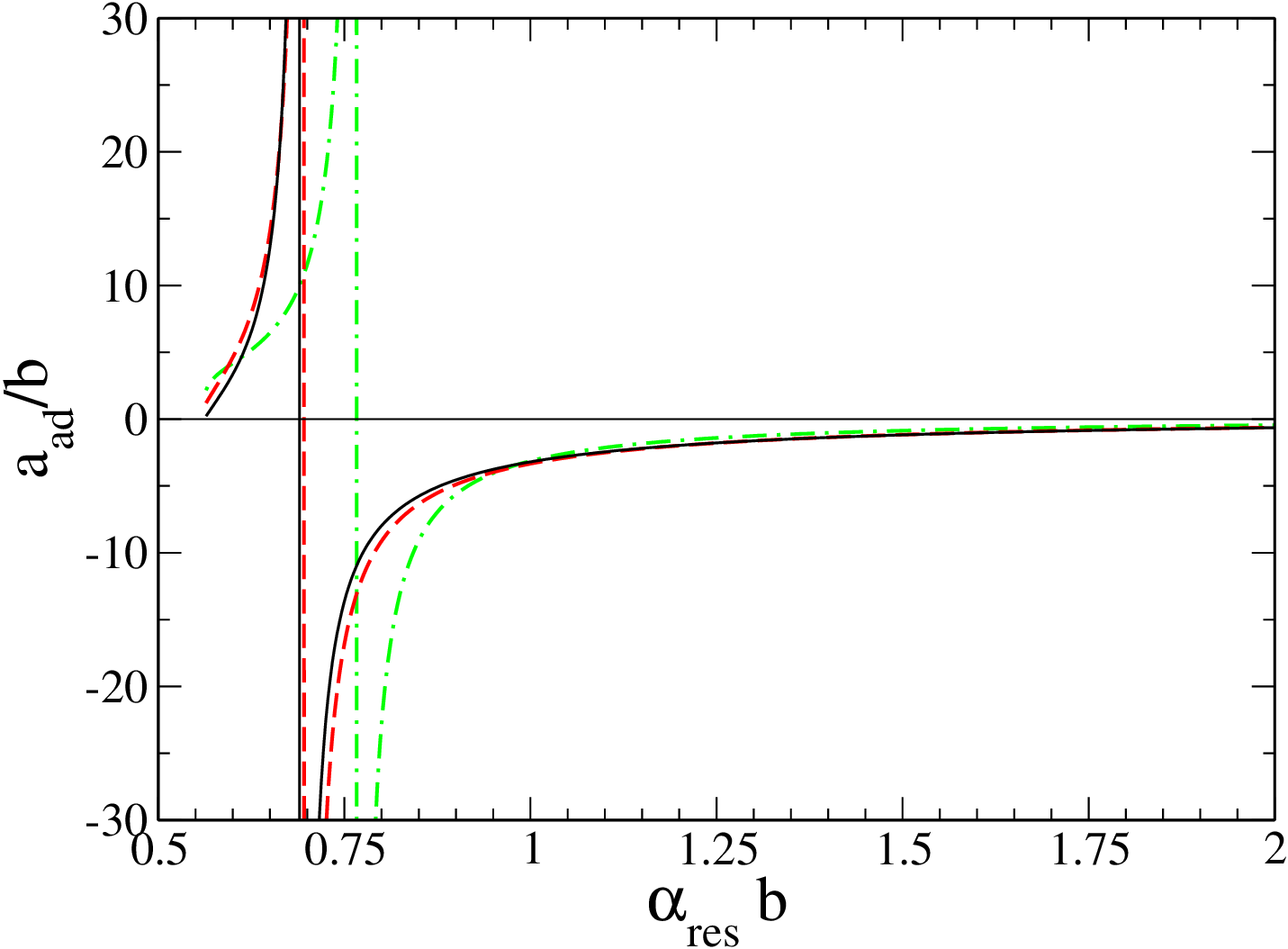}
\caption{(Color online)
Atom-dimer scattering length $a_{\rm ad}$ as a function of $\alpha_{\rm res}$
for fixed values of the scattering volume $\mathcal{V}_s$,
$\mathcal{V}_s=10^6 b^3$ (black solid line), $\mathcal{V}_s=10^3 b^3$ (red dashed line),
$\mathcal{V}_s=10 b^3$ (green dotted-dashed line). 
The divergence of $a_{\rm ad}$ coincides with the threshold of existence of a trimer
in the odd sector. In the limit of a broad Feshbach resonance $\alpha_{\rm res}b\to
1/\pi^{1/2}$, $a_{\rm ad}$ tends to $\approx 0.2b$.
\label{fig:ad}
}
\end{figure}

What typical values of $a_{\rm ad}$ can we expect~?
For $\mathcal{V}_s>0$ and much larger than $b^3$, 
the scattering amplitude of two atoms has a modulus $\leq 1/\alpha_{\rm res}$,
which is a small value at most of the order of $b$. For $k\simeq q_{\rm dim} \ll 
\alpha_{\rm res}$,
one finds that $|f(k)|\simeq 1/\alpha_{\rm res}$.
One may then expect intuitively
that $a_{\rm ad}$ weakly depends on $\mathcal{V}_s$,
and is at most of the order of $1/\alpha_{\rm res}$, that is at most $\simeq b$. 
This expectation is correct, see Fig.\ref{fig:ad}, except close to the 
threshold for the existence of a trimer in the odd sector, where $a_{\rm ad}$
diverges.

We now turn to a more rigorous analysis of the integral equation (\ref{eq:ieodd}).
The key ingredient is the low $K$ behavior 
of the various coefficients for $q_{\rm dim}>0$.
Consider the diagonal term $D(K)$. As we have already mentioned, $D(K)$ vanishes
in $K=0$; since here $k_{\rm rel}= i (q_{\rm dim}^2+3K^2/4)^{1/2}$, we see that
$k_{\rm rel}$ is an expandable function of $K$ which varies to second order in $K$.
The same conclusion holds for $D(K)$, which therefore vanishes quadratically 
in $K=0$; in the limit $q_{\rm dim} b \ll 1$ we find the simple result
\be
\lim_{K\to 0} \frac{D(K)}{K^2} \simeq \frac{\alpha_{\rm res}}{4}.
\label{eq:limDsurK2}
\ee
Consider next the coefficients of the matrix $M(K,k)$. From the explicit expressions
given in the appendix \ref{appen:ad}, we obtain for a fixed $k$:
\be
\lim_{K\to 0} M(K,k) = \frac{k^2/6}{k^2+q_{\rm dim}^2} 
\left(\begin{array}{rr} 1 & -1 \\ 0 & 0\end{array}\right).
\label{eq:limM}
\ee
Assuming that the functions $k^2 B_{L}^{\rm out}(k)$ are bounded, we find that 
$D(K) B_{L=0}^{\rm out}(K)$ has a finite limit in $K=0$, obeying the exact relation
\bea
&& \lim_{K\to 0}  D(K) B_{L=0}^{\rm out} (K) = \nonumber \\
&& -\frac{2}{3\pi} \int_0^{+\infty} dk 
\frac{k^4e^{-\frac{5}{8}b^2k^2}}{k^2+q_{\rm dim}^2} [B_{L=0}^{\rm out}
-B_{L=2}^{\rm out}](k).
\label{eq:exact_relation}
\eea
On the contrary, we find that the second line of the matrix $M(K,k)$ vanishes quadratically
for $K\to 0$, and the source term also vanishes quadratically in $K$,
so that $B_{L=2}^{\rm out}(K)$ has indeed a finite limit in $K=0$.

The existence of a well defined limit for $a_{\rm ad}$ in the large scattering volume 
can also be argued in simple terms.
All the coefficients in the integral equation (\ref{eq:ieodd}) have a well
defined limit for $q_{\rm dim}\to 0$. In particular, the diagonal
term in this limit assumes the simple form
\be
\lim_{\mathcal{V}_s/b^3\to +\infty} D(K) = \frac{K^2}{4}\left[\alpha_{\rm res} -h(K)\right]
\label{eq:atsf}
\ee
where $h(K)= qe^{q^2b^2}\mathrm{erfc}(qb)$, with $q=\sqrt{3}K/2$, varies
monotonically from zero to $1/(\pi^{1/2}b)$; since $\alpha_{\rm res}>1/(\pi^{1/2}b)$,
the expression in between square
brackets cannot vanish.  Taking as new functions $G_0(K)=K^2 B_{L=0}^{\rm out}(K)$
and $G_2(K)=K^2 B_{L=2}^{\rm out}(K)$, one faces 
for $\mathcal{V}_s\to +\infty$ an integral equation of the
form
\be
\frac{\alpha_{\rm res}}{4} \mathbf{G}(K) - \mathbf{O}[\mathbf{G}] 
= \mathbf{S}(K)
\label{eq:ieinf}
\ee
where the source term is the infinite $\mathcal{V}_s$ limit of the right hand side
of Eq.(\ref{eq:ieodd}), and $\mathbf{O}$ is a bounded operator depending
on $b$ but not on $\alpha_{\rm res}$. The value $G_0(0)$ is
finite for $\mathcal{V}_s=+\infty$, so is the atom-dimer scattering length.
As shown in the appendix, a simple transformation can make the operator $\mathbf{O}$
hermitian; numerically, one finds that the positive part of the spectrum 
of $\mathbf{O}$ consists of 
a continuum extending from zero to $1/(4\pi^{1/2}b)$, and of a discrete state of energy
above the continuum.
We see that $\alpha_{\rm res}/4$ cannot match an eigenvalue of the continuum,
but can indeed match the discrete eigenvalue, for
\be
\alpha_{\rm th}^{\rm odd} \simeq 0.69208/b.
\label{eq:ath_odd}
\ee
This particular value of $\alpha_{\rm res}$
corresponds to the threshold for the formation of an odd trimer
at $\mathcal{V}_s=\infty$, and the corresponding eigenvector was plotted
in Fig.\ref{fig:fot}a. For $\alpha_{\rm res}$ close to the threshold value,
the solution of (\ref{eq:ieinf}) acquires a diverging component on this eigenvector;
since the eigenvector has a value $G_0(0)\neq 0$ in $K=0$, 
this leads to an atom-dimer scattering length $a_{\rm ad}$ diverging as
$1/(\alpha_{\rm res}-\alpha_{\rm th}^{\rm odd})$.

The writing (\ref{eq:ieinf}) also makes it clear that asymptotic expressions 
can be obtained in the narrow resonance limit $\alpha_{\rm res} b \gg 1$: 
in this limit, the term proportional to $\alpha_{\rm res}$ dominates over
the bounded operator $\mathbf{O}$, which can thus be treated as a perturbation.
To leading order, $(G_0(K),G_2(K))=4\mathbf{S}(K)/\alpha_{\rm res}$,
which, injected into (\ref{eq:exact_relation}), 
gives the asymptotic equivalent
\be
a_{\rm ad} \stackrel{\alpha_{\rm res}b \gg 1}{\sim}
 -\frac{32}{3(5\pi)^{1/2} \alpha_{\rm res}^2b}.
\label{eq:asympt_equiv_aad}
\ee
valid in the limit of large $\mathcal{V}_s/b^3$ and large
$\alpha_{\rm res}b$.
We have checked that this relation is obeyed by the numerical results.
It is important physically to point out that,
as we shall see in section \ref{sec:enri}, this asymptotic result 
no longer holds in presence of direct interaction between atoms
in the open channel.
Anyway, it clearly shows that
$a_{\rm ad}$ depends not only on the effective range parameter
$\alpha_{\rm res}$ but also on the range $b$, which is
sensitive to the microscopic
details of the model interaction. In this sense, the large scattering
volume limit of $a_{\rm ad}$ is not a `universal' quantity.

This differs from the bosonic case on a narrow Feshbach resonance,
where the atom-dimer scattering length is a function of the scattering length
$a$ and the effective range $r_e$ only, as soon as $a$ greatly exceeds the
range of the potential; furthermore, this function is not bounded in the
large $a$ limit, but rather exhibits, on top of an overall linear growth
with $a$, a series of
divergences for values of $a/r_e$ corresponding to a threshold
for the formation of an Efimov trimer \cite{Petrovbb}.

\noindent{\sl $P$-wave atom-dimer scattering:}
We now assume that the incoming atom-dimer relative wave is a $p$-wave, that 
is it has a unit orbital momentum $L=1$.
In the subspace with one atom and one closed-channel molecule, 
the corresponding orbital wavefunction is obtained in momentum
space from the low-$K_0$ expansion of the Dirac distribution
corresponding to a molecule of wavevector $\mathbf{K}_0$ impinging on an atom
of wavevector $-\mathbf{K}_0$:
\be
(2\pi)^3 \delta(\mathbf{K}-\mathbf{K}_0) = (2\pi)^3
\left[\delta(\mathbf{K}) - \mathbf{K}_0\cdot (\mathbf{grad}\,\delta)(\mathbf{K})
+\ldots\right],
\ee
and one may take exactly $E=-E_{\rm dim}$ at this order.
Since the molecule has a spin $S_{\rm mol}=1$
this may correspond to a total spin $J=0$, $1$ or $2$.
The present work is restricted to a total spin $J=1$, and 
the corresponding ansatz
turns out to be in the even sector:
\bea
\betab(\mathbf{K}) &=& (2\pi)^3 (-K_0) p_{\rm closed}^{1/2}
\times \nonumber \\
&\times &
\{
[\mathbf{e}_z\cdot(\mathbf{grad}\,\delta)(\mathbf{K})]\mathbf{e}_y
-
[\mathbf{e}_y\cdot(\mathbf{grad}\,\delta)(\mathbf{K})]\mathbf{e}_z
\} \nonumber \\
&+&\frac{B_{\rm L=1}^{\rm out}(K)}{K}
\left[(\mathbf{K}\cdot\mathbf{e}_z) \mathbf{e}_y -(\mathbf{K}\cdot\mathbf{e}_y) \mathbf{e}_z\right].
\label{eq:ansatz_volad}
\eea
We insert this ansatz in the integral equation (\ref{eq:integrale}), keeping in mind that here
$A_0\equiv 0$. The part of the ansatz involving the gradient of the Dirac distribution
gives a vanishing contribution in the diagonal term of the equation (since $1/f(k_{\rm rel})$
vanishes quadratically in $K=0$ for the total energy $E=-E_{\rm dim}$), but gives
a non-zero, smooth contribution in the integral term, serving as a source term for
the scattered wave $B_{L=1}^{\rm out}$. Performing the angular average as detailed in
the appendix \ref{appen:ad} we obtain
\begin{widetext}
\be
D(K) B_{\rm L=1}^{\rm out}(K) - \frac{2}{\pi} \int_0^{+\infty} dk \, 
k^2 [C_0(K,k)-C_2(K,k)] B_{\rm L=1}^{\rm out}(k) 
e^{-5(K^2+k^2)b^2/8}=
(-K_0) p_{\rm closed}^{1/2} \frac{(-8\pi K)}{K^2+q_{\rm dim}^2},
\label{eq:intadp}
\ee
\end{widetext}
where the functions $C_0$ and $C_2$ are defined in the appendix taking $q=q_{\rm dim}$,
and $D(K)$ is given by (\ref{eq:defD}).

\begin{figure}
\includegraphics[width=8cm,clip]{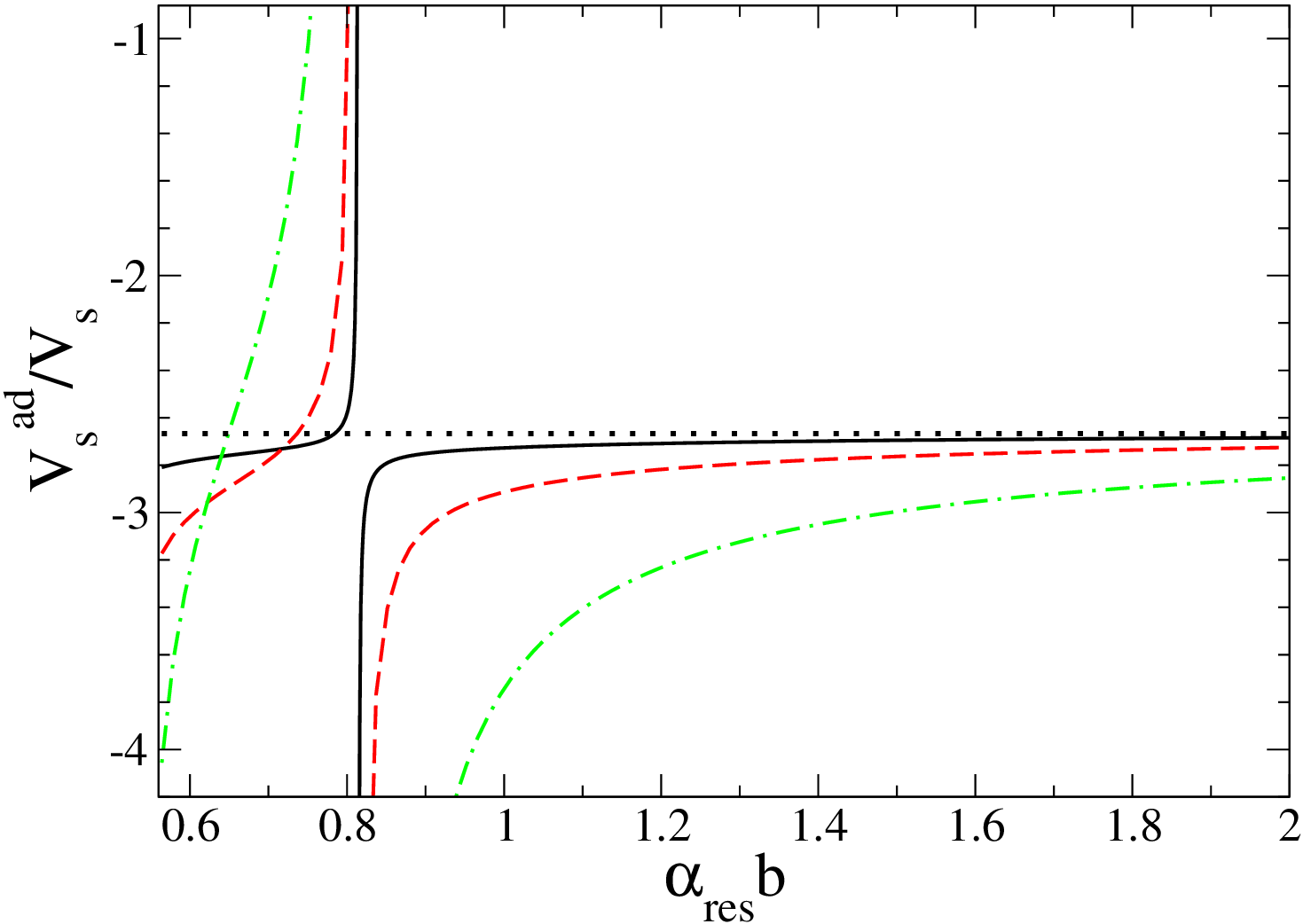}
\caption{(Color online) Atom-dimer scattering volume $\mathcal{V}_s^{\rm ad}$ for a total spin $J=1$
(see text) as a function of $\alpha_{\rm res}b$, for a fixed value of the atom-atom scattering volume
$\mathcal{V}_s/b^3=100$ (dashed-dotted green line), $\mathcal{V}_s/b^3=10^3$ (dashed red line), $\mathcal{V}_s/b^3=10^4$ (solid black line).
To reveal the scaling of $\mathcal{V}_s^{\rm ad}$ with $\mathcal{V}_s$ close to the Feshbach resonance,
$\mathcal{V}_s^{\rm ad}$ is expressed in units of $\mathcal{V}_s$. Dotted horizontal line: analytical prediction
(\ref{eq:asympt_volad})
in the limit $\mathcal{V}_s\to +\infty$.}
\label{fig:volad}
\end{figure}

The analysis performed for the atom-dimer $s$-wave scattering is readily extended
to the present $p$-wave scattering. Since the inhomogeneous term in the right hand side
of (\ref{eq:intadp}) vanishes linearly in $K=0$ and the diagonal part $D(K)$ vanishes 
quadratically, $B_{\rm L=1}^{\rm out}(K)$ diverges as $1/K$.
Such a low-$K$ behavior was expected: From Eq.(\ref{eq:scat_state}) expanded to first order
in the incoming wavevector, here called $\mathbf{K}_0$ rather than $\mathbf{k}$, one obtains 
for the wavefunction at large distances
\be
\psi_{\mathbf{K}_0}(\mathbf{r}) \simeq i \mathbf{K}_0\cdot \mathbf{r} 
\left[1- \frac{3\mathcal{V}_s^{\rm ad}}{r^3}\right].
\ee
Taking the Fourier transform with respect to the relative atom-molecule coordinates
$\mathbf{r}=\mathbf{r}_{\rm mol}-\mathbf{r}_{\rm at}$ leads to the low-$K$ behavior
\be
\tilde{\psi}_{\mathbf{K}_0}(\mathbf{K}) 
\simeq -(2\pi)^3 (\mathbf{K}_0\cdot\mathbf{grad})\delta(\mathbf{K})
-12\pi\mathcal{V}_s^{\rm ad} \frac{\mathbf{K}\cdot\mathbf{K}_0}{K^2}.
\ee
So we conclude that
\be
B_{L=1}^{\rm out} \sim p_{\rm closed}^{1/2} (-K_0) \frac{12\pi\mathcal{V}_s^{\rm ad}}{K}.
\ee
From the numerical solution of (\ref{eq:intadp}),
the atom-dimer scattering volume $\mathcal{V}_s^{\rm ad}$ seems to scale as the atom-atom scattering
volume $\mathcal{V}_s$ itself close to the Feshbach resonance.
So we plot in Fig.\ref{fig:volad} the ratio $\mathcal{V}_s^{\rm ad}/\mathcal{V}_s$ 
as a function of $\alpha_{\rm res}$ for increasing
values of $\mathcal{V}_s$. Another interesting feature is the divergence of $\mathcal{V}_s^{\rm ad}$
at the threshold for a trimer formation in the even sector.

The same analytical techniques as in the case of $s$-wave atom-dimer scattering may be used
to predict the scaling of $\mathcal{V}_s^{\rm ad}$ with $\mathcal{V}_s$.
First we divide (\ref{eq:intadp}) by $K$ and we take the limit $K\to 0$. As discussed in the
$s$-wave atom-dimer scattering case, $D(K)/K^2$ has a finite limit, so does $D(K) B_{L=1}^{\rm out}(K)/K$. 
Furthermore one can show from (\ref{eq:Cn}) (with $q=q_{\rm dim}$) that
\be
\lim_{K\to 0} \frac{C_0(K,k)-C_2(K,k)}{K} = \frac{2}{3} \frac{k}{k^2+q_{\rm dim}^2}.
\ee
We thus obtain the exact relation:
\bea
&&\lim_{K\to 0} \frac{D(K)}{K^2} 3\mathcal{V}_s^{\rm ad} 
-\frac{4}{3\pi} \int_0^{+\infty}\!\!\!\! dk\, \frac{k^3 e^{-5k^2b^2/8}}{k^2+q_{\rm dim}^2} \frac{B_{L=1}^{\rm out}(k)}{4\pi (-K_0) p_{\rm closed}^{1/2}}
\nonumber \\
&&= -\frac{2}{q_{\rm dim}^2}.
\label{eq:exact_volad}
\eea
Next we take the limit of an infinite scattering volume in (\ref{eq:intadp}), that is we take $q_{\rm dim}\to 0$.
The source term now diverges as $1/K$ in $K=0$; since $D(K)$ vanishes as $K^2$, we expect that the function
\be
B_\infty (K) \equiv \lim_{\mathcal{V}_s\to+\infty} B_{L=1}^{\rm out}(K)
\ee
diverges as $1/K^3$ in $K=0$. To check the existence of $B_\infty$ as a limit, one thus has to check that the integral 
in (\ref{eq:intadp}) does not have a divergence in $k=0$ for such a $1/k^3$ behavior of the $B(k)$ function: 
the factor $k^2$ of three-dimensional integration and the fact that $C_0(K,k)-C_2(K,k)$ vanishes
linearly with $k$ indeed bring an overall $k^3$ factor that compensates the divergence.
As a consequence it is reasonable to assume that there exists a constant $C$ such that
\be
|B_{L=1}^{\rm out}(K)| \leq \frac{C}{K^3}
\label{eq:raisonnable}
\ee
uniformly in $K$ and $\mathcal{V}_s$. This allows to show that the integral term in (\ref{eq:exact_volad})
is $O(1/q_{\rm dim})$ and is thus negligible as compared to $1/q_{\rm dim}^2$.
Using (\ref{eq:limDsurK2}) and $q_{\rm dim}\sim 1/(\alpha_{\rm res}\mathcal{V}_s)^{1/2}$ we obtain
\be
\mathcal{V}_s^{\rm ad} \sim -\frac{8}{3} \mathcal{V}_s.
\label{eq:asympt_volad}
\ee
This result corresponds to the dotted line in Fig.\ref{fig:volad}. 
Strictly speaking, it asymptotically holds
for all values of $\alpha_{\rm res}b$ except right on the threshold for the even trimer formation, for reasons
that are explained in subsection \ref{subsec:recomb}.
Away from this threshold we thus reach the important conclusion that,
very close to the Feshbach resonance, the atom-dimer scattering
volume for a total angular momentum $J=1$ is a `universal' quantity in
the sense that it does not depend on the range $b$ of the interaction,
but only on the atom-atom scattering volume.

\subsection{Scattering of three atoms: recombination rate}
\label{subsec:recomb}

In this subsection, we consider the case of three incoming atoms, 
in the form of plane waves of wavevectors $\mathbf{k}_1^{0}$,
$\mathbf{k}_2^{0}$ and $\mathbf{k}_3^{0}$.
Without loss of generality we move to the center of mass frame
and assume a vanishing total momentum.
We consider the case of a positive scattering volume $\mathcal{V}_s>0$,
so that there exists a dimer state in the two-body problem, that may be
populated by the collision of three atoms. The goal here is to determine
the rate with which such a dimer state is formed, the so-called recombination
rate.

For this physical situation, the total energy is positive
so $A_0$ in Eq.(\ref{eq:A})
does not vanish, but rather defines the state of the three
incoming fermions,
\be
|\Psi_0\rangle = \int \frac{d^3kd^3K}{(2\pi)^6}
A_0(\mathbf{K},\mathbf{k})
a^\dagger_{\frac{1}{2}\mathbf{K}+\mathbf{k}}
a^\dagger_{\frac{1}{2}\mathbf{K}-\mathbf{k}}
a^\dagger_{-\mathbf{K}} |0\rangle.
\label{eq:Psi0}
\ee
Setting $\mathbf{K}_0= \mathbf{k}_1^{0}+\mathbf{k}_2^{0}$
and $\mathbf{k}_0= (\mathbf{k}_1^{0} - \mathbf{k}_2^{0})/2$, one has
\be
\label{eq:incoming}
A_0(\mathbf{K},\mathbf{k}) = (2\pi)^6 \delta(\mathbf{K}-\mathbf{K}_0)
\frac{1}{2}\left[\delta(\mathbf{k}-\mathbf{k}_0) - \delta(\mathbf{k}+\mathbf{k}_0)\right].
\ee
To derive
a simplified expression in the low incoming kinetic energy limit,
\be
K_0,k_0 \ll q_{\rm dim}, \frac{1}{b},
\label{eq:low_kin}
\ee
which implies $E\ll E_{\rm dim}$,
one expands $A_0$ in powers of $k_0$ and $K_0$. The expression between square brackets
gives $k_0$ times a gradient of delta $+O(k_0^3)$. The expansion in powers of $K_0$
gives 
\bea
A_0(\mathbf{K},\mathbf{k})  &=& (2\pi)^6  \left[\delta(\mathbf{K}) -\mathbf{K}_0\cdot
(\mathbf{grad}\,\delta)(\mathbf{K}) + O(K_0^2)\right] \nonumber \\
&\times& \left[-\mathbf{k}_0 \cdot (\mathbf{grad}\,\delta)(\mathbf{k}) + O(k_0^3)\right].
\eea
One has to keep the leading order in $k_0 \sim K_0$ giving a non-zero value for 
the incoming state $|\Psi_0\rangle$.
Keeping the first term in the first factor gives a vanishing contribution
so that one has to keep the second term:
\be
A_0(\mathbf{K},\mathbf{k}) \simeq (2\pi)^6 \left[\mathbf{K}_0\cdot
(\mathbf{grad}\,\delta)(\mathbf{K})\right]
 \, \left[\mathbf{k}_0 \cdot (\mathbf{grad}\,\delta)(\mathbf{k})\right].
\ee
This choice for $A_0$ corresponds to the limit of a vanishing total energy,
so that we now take $E=0$.

This expression for $A_0$, when inserted in Eq.(\ref{eq:integrale}), gives in the right-hand
side the source term
\bea
&&-\frac{6\pi \hbar^2}{m\Lambda} \, (2\pi)^3 \times \nonumber \\
&& \left\{
\left[\mathbf{K}_0\cdot \left(\mathbf{grad}\,\delta\right)(\mathbf{K}) \right]\, \mathbf{k}_0
-
\left[\mathbf{k}_0\cdot \left(\mathbf{grad}\,\delta\right)(\mathbf{K}) \right]\, \mathbf{K}_0
\right\}. 
\label{eq:soso}
\eea
Since the gradient of $\delta$ can be seen as the product of $\mathbf{K}$ with an isotropic distribution,
one finds that this source term is in the even sector, of the form Eq.(\ref{eq:even}), where
$\mathbf{e}_z$ is taken along the direction of $\mathbf{K}_0$ and $\mathbf{e}_y$ is taken
along the direction of $\mathbf{k}_{0}^\perp$,
the component of $\mathbf{k}_0$ in the plane orthogonal to $\mathbf{K}_0$.
We take for $\betab$ the even ansatz with a specific form adapted to the present physical
situation,
\bea
&&\betab(\mathbf{K}) = \nonumber \\
&&\mathcal{G}(K) \left\{
\left[\mathbf{K}_0\cdot \left(\mathbf{grad}\,\delta\right)(\mathbf{K}) \right]\, \mathbf{k}_0
-
\left[\mathbf{k}_0\cdot \left(\mathbf{grad}\,\delta\right)(\mathbf{K}) \right]\, \mathbf{K}_0
\right\} \nonumber \\
&&+\frac{4\pi g(K)}{K^2-K_{\rm dim}^2-i0^+} \frac{1}{K} 
\left[(\mathbf{K}\cdot\mathbf{e}_z) \mathbf{e}_y -(\mathbf{K}\cdot\mathbf{e}_y) \mathbf{e}_z\right].
\label{eq:ansatz_merveilleux}
\eea
The first term in the right-hand side is motivated by the fact that the source term contains
a gradient of $\delta$, so that $\betab$ has also to contain a gradient of $\delta$.
In the second term, we have pulled out explicitly a singularity with a pole at $K = K_{\rm dim}+i0^+$,
where 
\be
K_{\rm dim}= \frac{2}{\sqrt{3}} q_{\rm dim}
\ee
which is the value of $K$ given by Eq.(\ref{eq:Erel}) when $k_{\rm rel}= i q_{\rm dim}$, keeping in mind
that the total energy is here $E\simeq 0$.
Physically $K_{\rm dim}$ is the value $K_{\rm out}$
of $K$ corresponding to the motion in opposite directions
of a flying atom and a flying dimer formed by the three-atom collision, and the term $i0^+$ in the denominator of the ansatz ensures
that this relative motion is a purely outgoing wave.
The conservation of energy indeed imposes $E=3\hbar^2 K_{\rm out}^2/4m
-E_{\rm dim}$, that is $K_{\rm out} \simeq K_{\rm dim}$ since
we assumed $E\ll E_{\rm dim}$.

We now inject the ansatz Eq.(\ref{eq:ansatz_merveilleux}) in the 
integral equation Eq.(\ref{eq:integrale}).
The bit in gradient of $\delta$ in the ansatz, when injected in the 
diagonal term of (\ref{eq:integrale}), gives a contribution which
is a distribution of the same structure as the source term
(\ref{eq:soso}) created by $A_0$; 
the function $\mathcal{G}(K)$ is adjusted to have an exact cancellation:
\be
\mathcal{G}(K) = -(2\pi)^3\,\frac{6\pi\hbar^2}{m \Lambda} \frac{3 f(k_{\rm rel}) }{k_{\rm rel}^2 e^{-k_{\rm rel}^2 b^2}}
\label{eq:valG}
\ee
where $k_{\rm rel}$ is defined in Eq.(\ref{eq:Erel}) and
is equal here to $i(\sqrt{3}/2)K$.
When injected in the integral on the left hand side of Eq.(\ref{eq:integrale}), the bit in gradient
of $\delta$ in the ansatz gives rise to a smooth function of 
$\mathbf{K}$ (not a distribution).
After lengthy calculations and angular averages detailed in appendix \ref{appen:ad}, one finds
an inhomogeneous integral equation for $g(K)$:
\begin{widetext}
\bea
-\frac{K^2 e^{3 K^2 b^2/4}}{4 f\left(i\frac{\sqrt{3}}{2}K\right)}
\frac{g(K)}{K^2-K_{\rm dim}^2-i0^+}
-\frac{2}{\pi} \int_0^{+\infty} dK'\, K'^2 \left[C_0(K,K')-C_2(K,K')\right]
\frac{g(K')}{K'^2-K_{\rm dim}^2-i0^+}
e^{-5 b^2 (K^2+K'^2)/8}
&=& \nonumber  \\
 -\frac{36\pi\hbar^2}{m \Lambda} K_0 k_0^\perp \mathcal{V}_s \frac{e^{-5 K^2 b^2/8}}{K}, &&
\label{eq:intevensec}
\eea
\end{widetext}
where $C_0$ and $C_2$ are given by (\ref{eq:C0}) and (\ref{eq:C2})
with $k=K'$ and $q=0$.
It remains to solve this integral equation; one notes that there is no delta distribution
arising in the first term of this equation, since $1/f(k_{\rm rel})=1/f(i q_{\rm dim})=0$ for $K=K_{\rm dim}$,
so that $i0^+$ may be omitted in the denominator of this first term
\cite{technical}.

To obtain the recombination rate from the solution $g(K)$ of
Eq.(\ref{eq:intevensec}), we proceed in two steps.
First, we calculate the rate of dimer formation, that is the
recombination rate, in terms
of $g(K_{\rm dim})$,
after having enclosed the three atoms in a fictitious 
cubic box of size $L$. Second, 
we construct an operator $\hat{O}$ such that its expectation value
in the unperturbed incoming state (\ref{eq:Psi0}) of the three atoms
gives the recombination rate; 
calculating the expectation value of this operator
for a Fermi sea in the thermodynamic limit then gives the recombination
rate for a macroscopic gas.

\noindent {\sl Recombination rate for three atoms:}
Enclosing the three atoms in a arbitrarily large cubic box with
periodic boundary conditions introduces the following
 normalization factor in the state vector,
\begin{equation}
|\Psi^{\rm box}\rangle \simeq \frac{1}{L^{9/2}} | \Psi\rangle.
\label{eq:psi_box}
\end{equation}
This is directly seen on the incoming state vector (\ref{eq:Psi0}):
each of the three atoms is in a plane wave, with a wavefunction in the
box differing from the free space one by the normalization factor
$1/L^{3/2}$.

To calculate the probability flux of dimer formation, the most convenient
is to perform the reasoning in the subspace of (\ref{eq:ansatz})
with one atom and one closed-channel molecule, where the formation of a dimer
manifests itself by an outgoing wave of the molecule of momentum $K_{\rm dim}$
and an outgoing wave of the atom with the same momentum in the opposite 
direction. 
In momentum space, this outgoing wave results
from the existence of a pole of $\betab(\mathbf{K})$ in $K=K_{\rm dim}$,
as was made apparent in the ansatz (\ref{eq:ansatz_merveilleux}).
In position space, taking the Fourier transform of $\betab(\mathbf{K})$
and writing $g(K)=g(K_{\rm dim})+[g(K)-g(K_{\rm dim})]$, we isolate the
outgoing wave, and we obtain in the limit of a 
large atom-molecule separation:
\bea
\left[\Psib_{\rm mol}^{\rm box}\right]_{\rm out} (\mathbf{r}_{\rm mol};
\mathbf{r}_{\rm at}) \simeq \frac{g(K_{\rm dim})}{L^{9/2}}
\frac{e^{i K_{\rm dim} r}}{r} \mathbf{e}_r\wedge\mathbf{e}_x \nonumber \\
\eea
with $\mathbf{r}=r\, \mathbf{e}_r
\equiv \mathbf{r}_{\rm mol}-\mathbf{r}_{\rm at}$ is the
position of the relative particle.
The associated probability current for the relative particle
of reduced mass $2m/3$ is then
\bea
\mathbf{j}_{\rm out}&=&\sum_\gamma \frac{\hbar}{2m/3}\mbox{Im}\,
\left[\Psi_{\gamma,\mathrm{out}}^* 
\partial_{\mathbf{r}} \Psi_{\gamma,\mathrm{out}}\right] \\
&\simeq& \frac{3\hbar K_{\rm dim}}{2mr^2} |g(K_{\rm dim})|^2
\left[(\mathbf{e}_r\cdot\mathbf{e}_z)^2 
+(\mathbf{e}_r\cdot\mathbf{e}_y)^2\right] \mathbf{e}_r. \nonumber
\eea
One then calculates the total flux of the current through $4\pi$ steradian
and one integrates
over the center of mass position. Since the flying dimer has a probability
amplitude $p_{\rm closed}^{1/2}$ to be in the form of a molecule in the
closed channel, it remains to divide the total flux by $p_{\rm closed}$
to get the rate of dimer formation for three atoms in the box,
\begin{equation}
\label{eq:rec3}
\frac{d}{dt} N_{\rm dim}^{\rm box} = 
4\pi \frac{\hbar K_{\rm dim}}{m} \frac{|g(K_{\rm dim})|^2}{p_{\rm closed}}
\frac{1}{L^6}.
\end{equation}

\noindent {\sl Recombination rate for a macroscopic gas:}
To extend (\ref{eq:rec3}) to a macroscopic number of atoms, we heuristically generalize
to fermions an operatorial expression derived in \cite{Kagan} for bosons:
In the bosonic case, the recombination rate in a macroscopic gas
is expressed in terms of $\langle \left[\hat{\psi}^\dagger(\mathbf{R})\right]^3
\left[\hat{\psi}(\mathbf{R})\right]^3\rangle_0$, where $\hat{\psi}$
is the bosonic field operator and the expectation value 
$\langle \ldots\rangle_0$ is taken 
in a mean-field state for the bosons not including the short
range microscopic correlations induced by the interaction potential
\cite{Markus}. Such a local formula results from the
assumption that the size of a produced dimer is much smaller
than the macroscopic correlation lengths of the gas, such as
the healing length and the thermal de Broglie wavelength \cite{Kagan}.

In the case of fermions, one has to rederive the formula since $\hat{\psi}^3=0$.
This is done in the appendix \ref{appen:dim_rate} and leads to the
following prescription for the recombination rate:
\bea
\frac{d}{dt} N_{\rm dim} = \mathcal{K}_{\rm rec} 
\int d^3R \sum_{(\alpha,\beta)\in \{(x,y),(x,z),(y,z)\}} \nonumber \\
\times \langle \left(\partial_{R_\beta}\hat{\psi}\right)^\dagger 
\left(\partial_{R_\alpha}\hat{\psi}\right)^\dagger
\hat{\psi}^\dagger 
\hat{\psi}
\left(\partial_{R_\alpha}\hat{\psi}\right)
\left(\partial_{R_\beta}\hat{\psi}\right)
\rangle_0,
\label{eq:la_recette}
\eea
where the field operator and its derivatives are all evaluated
in $\mathbf{R}$.
This expression involves as a factor the recombination constant $\mathcal{K}_{\rm rec}$,
not to be confused with the recombination rate.
In the considered limit of a fermionic kinetic energy smaller than the dimer binding
energy (\ref{eq:low_kin}) we indeed expect $\mathcal{K}_{\rm rec}$ to be a constant,
that is not to depend on the fermionic kinetic energy. On the contrary, the recombination
rate $dN_{\rm dim}/dt$ will involve a factor proportional to the square of the
kinetic energy of the fermions, as predicted in \cite{Greene} with a different approach.

To illustrate this point, let us consider the case of a spatially homogeneous 
weakly interacting zero temperature Fermi gas.
The condition of low kinetic energy is then that the
Fermi energy $\hbar^2 k_F^2/(2m)$ is smaller than the 
dimer binding energy $E_{\rm dim}$.
It remains to calculate the expectation value $\langle \ldots\rangle_0$ 
of Eq.(\ref{eq:la_recette}) in the Fermi
sea of the ideal Fermi gas of density $n=k_F^3/(6\pi^2)$, to get
\begin{equation}
\frac{d}{dt} N_{\rm dim} = \frac{3 k_F^4}{25} \mathcal{K}_{\rm rec} N\, n^2. 
\end{equation}
The factor $k_F^4$ reveals the expected kinetic energy dependence of the
recombination rate. The recombination rate will weakly depend on temperature
as long as the gas remains strongly degenerate, $k_B T \ll \hbar^2 k_F^2/(2m)$.

\noindent{\sl Value of the recombination constant:} 
We obtain  $\mathcal{K}_{\rm rec}$ 
by applying Eq.(\ref{eq:la_recette}) to our solution
of the three-body problem. In this case, the uncorrelated state $|\Psi_0\rangle$
over
which to average in the expectation value $\langle \ldots\rangle_0$ 
is a Slater determinant with three atoms in plane waves
of wavevectors $\mathbf{k}_1^0=\mathbf{k}_0+\mathbf{K}_0/2$,
$\mathbf{k}_2^0=-\mathbf{k}_0+\mathbf{K}_0/2$ and 
$\mathbf{k}_3^0=-\mathbf{K}_0$, respectively. Using Wick's theorem 
\cite{aide_wick}, we obtain
\be
\frac{d}{dt} N_{\rm dim}^{\rm box} = 
9 \mathcal{K}_{\rm rec}  \frac{(\mathbf{k}_0\wedge\mathbf{K}_0)^2}{L^6}.
\label{eq:bout_manquant}
\ee
Equating this expression to Eq.(\ref{eq:rec3})
we obtain
\begin{equation}
\mathcal{K}_{\rm rec} = \frac{4\pi}{9} \frac{\hbar K_{\rm dim}}{m}
\frac{|g(K_{\rm dim})|^2}{(\mathbf{K}_0\times\mathbf{k}_0)^2\, p_{\rm closed}}. 
\label{eq:Krec_e}
\end{equation}
We find that, as expected, this recombination constant 
does not depend on the incoming energy,
that is on the norms $K_0$ and $k_0$, in the present limit of vanishing incoming energy:
$g(K)$ is indeed proportional to $K_0 k_0^\perp = ||\mathbf{K}_0\times\mathbf{k}_0||$,
as the source term in the linear equation (\ref{eq:intevensec}) is.

\begin{figure}
\includegraphics[width=8cm,clip]{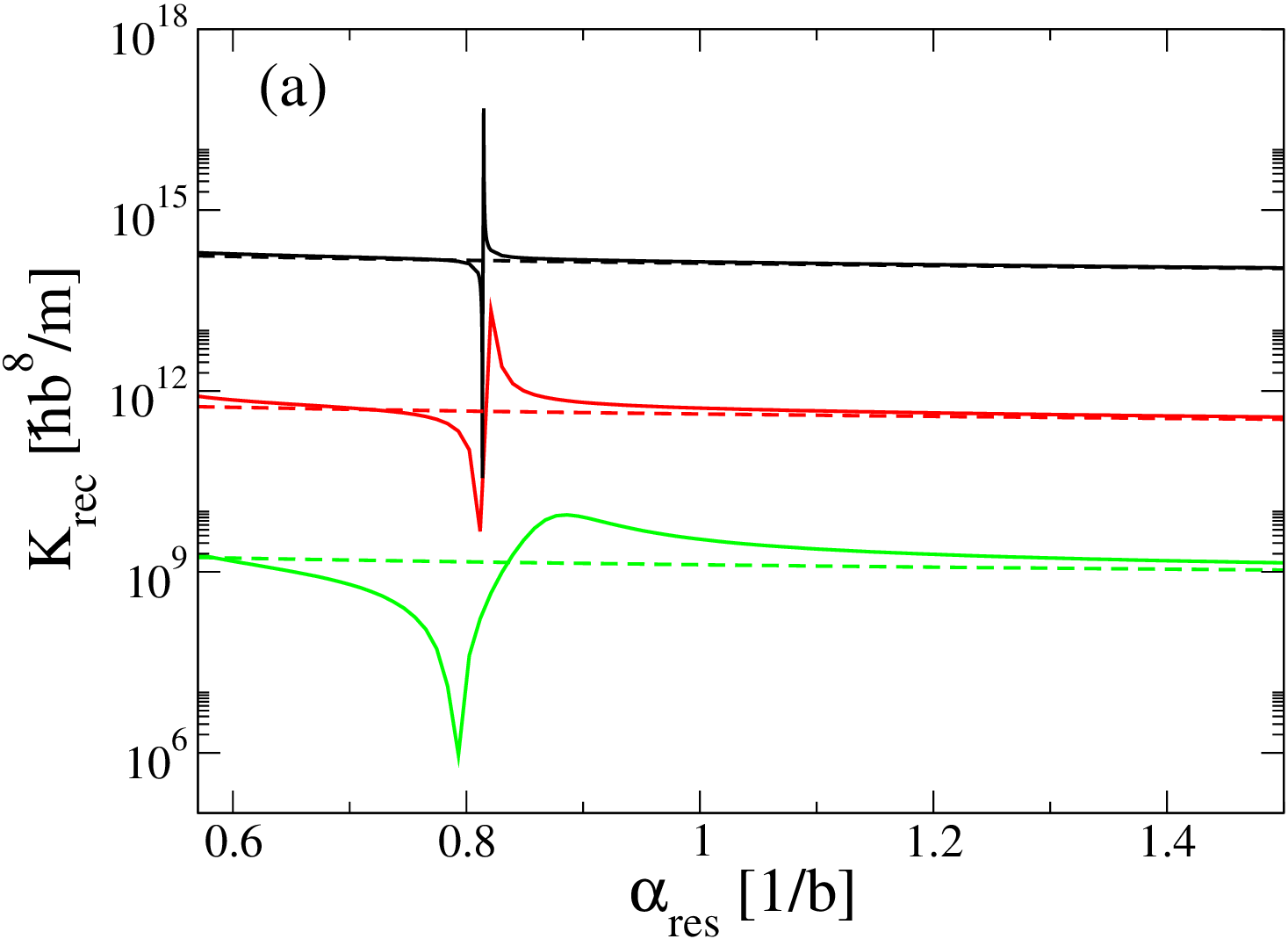}
\includegraphics[width=8cm,clip]{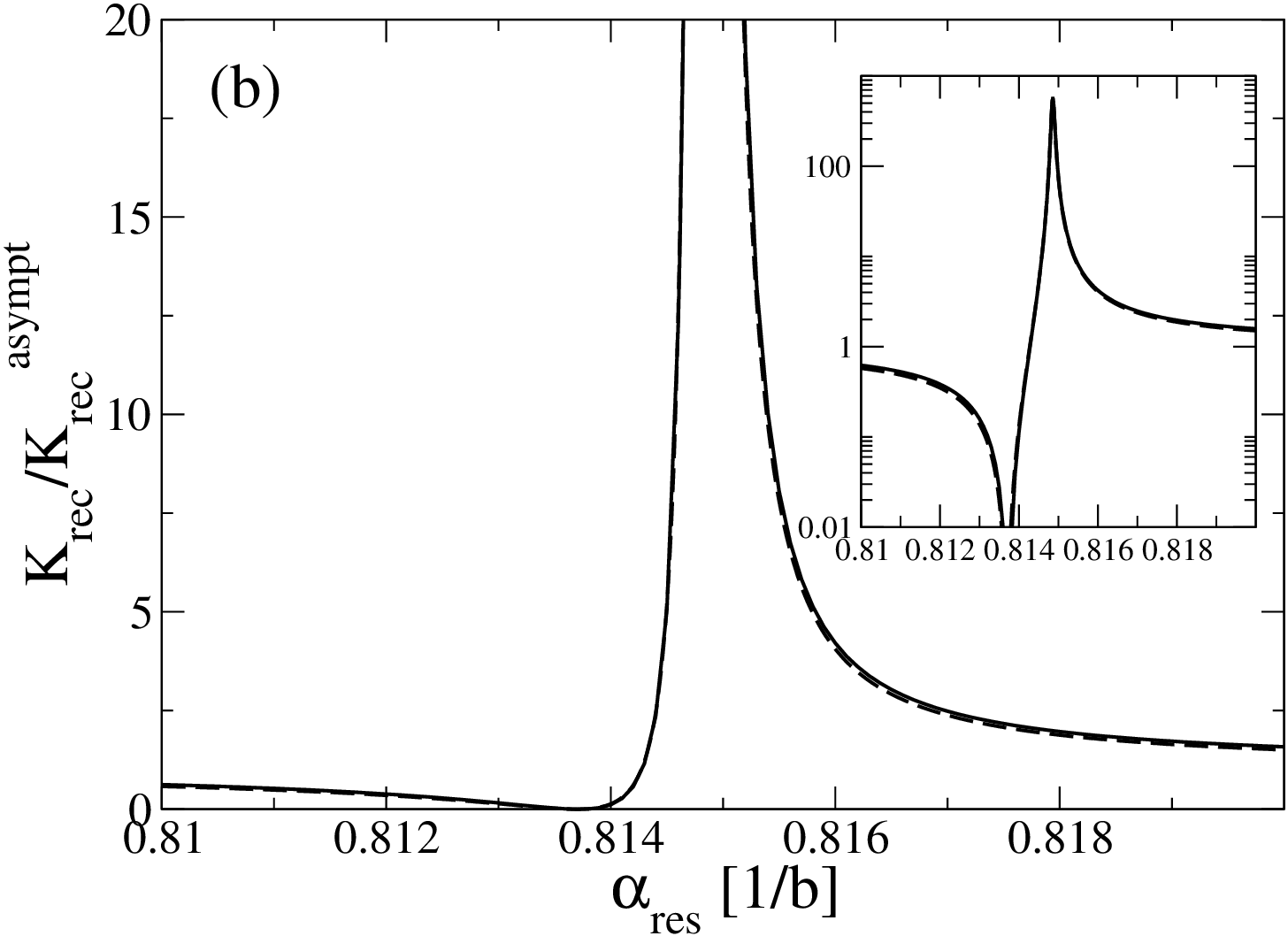}
\caption{(Color online) Recombination constant $\mathcal{K}_{\rm rec}$ 
appearing in the expression (\ref{eq:la_recette}) giving the rate
of formation of weakly bound dimers when three low energy atoms are
colliding, as a function of $\alpha_{\rm res}$ for a fixed value
of the scattering volume. (a) $\mathcal{K}_{\rm rec}$ in units of
$\hbar b^8/m$.  The scattering volume is, 
from bottom to top, $\mathcal{V}_s=100 b^3$ (green line),
$\mathcal{V}_s=1000 b^3$ (red line) and $\mathcal{V}_s=10^4 b^3$ (black line).
Solid lines: numerical solution. Dashed lines: asymptotic formula
(\ref{eq:Kasympt}).
(b) Ratio of $\mathcal{K}_{\rm rec}$ to the asymptotic
formula (\ref{eq:Kasympt}), for $\mathcal{V}_s=10^4 b^3$.
Solid line: numerical solution. Dashed line: analytically predicted Fano profile
(\ref{eq:fano}).
The insert is exactly the same figure but with a log scale on the 
vertical axis.
}
\label{fig:recomb}
\end{figure}

One solves the integral equation (\ref{eq:intevensec}) numerically,
to access $g(K_{\rm dim})$.
The corresponding values of the recombination constant are given as functions of
$\alpha_{\rm res}$ in Fig.\ref{fig:recomb}a, for three values of the scattering volume.
As expected, a rapid rise of the recombination constant is observed when one gets closer
to the Feshbach resonance, that is for increasing values of $\mathcal{V}_s$.
For a fixed $\mathcal{V}_s$, one observes a smooth dependence of $\mathcal{K}_{\rm rec}$
with $\alpha_{\rm res}$, except in the vicinity of $\alpha_{\rm res} b=0.8$: 
both a dip and a peak in
$\mathcal{K}_{\rm rec}$ are observed; this singular structure becomes extremely 
narrow in the large $\mathcal{V}_s$ limit, both the distance between the dip and the peak,
and the width of the peak, apparently tending to zero.
These features can be obtained analytically as follows, 
by investigating the large $\mathcal{V}_s$
limit of (\ref{eq:intevensec}). 

Let us examine first the diagonal term in the left
hand side of the equation (\ref{eq:intevensec}). 
At low values of $K$, much below $1/b$, one can approximate
the inverse scattering amplitude as
$1/f(iq)\simeq -1/(q^2\mathcal{V}_s)+\alpha\simeq \alpha (q^2-q_{\rm dim}^2)/q^2$,
where we used Eq.(\ref{eq:qdim_approx}) since we are close to resonance.
Setting $q=\sqrt{3}K/2$, 
we then see that this diagonal term at low energy is close to $\alpha g(K)/4$,
so it is very smooth. At high values of $K$, of the order of $1/b$ or larger, one
can directly set $\mathcal{V}_s=+\infty$, and one sees that the factor of $g(K)$
in this diagonal term decreases smoothly from $\alpha_{\rm res}/4$ to the positive
quantity $[\alpha_{\rm res} -1/(\pi^{1/2}b)]/4$ when $K$ increases to infinity.

Let us now turn to the integral term. The value of $K'$ is cut to values at most
of the order of $1/b$ by the Gaussian factor. For low values of $K$, 
below $1/b$, an approximate expression of the kernel 
can be obtained \cite{pdd},
\be
C_0(K,K') - C_2(K,K') \simeq \frac{2 K K'}{3(K^2+K'^2)}.
\label{eq:diff_approx}
\ee 
This shows that the kernel of the integral part is smooth and bounded, even in the low
$K$ and $K'$ limit. Neglecting $K_{\rm dim}$ in the denominator of the integral
term gives a diverging factor $1/K'^2$ which is however exactly compensated
by the $K'^2$ Jacobian term of three-dimensional integration.

The only source of singularity in the solution $g(K)$ 
may thus be the source term, in the right hand side of
(\ref{eq:intevensec}). The presence of a factor $\mathcal{V}_s$ will cause
$g(K)$ to diverge at high $\mathcal{V}_s$, by linearity of the equation, and the
$1/K$ divergence of the source will lead to a singular behavior of $g(K)$ in $K=0$.
These two problems can be eliminated by taking as unknown function
\be
F(K) = \frac{K g(K)}{\mathcal{V}_s}.
\ee
We multiply (\ref{eq:intevensec}) by $K/\mathcal{V}_s$.
The kernel of the integral term for $F(K)$ taken in the limit $K_{\rm dim}=0$
(thus neglecting $K_{\rm dim}^2$ in the denominator)
now behaves at low momenta as $(K/K')[C_0(K,K)-C_2(K,K')]\simeq (2K^2/3)/(K^2+K'^2)$,
which remains a bounded quantity. Having eliminated the singularity in the source term,
we can suppose that
\be
F_\infty(K) = \lim_{\mathcal{V}_s\to + \infty} F(K)
\label{eq:deffinf}
\ee
is a regular and bounded function, $|F_\infty(K)|\leq |F|_{\rm max}$.
Then, if one uses the approximate expression (\ref{eq:diff_approx}),
one easily sees that the integral term in the equation for $F_\infty(K)$ is bounded 
by
\be
\frac{2}{\pi} \int_0^{+\infty} dK' \frac{2 K^2}{3(K^2+K'^2)} |F|_{\rm max}=
\frac{2}{3} K |F|_{\rm max},
\ee
so that it tends to zero in $K=0$. Since the source term for $F_\infty$ is non-zero
for $K=0$ we conclude that
\be
F_\infty(0) = -\frac{144 \pi \hbar^2}{m\Lambda \alpha_{\rm res}} K_0 k_0^\perp.
\label{eq:Finf0}
\ee
Using the useful expression (\ref{eq:p_closed_infini}) for the
probability to find the dimer in the closed channel,
we thus obtain analytically the asymptotic value of the recombination
constant for large $\mathcal{V}_s/b^3$ \cite{theoreme}:
\be
\mathcal{K}_{\rm rec} \sim \mathcal{K}_{\rm rec}^{\rm asympt} =
\frac{\hbar}{m} (48\pi)^2 \left(\frac{\mathcal{V}_s^5}{3\alpha_{\rm res}}\right)^{1/2}.
\label{eq:Kasympt}
\ee
We first note that this result, contrarily to the atom-dimer
scattering length, is `universal', that is it does not depend
on the potential range $b$ but only on the parameters $\mathcal{V}_s$ and
$\alpha$ entering in the low-$k$ expansion of the two-body
scattering amplitude. In particular, (\ref{eq:Kasympt}) is not specific
to our choice of a Gaussian cut-off function in $\chib(\mathbf{k})$, 
as we have checked for a general cut-off function
that is derivable with respect to $\mathbf{k}$.
Second, the exponent governing the dependence in $\mathcal{V}_s$
is the same for a broad or a narrow Feshbach resonance. It may thus make sense
to compare this prediction to the earlier work of \cite{Greene},
where a numerical calculation was performed for a resonant interaction
in a single
channel model: the recombination rate was found to increase as a power law
in $\mathcal{V}_s$, with an exponent argued to be equal to $8/3$.
Since $8/3$ and $5/2$ differ by about 6\% only, it seems difficult to see this
difference from the numerical results of \cite{Greene}.

To see how our numerical results approach the asymptotic prediction
(\ref{eq:Kasympt}), we have 
plotted in dashed lines in Fig.\ref{fig:recomb}a the asymptotic
value $\mathcal{K}_{\rm rec}^{\rm asympt}$, as a function
of $\alpha_{\rm res}$, for the considered values
of the scattering volume. For increasing values of $\mathcal{V}_s$, we indeed observe convergence
of the ratio $\mathcal{K}_{\rm rec}/\mathcal{K}_{\rm rec}^{\rm asympt}$ to unity, 
but this convergence is not uniform in $\alpha_{\rm res}$: the
singular structure already apparent in Fig.\ref{fig:recomb}a becomes narrower and narrower 
for increasing $\mathcal{V}_s$, but e.g.\ the peak in this singular structure
leads to increasing deviation from unity
of the ratio $\mathcal{K}_{\rm rec}/\mathcal{K}_{\rm rec}^{\rm asympt}$.

The existence of this singular structure and the dependence of the recombination
rate on $\alpha_{\rm res}$ within this structure can be obtained analytically as follows.
First, we formally write the integral equation obtained for $F_\infty(K)$
in the limit $\mathcal{V}_s\to +\infty$ for a fixed value of $\alpha_{\rm res}$:
\be
\frac{\alpha_{\rm res}}{4} F_\infty(K) - I_0[F_\infty](K) = S(K)
\label{eq:I0corps}
\ee
where the source term $S(K)$ is obtained by multiplication of the right-hand side
of (\ref{eq:intevensec}) by $K/\mathcal{V}_s$, and $I_0$ is a linear operator,
given explicitly in the appendix \ref{appen:singular}.
We find numerically that the spectrum of $I_0$ consists of a continuum
extending from $0$ to $1/(4\pi^{1/2}b)$, and of one discrete eigenvalue
above the continuum. If one remembers that, from (\ref{eq:alpha_res}),
$\alpha_{\rm res}>1/(\pi^{1/2}b)$,
it becomes clear that the homogeneous equation obtained
by replacing $S$ with zero will admit a non-zero solution $u_0(K)$ only
for $\alpha_{\rm res}=\alpha_{\rm th}^{\rm even}$, where
mathematically $\alpha_{\rm th}^{\rm even}/4$ is the discrete
eigenvalue of $I_0$, and
physically 
$\alpha_{\rm th}^{\rm even}$ is the 
threshold value of $\alpha_{\rm res}$ for the existence on resonance
of an even trimer. The operator appearing
in (\ref{eq:I0corps}) is indeed the infinite scattering volume limit 
and the zero energy limit of the operator $L(E)$ of the subsection \ref{subsec:ewbt}
on trimers, restricted to the even sector. Numerically we find
\be
\alpha_{\rm th}^{\rm even} \simeq  0.81408/b.
\label{eq:ath_even}
\ee
In presence of the source term $S$, and for a value of $\alpha_{\rm res}$
slightly deviating from $\alpha_{\rm th}^{\rm even}$, one realizes that a component
of $F_\infty(K)$, proportional to $u_0(K)$, may diverge as $1/(\alpha_{\rm res}-
\alpha_{\rm th}^{\rm even})$. The appearance of such a small denominator implies
that $F_\infty(K)$ is not uniformly bounded in $\alpha_{\rm res}$ in the vicinity
of $\alpha_{\rm th}^{\rm even}$, so that the asymptotic law (\ref{eq:Kasympt}) may
not hold uniformly in $\alpha_{\rm res}$.

This very simply reveals that the singular structure in the recombination coefficient
is a consequence of the existence of a weakly bound trimer.
Quantitatively, as shown in the appendix \ref{appen:singular}, 
by going beyond the $\mathcal{V}_s=\infty$
approximation, one can calculate analytically the contribution to $F(K)$ which becomes
large for $\alpha_{\rm res}$ close to $\alpha_{\rm th}^{\rm even}$. This leads to a Fano profile
\cite{Fano}
\be
\mathcal{K}_{\rm rec} \simeq \mathcal{K}_{\rm rec}^{\rm asympt}\,
\frac{(\alpha_{\rm res}-\alpha_0)^2}{(\alpha_{\rm res}-\alpha_1)^2+\Delta\alpha^2}
\label{eq:fano}
\ee
with the following low-$q_{\rm dim}$ expansions,
\bea
\label{eq:alpha0}
\alpha_0 &\simeq& \alpha_{\rm th}^{\rm even} - 3.2\, q_{\rm dim}^2b \\
\label{eq:alpha1}
\alpha_1 &\simeq& \alpha_{\rm th}^{\rm even} + 6.294\, q_{\rm dim}^2b \\
\label{eq:Da}
\Delta\alpha &\simeq& 35.89\, q_{\rm dim}^3 b^2 \\
\label{eq:amus}
\frac{\Delta \alpha}{\alpha_1-\alpha_0} &\simeq& \frac{16}{3\sqrt{3}} 
\frac{q_{\rm dim}}{\alpha_{\rm th}^{\rm even}}
\eea
where $q_{\rm dim}\simeq 1/\sqrt{\alpha_{\rm th}^{\rm even} \mathcal{V}_s}$.
This is in agreement with the numerical results at finite but 
large $\mathcal{V}_s$, see Fig.\ref{fig:recomb}b.

Note that such a Fano profile in the recombination constant as a function 
of the width of the Feshbach resonance does not occur in the case 
of bosons in the large scattering length limit: 
In the bosonic case, when the scattering length
$a$ becomes much larger than the range and the effective range of
the interaction potential, the recombination constant, apart from an asymptotic
$a^4$ factor, has only a bounded oscillatory behavior as 
a function of $a$ or of the potential range
\cite{Bedaque,Ovchinnikov}. This is a consequence
of the fact that trimers of bosons always exist for large enough $a$, whatever
the width of the Feshbach resonance, and they exist in arbitrarily large
numbers for $a$ arbitrarily large, the so-called Efimov effect;
the oscillatory behavior in the recombination constant then results from
the successive entrances of new Efimov trimers, as $a$ grows.

Finally, turning back to the fermionic case,
we note that, in reality, there may exist deeply bound dimers even
in the limit $|\mathcal{V}_s|\gg b^3$, which may be formed by the collision
of three incoming atoms, in competition with the weakly bound dimer.
This effect, beyond our model Hamiltonian, is discussed in subsection
\ref{subsec:loss_recomb}, where
the corresponding recombination constant towards deeply bound dimers 
is estimated.

\section{Estimation of losses due to deeply bound dimers}
\label{sec:loss}

Very close to the Feshbach resonance,
the model Hamiltonian that we have considered in this work
can support only a weakly bound dimer in the two-body problem, 
that is with an energy much smaller than $\hbar^2/mb^2$.

In real experiments, with alkali atoms, the van der Waals interaction potential
is very deep and support several deeply bound dimers in the two-body problem.
These deeply bound dimers can be formed by three-body collisions, and liberation
of a huge binding energy, leading to particle losses.
As a consequence, the trimers will acquire a finite
lifetime, the atom-dimer scattering will not be purely elastic, and the scattering
of three atoms will lead to recombination not only to the weakly bound dimer,
but also to the deeply bound dimers.

A first possibility to estimate these inelastic contributions is to modify
the Hamiltonian, so as to have deeply bound dimers in the model,
e.g.\  by including a separable interaction potential in
the open channel \cite{Koehler1,Koehler2}, or by considering a true potential
and introducing the adiabatic potential curves in hyperspherical coordinates
\cite{Efimov_hypersp,Esry}.
This however modifies the mathematical structure 
of the problem and is beyond the scope of the present paper.

Fortunately one may easily estimate the loss rate, that is the rate of formation
of deeply bound dimers, by the following recipe, 
expected to be accurate
within an unknown approximately constant factor \cite{Petrov4F,Werner}:
\be
\Gamma_{\rm loss} = \frac{\hbar}{m b^2} P_{<b}
\label{eq:gamma_loss}
\ee
where $P_{< b}$ is essentially the probability that the
three atoms be all within a volume 
of the order of $b^3$. More precisely, since we are using a two-channel model,
this probability can be split in two contributions, one coming from the purely atomic
component (all three particles in the open channel),
\be
P_{<b}^{\rm at} = \int_{\rho < b} d^3r_1 d^3r_2 d^3r_3 
|\Psi_{\rm at}^{\rm norm}(\mathbf{r}_1,\mathbf{r}_2,\mathbf{r}_3)|^2,
\label{eq:Pat}
\ee
where $\rho$ is the hyperradius (\ref{eq:hyperradius}),
and the other contribution coming from the molecular component 
(with one open channel atom and one closed channel molecule),
\be
P_{<b}^{\rm mol} = \int_{|\mathbf{r}_{\rm mol}-\mathbf{r}_{\rm at}|<b} 
d^3r_{\rm mol} d^3r_{\rm at}  
|\Psib_{\rm mol}^{\rm norm}(\mathbf{r}_{\rm mol}; \mathbf{r}_{\rm at})|^2.
\label{eq:Pmol}
\ee
The wavefunctions are here
normalized, hence the apex ``norm", as we shall explain case by case.
It remains to apply a Fourier transform to $\betab$ and to an anti-symmetrized 
version of $A$ in (\ref{eq:ansatz}) 
to calculate numerically the corresponding wavefunctions 
$\Psi_{\rm mol}^{\rm norm}$ and $\Psi_{\rm at}^{\rm norm}$. But one can also
have analytic estimates, by approximating the wavefunctions by their 
small-radius
expansions $|\mathbf{r}_i-\mathbf{r}_j|\ll b$, as we shall see.

\subsection{Lifetime of the trimers}
\label{subsec:loss_trim}

In the case of trimers, the state vector can be normalized in the center of mass
frame, as detailed in the appendix \ref{appen:ad}.
One may then calculate the probabilities in Eqs.(\ref{eq:Pat},\ref{eq:Pmol}) 
numerically; the corresponding rate $\Gamma_{\rm loss}$ 
then represents the inverse lifetime of the trimer due to
spontaneous decay into a deeply bound dimer and a free atom.
We recall that (\ref{eq:gamma_loss}) contains an unknown numerical
factor that depends on the microscopic details of the interaction,
so the values of the lifetimes that we shall obtain are only
indicative.

\begin{figure}
\includegraphics[width=8cm,clip]{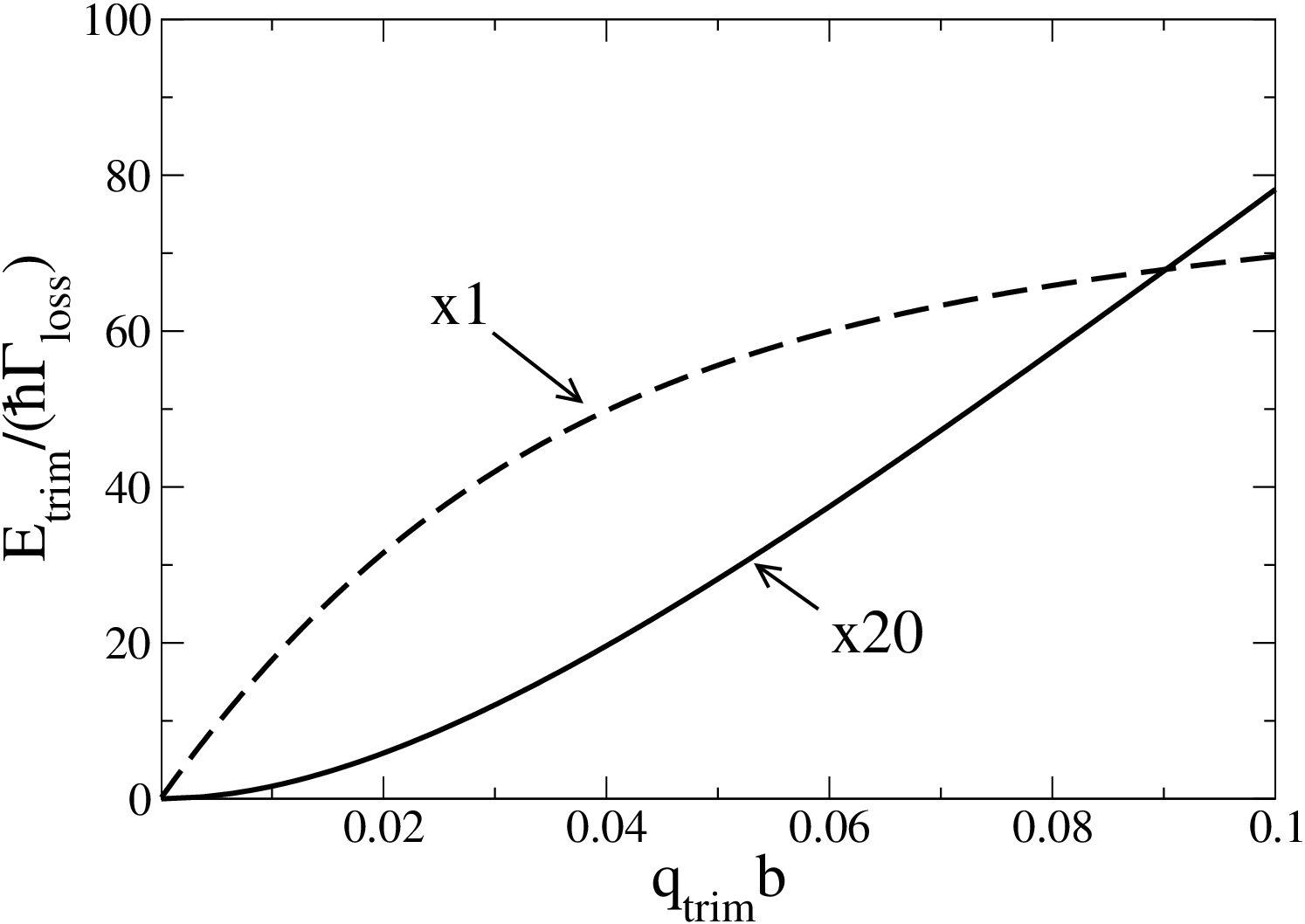}
\caption{For an infinite scattering volume,
quality factor of the trimers, as a function of $q_{\rm trim}b$,
for the even sector (solid line) and for the odd sector (dashed line). 
For clarity, the quality factor in the even sector was multiplied by 20.
The quality factor is defined as the ratio of the trimer binding energy,
here equal to $E_{\rm trim}=\hbar^2 q_{\rm trim}^2/m$ since $\mathcal{V}_s=\infty$,
and of $\hbar$ times the spontaneous decay rate of the trimer
due to the formation of deeply bound dimer and a free atom
as estimated by the simple recipe (\ref{eq:gamma_loss}).}
\label{fig:loss_trim}
\end{figure}

As a consequence, it seems more interesting physically to obtain 
the scaling laws of the trimer lifetime close to the trimer formation
threshold, that is when the trimer binding energy is 
$\ll \hbar^2/(m b^2)$: Does the trimer decay rate tend
to zero 
on the threshold~? Even if this is the case, one cannot immediately
conclude that the weakly bound trimers are long-lived, because their 
binding energies
also tend to zero on the threshold. One rather has to see if the decay rate
tends to zero faster or not than the binding energy 
$E_{\rm trim}^{\rm bind}$ of the trimer.
To this end, we form what we call the quality factor of the trimer:
\be
Q=\frac{E_{\rm trim}^{\rm bind}}{\hbar\Gamma_{\rm loss}}.
\ee
This quality factor is 
shown as a function of $q_{\rm trim}$ 
in Fig.\ref{fig:loss_trim} for an infinite scattering volume, that is
in practice for $1/|\alpha_{\rm res}\mathcal{V}_s|\ll q_{\rm trim}^2$.
One sees that the quality factor $Q$ tends to zero at the threshold
for trimer formation, which is not a positive result. 
The odd sector is however much more favorable
(keeping in mind that the quality factor in the even sector was multiplied by a factor
20 for clarity in the figure): Values of $Q$ much larger than unity are obtained
already for moderately small values of $q_{\rm trim} b$. This is due to the fact
that $Q$ vanishes more slowly in the odd than in the even sector: on the figure,
$Q_{\rm odd}$ seems to vanish linearly whereas $Q_{\rm even}$ seems to
vanish quadratically.

The scaling of the quality factor with $q_{\rm trim}$ close to the trimer formation threshold
can be obtained analytically from the low-$K$ dependence of $\betab(K)$, considering
again an infinite scattering volume.

In the even sector, we have seen in Fig.\ref{fig:fot}a that the function
$B_{L=1}(K)$ right on the threshold diverges at low $K$ as $1/K$ only.
This can be shown, as done in appendix \ref{appen:singular},
using the fact that the kernel $(C_0-C_2)(K,k)$ vanishes linearly in $K$,
whereas the diagonal part of the integral equation vanishes quadratically
as $K^2 \alpha_{\rm res}/4$ for $q_{\rm trim}=0$.
As a consequence, the function $\mathcal{B}(K)=K B_{L=1}(K)$ is bounded,
and the unnormalized state vector $|\Psi\rangle$ of the trimer is square integrable in the
center of mass frame, as is apparent on Eqs.(\ref{eq:norm_mol_even},\ref{eq:norm_at_even}).
We thus face the same phenomenon as in the two-body case: at threshold, the 
normalized
trimer wavefunction is non-zero. As a consequence, the probability of finding
the particles at relative distances less than $b$ tends to a non-zero value at threshold,
$\Gamma_{\rm loss}$ does not vanish,
\be
\lim_{q_{\rm trim}b\to 0} \Gamma_{\rm loss}^{\rm even} > 0,
\ee
and the quality factor $Q_{\rm even}$ tends to zero as $(q_{\rm trim}b)^2$.

In the odd sector, the situation is more favorable. As we have seen in Fig.\ref{fig:fot}b,
for the arbitrary normalization chosen in that figure \cite{laquelle},
the function $B_{L=0}(K)$ right on threshold diverges as $1/K^2$ [whereas the $B_{L=2}(K)$
is $O(1/K)$]. As a consequence, the functions $\betab(\mathbf{K})$ and thus
$\Psi_{\rm mol}$ are not square integrable,
and $\langle \Psi_{\rm at}|\Psi_{\rm at}\rangle$ is also infinite,
see Eqs.(\ref{eq:norm_mol_odd},\ref{eq:norm_at_odd}).
This means that the normalization factor $\mathcal{N}_t$ linking
the correctly normalized state vector to the unnormalized one $|\Psi\rangle$,
\be
|\Psi^{\rm norm}\rangle = \mathcal{N}_t |\Psi\rangle, 
\ee
vanishes for $q_{\rm trim}b$ tending to zero. As a consequence, $P_{<b}$ and $\Gamma_{\rm loss}$
also vanish.

We now determine the corresponding scaling law.
The functions $B_{L=0}$ and $B_{L=2}$ solve a homogeneous integral equation, corresponding
to the homogeneous part of (\ref{eq:ieodd}) [that is with the source term 
set to zero in the right-hand side] written for an energy $E=-\hbar^2 q_{\rm trim}^2/m$.
One can then recycle the reasoning performed below Eq.(\ref{eq:ieodd}). 
At $K\ll 1/b$, the diagonal term 
\be
D(K)\simeq \alpha_{\rm res}(K^2+K_{\rm trim}^2)/4,
\ee
where $K_{\rm trim}=2 q_{\rm trim}/\sqrt{3}$. On the other hand, the first line
of the matrix $M(K,k)$ has a non-zero limit for $K\to 0$, see (\ref{eq:limM}).
For a choice of normalization of the $B$'s such
that the functions $k^2 |B_{L=0,2}|(k)$ are uniformly bounded for 
$K_{\rm trim}$ tending to zero,
we reach the form
\be
B_{L=0}(K) = \frac{\mathcal{E}(K)}{K^2+K_{\rm trim}^2}
\label{eq:form_pour_B0}
\ee
where the envelope function $\mathcal{E}(K)$ has a finite but non-zero limit in $K=0$,
and is uniformly bounded as a function of $K$ and $K_{\rm trim}$
\cite{tres_technique}.
Since the second line of the matrix $M(K,k)$ vanishes for $K\to 0$, we find that 
$B_{L=2}$  is dominated by $B_{L=0}$ at low $K,K_{\rm trim}$ and can be neglected
\cite{justif}. 
Inserting the form (\ref{eq:form_pour_B0}) in the normalization integrals 
(\ref{eq:norm_mol_odd},\ref{eq:norm_at_odd}) we obtain the asymptotic results
in the $q_{\rm trim}\to 0$ limit:
\bea
\langle \Psi_{\rm mol} | \Psi_{\rm mol}\rangle 
&\sim & 
\frac{\sqrt{3}}{16\pi} \frac{\mathcal{E}^2(0)}{q_{\rm trim}} \\
\langle \Psi_{\rm at} | \Psi_{\rm at}\rangle  &\sim&
\frac{1-p_{\rm closed}^{\rm res}}{p_{\rm closed}^{\rm res}}
\langle \Psi_{\rm mol} | \Psi_{\rm mol}\rangle
\eea
where $p_{\rm closed}^{\rm res}=0.185\ldots$ 
here at the odd trimer formation threshold.
This shows that $|\mathcal{N}_t|^2$ scales as $q_{\rm trim}b$.
In calculating the probability $P_{<b}$ to have the particles `inside' the interaction
potential for a non-zero but small $q_{\rm trim}$, 
we can take directly the unnormalized wavefunctions $\Psi$ for
$q_{\rm trim}=0$, so that $P_{<b}$ scales as $|\mathcal{N}_t|^2$ and
\be
\Gamma_{\rm loss}^{\rm odd} \propto \frac{\hbar q_{\rm trim}}{mb}.
\ee
The quality factor $Q_{\rm odd}$ thus
 vanishes linearly in $q_{\rm trim}b$.

\subsection{Losses in atom-dimer scattering}
\label{subsec:loss_ad}

We now estimate the loss rate in the collision of an atom with a weakly bound
dimer. We enclose the atom and the dimer in a fictitious cubic box of volume $L^3$,
so that the state vector in the box can be normalized to unity.
The box has a size $L\gg b,|a_{\rm ad}|$ so that this state vector
differs from the free space one by a normalization factor only,
\be
|\Psi^{\rm norm}\rangle \simeq \frac{1}{L^3} |\Psi\rangle.
\ee
One has then indeed, in the subspace with one atom and one molecule at large distances
:
\be
\Psib_{\rm mol}^{\rm norm}(\mathbf{r}_{\rm mol}; \mathbf{r}_{\rm at}) 
\simeq \frac{p_{\rm closed}^{1/2}}{L^3} 
\left[1-\frac{a_{\rm ad}}
{|\mathbf{r}_{\rm mol}-\mathbf{r}_{\rm at}|}\right]
\mathbf{e}_z ,
\ee
and in the subspace with three atoms, 
for the position of the third atom going to infinity for fixed positions
of atoms one and two \cite{aide,pc},
\be
\Psi_{\rm at}^{\rm norm}(\mathbf{r}_1,\mathbf{r}_2,\mathbf{r}_3) \simeq
\frac{1}{\sqrt{3}L^3}
\phi(\mathbf{r}_1-\mathbf{r}_2) 
\left[1-\frac{a_{\rm ad}}{|\mathbf{r}_3
-(\mathbf{r}_1+\mathbf{r}_2)/2|}\right],
\ee
so that, after spatial integration of the modulus square of these two
wavefunctions in the box, using $\int d^3r\, |\phi(\mathbf{r})|^2=1-p_{\rm closed}$,
one finds that $|\Psi^{\rm norm}\rangle$ is normalized to unity in the cubic box,
to zeroth order in $a_{\rm ad}/L$ and $b/L$.

\begin{figure}
\includegraphics[width=8cm,clip]{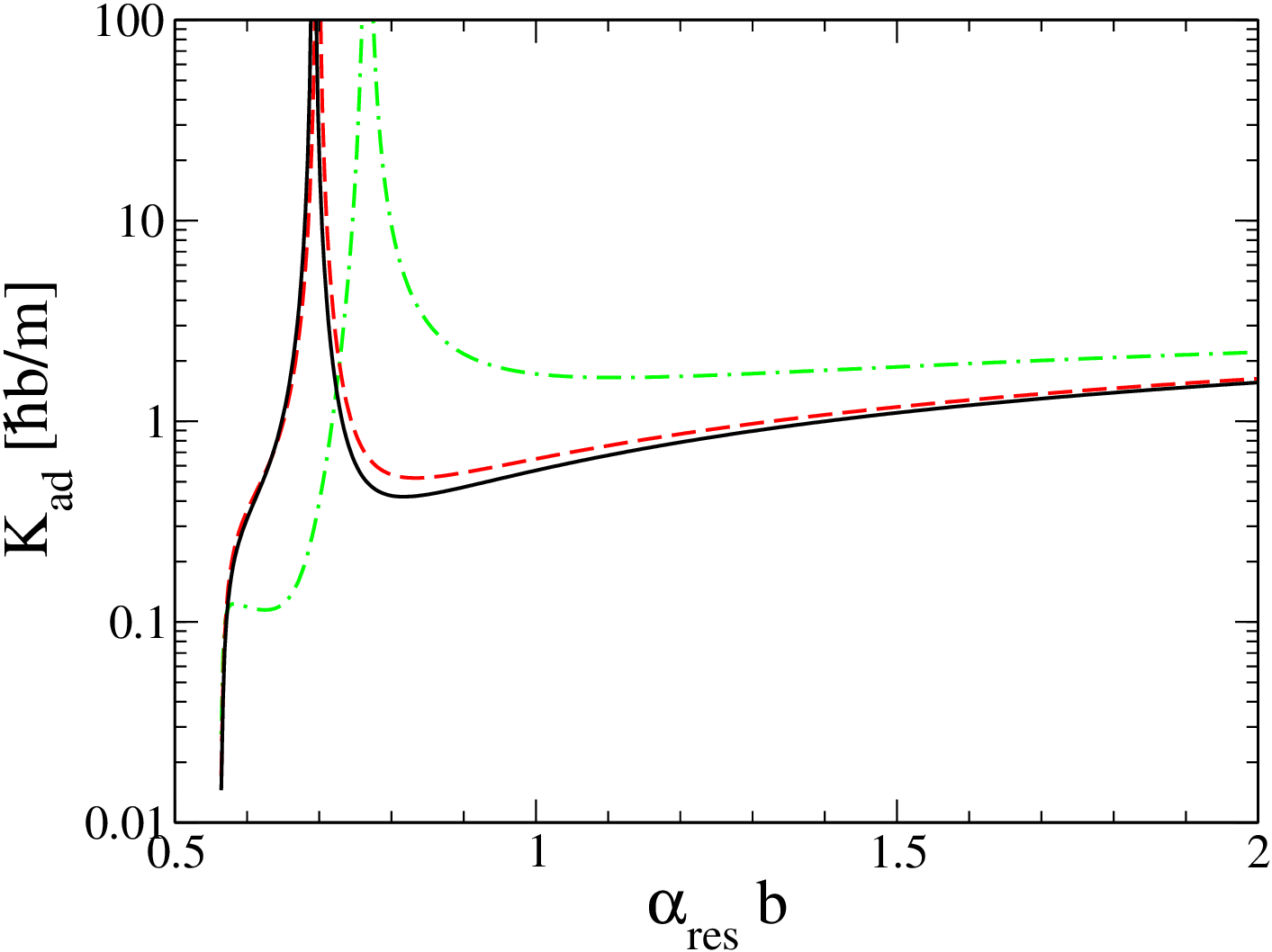}
\caption{Loss constant $K_{\rm ad}$ due to formation of deeply bound dimers 
in the atom-dimer collision, as estimated by the recipe Eq.(\ref{eq:gamma_loss})
and the relation Eq.(\ref{eq:Kad}),
as a function of $\alpha_{\rm res}$ for
$\mathcal{V}_s=10^6 b^3$ (black solid line), $\mathcal{V}_s=10^3 b^3$ (red dashed line),
$\mathcal{V}_s=10 b^3$ (green dotted-dashed line). 
$K_{\rm ad}$ is in units of $\hbar b/m$. 
A divergence of $K_{\rm ad}$ occurs at the threshold of odd trimer formation.
%In reality we expect that $K_{\rm ad}$ should reach a finite value, due to the finite
%lifetime of the trimer; this is beyond the reach of the simple recipe Eq.(\ref{eq:gamma_loss}).
\label{eq:adl}}
\end{figure}

To link this calculation 
with an experimentally relevant quantity, we consider a low density
mixture of 
$N_{\rm at}$ atoms and $N_{\rm dim}$ dimers in
a volume $L^3$. The loss rate will be
\be
\frac{d}{dt} N_{\rm at}= \frac{d}{dt} N_{\rm dim} = -K_{\rm ad}\frac{N_{\rm at} N_{\rm dim}}{L^3}.
\ee
The loss constant $K_{\rm ad}$ is related to Eq.(\ref{eq:gamma_loss}) by setting $N_{\rm at}
=N_{\rm dim}=1$ in the above equation,
\be
K_{\rm ad} = L^3 \Gamma_{\rm loss}
\label{eq:Kad}
\ee
which can be checked to be independent of $L^3$.
The resulting atom-dimer loss constant is plotted in Fig.\ref{eq:adl} as a function of
$\alpha_{\rm res}$. We see that, for $\mathcal{V}_s\gg b^3$,
\be
K_{\rm ad} \propto \frac{\hbar b}{m}
\label{eq:Kadtyp}
\ee
except close to the trimer formation threshold where, within the simple recipe,
$K_{\rm ad}$ diverges. We also see a drop of $K_{\rm ad}$ to a smaller but non
zero value in the limit of broad Feshbach resonances, equal to
$\hbar b/m$ within a numerical factor: this drop is due to the fact that 
$P_{<b}^{\rm mol}$ tends to zero in this limit, so that $P_{<b}$ reduces to the atomic
contribution $P_{<b}^{\rm at}$, which is 
elsewhere dominated by $P_{<b}^{\rm mol}$.

The property (\ref{eq:Kadtyp}) can be understood analytically.
E.g.\ to estimate $P_{<b}^{\rm mol}$, one can approximate $\Psi_{\rm mol}^{\rm norm}$
in (\ref{eq:Pmol}) by its value in $\mathbf{r}_{\rm mol}=\mathbf{r}_{\rm at}$, 
which is generically
non-zero for the odd ansatz. Taking $\mathcal{V}_s=\infty$ gives a finite value
for the wavefunction,
since $B_{L=0}^{\rm out}(K)$ 
diverges as $1/K^2$, see (\ref{eq:we_set}), and this is integrable on a vicinity
of $\mathbf{K}=\mathbf{0}$ in three dimensions. This explains the weak $\mathcal{V}_s$ 
dependence of $K_{\rm ad}$ for large scattering volumes. For $\alpha_{\rm res}b$ of the order
of unity, Eq.(\ref{eq:Kadtyp}) then holds from dimensional analysis, 
apart from the divergence 
close to the trimer formation threshold. At large $\alpha_{\rm res}b$, we see
from (\ref{eq:ieinf}) that the scattered wave $B_{L=0,2}^{\rm out}(K)$ is $O(1/\alpha_{\rm res})$
and is dominated by the contribution of the incoming wave $\propto \delta(\mathbf{K})$.
We thus find again (\ref{eq:Kadtyp}), with a dominant contribution $P_{<b}^{\rm mol}$
from the molecular sector and an atomic sector contribution which is about
$1-p_{\rm closed}^{\rm res}
=O[1/(\alpha_{\rm res}b)]$ times smaller.

To make the discussion more complete, we also estimate the rate of formation of deeply bound
dimers when the atom and the dimer scatter in the $p$-wave, each with a momentum of modulus $K_0$.
In this case the wavefunction for distances between the particles $\ll 1/K_0$
is obtained from the ansatz (\ref{eq:ansatz_volad}), containing the overall factor $K_0$. 
On the contrary, if one takes a box size $L\gg 1/K_0$, the normalization factor linking
$|\Psi^{\rm norm}\rangle$ to $|\Psi\rangle$ remains $\simeq 1/L^3$.
As a consequence, the $p$-wave loss constant $K_{\rm ad}^p=L^3\Gamma_{\rm loss}$ will be proportional
to $K_0^2$. Furthermore, to estimate $P_{<b}^{\rm mol}$, we can expand the wavefunction
$\Psib_{\rm mol}(\mathbf{r}_{\rm mol};\mathbf{r}_{\rm at})$, which is now in the even sector,
to leading order in $\mathbf{r}=\mathbf{r}_{\rm mol}-\mathbf{r}_{\rm at}$, that is to first order,
which amounts in the Fourier transform of $\betab(\mathbf{K})$ to replacing $\exp(i\mathbf{K}\cdot\mathbf{r})$
with $i\mathbf{K}\cdot\mathbf{r}$.
Taking directly the limit of an infinite scattering volume, we see from (\ref{eq:raisonnable}) that this first
order estimate remains finite, because $K/K^3=1/K^2$ is integrable around $K=0$
in 3D. As a consequence, we expect $K_{\rm ad}^{p}$ to be of the order of $\hbar K_0^2 b^3/m$,
except close to the even trimer formation threshold, where it diverges within the present formalism.
This expectation is confirmed by the numerical calculation (not shown).
Physically, this means that, away from the even trimer formation threshold, $K_{\rm ad}^p$ will be smaller
than $K_{\rm ad}$ by a factor of the order of $(K_0 b)^2$, that is at least by a factor $b^3/\mathcal{V}_s$
since we assume here $K_0 \ll q_{\rm dim}$.

\subsection{Recombination to deeply bound dimers}
\label{subsec:loss_recomb}

To complete this section, we now evaluate the rate of formation of deeply bound
dimers in the collision of three asymptotically free atoms.
We again enclose the atoms in a fictitious cubic box of size $L$ with periodic
boundary conditions, so that the normalized state vector in the box is
related to the free space one in the large $L$ limit by (\ref{eq:psi_box}).
Reproducing the reasoning of the appendix \ref{appen:dim_rate}
leading to the prescription (\ref{eq:la_recette}), we can now
define a recombination constant towards deeply bound dimers,
that we call $\mathcal{K}_{\rm rec}^{\rm deep}$.
Applying the resulting prescription to three atoms in the box,
we obtain the equivalent of (\ref{eq:bout_manquant}) for the rate of
formation of deeply bound dimers, with
$\mathcal{K}_{\rm rec}$ replaced with 
$\mathcal{K}_{\rm rec}^{\rm deep}$.
On the other hand, the rate of formation of deeply bound dimers
is $\Gamma_{\rm loss}$ defined by the recipe (\ref{eq:gamma_loss}).
Equating the two expressions of this rate leads to
\be
\mathcal{K}_{\rm rec}^{\rm deep} = 
\frac{L^6 \Gamma_{\rm loss}}{9 (\mathbf{K}_0\wedge
\mathbf{k}_0)^2}.
\ee

We consider first the molecular sector. 
One has to perform the Fourier transform of $\betab(\mathbf{K})$
in Eq.(\ref{eq:ansatz_merveilleux}) to 
obtain the atom-molecule wavefunction.
The Fourier transform of
the first term of the ansatz can be calculated exactly: Since it is composed
of gradients of delta distribution in momentum space, it gives a wavefunction
varying linearly with the coordinates of $\mathbf{r}=\mathbf{r}_{\rm mol} 
-\mathbf{r}_{\rm at}$. 
It is found that this wavefunction is proportional to $\mathcal{V}_s$.
In the second term of the ansatz, we take the large scattering volume
limit, away from the trimer formation threshold, so that
we neglect $K_{\rm dim}^2$ in the denominator and we take
$g(K)\simeq  \mathcal{V}_s F_\infty(K)/K$, as discussed around
(\ref{eq:deffinf}). 
This gives again a contribution proportional to $\mathcal{V}_s$.
The Fourier transform cannot be calculated analytically, but we only
need the wavefunction for $r\lesssim b$ so that we can restrict
to a small-$r$ expansion of the atom-molecule wavefunction. 
The linear order in $\mathbf{r}$ is
the first non-zero one, and we obtain the contribution from the molecular sector
\bea
\left(\mathcal{K}_{\rm rec}^{\rm deep}\right)_{\rm mol} 
\propto \frac{\hbar}{m} b^3 \mathcal{V}_s^2 \alpha_{\rm res} 
p_{\rm closed}^{\rm res}
\nonumber \\
\times \left[
1+\frac{16}{3\pi\alpha_{\rm res}} \int_0^{+\infty} dK\,  F_\infty(K)/F_\infty(0) 
\right]^2.
\eea
The expression in between square brackets is not a slowly varying function
of $\alpha_{\rm res}b$, because it diverges in the vicinity 
of the trimer formation threshold (an artifact of the approximations 
performed here on the second term of the ansatz (\ref{eq:ansatz_merveilleux})).
We have checked numerically that
the expression can be well approximated by the fitting formula
\be
\left[\ldots \right] \simeq \frac{1+\frac{1.239}{\alpha_{\rm res}b-0.538}}
{1-\alpha_{\rm th}^{\rm even}/\alpha_{\rm res}}.
\label{eq:fit}
\ee

The same procedure can be applied in the atomic sector. To estimate the purely atomic wavefunction,
we expand it to leading order in the interatomic distances. The zeroth and first order vanish,
and we get to second order
\be
\Psi_{\rm at} (\mathbf{r}_1,\mathbf{r}_2,\mathbf{r}_3) \simeq
\frac{\sqrt{6}}{2} 
\int \frac{d^3k d^3K}{(2\pi)^6} A(\mathbf{K},\mathbf{k}) 
(\mathbf{x}\wedge\mathbf{y})\cdot
(\mathbf{k}\wedge\mathbf{K})
\ee
where we have introduced the Jacobi-like coordinates
$\mathbf{x}=\mathbf{r}_1-\mathbf{r}_2$
and $\mathbf{y}=\mathbf{r}_3-(\mathbf{r}_1+\mathbf{r}_2)/2$.
This leads to 
\be
\left(\mathcal{K}_{\rm rec}^{\rm deep}\right)_{\rm atom} \simeq
\left(\mathcal{K}_{\rm rec}^{\rm deep}\right)_{\rm mol} 
\frac{1-p_{\rm closed}^{\rm res}}
{p_{\rm closed}^{\rm res}} f_{\rm slow}
\ee
where the factor $f_{\rm slow}$ depends only on $\alpha_{\rm res}b$,
it is of the order of $0.005$ for a broad Feshbach resonance
and it increases by a factor $\simeq 5$ from
broad to narrow Feshbach resonances.

As a consequence, our estimate of the recombination constant
towards deeply bound dimers, away from the trimer formation threshold,
scales as 
\be
\mathcal{K}_{\rm rec}^{\rm deep} \propto
\frac{\hbar}{m} b^3 \mathcal{V}_s^2 \alpha_{\rm res},
\ee
for a given $\alpha_{\rm res}$, in the large scattering volume
limit.
We thus see from (\ref{eq:Kasympt}) that the formation of weakly bound dimers 
wins over the deeply bound ones in this limit.
Note that the estimate of $\mathcal{K}_{\rm rec}^{\rm deep}$ also
holds on the negative scattering volume side of the resonance, still
restricting to the low relative incoming atomic momenta
$k\ll 1/(\alpha_{\rm res} |\mathcal{V}_s|)^{1/2}$. 

\section{Effect of a non-resonant interaction in the open channel}
\label{sec:enri}

In real life there exists an attractive  van der Waals interaction between
atoms in the open channel, responsible for a residual
interaction in the $p$ wave even very far from the Feshbach resonance.
This residual interaction may be characterized by the
so-called background scattering volume 
$\mathcal{V}_s^{\rm bg}$.
Usually it is assumed that this residual interaction is weak,
$|\mathcal{V}_s^{\rm bg}|\approx b^3$ where the interaction
range $b$ is of the order of the van der Waals length, so that
it is neglected in the vicinity of the Feshbach resonance
as compared to the effect of the coupling to the closed channel
\cite{Chevy,Gurarie2}.
However, with the pure closed channel coupling Hamiltonian
(\ref{eq:hamil}) used in this paper, we found that several quantities
were depending not only on the low-$k$ scattering properties
parameterized by $\mathcal{V}_s$ and $\alpha$, but also
on the range $b$ of the potential, such as the threshold for trimer
formation, which raises the issue of their dependence with the
microscopic details of the model. Furthermore, we found that
the atom-dimer scattering length assumes values smaller than $b$
for broad Feshbach resonances, so that it is not evident that
the residual interaction is really negligible.

To address these questions, we model the residual interaction
by a separable potential of coupling
constant $g_0$ with the same cut-off function $\chib(\mathbf{k})$
as in the closed channel coupling. This amounts to adding to the Hamiltonian
(\ref{eq:hamil}) the open channel interaction
\bea
V_{\rm open} = \frac{g_0}{2} \int \frac{d^3K d^3k d^3k'}{(2\pi)^9}
\chib(\mathbf{k}')\cdot \chib^*(\mathbf{k}) \times \nonumber \\
\times a_{\frac{1}{2}\mathbf{K}-\mathbf{k}'}^\dagger 
a_{\frac{1}{2}\mathbf{K}+\mathbf{k}'}^\dagger 
a_{\frac{1}{2}\mathbf{K}+\mathbf{k}}
a_{\frac{1}{2}\mathbf{K}-\mathbf{k}}.
\eea
We determine $g_0$ by relating it to $\mathcal{V}_s^{\rm bg}$ 
from the solution of the two-body problem.
Then we solve the three-body problem again with the simultaneous
inclusion of the closed channel coupling and the open channel
interaction.

\subsection{Modification of the two-body problem}

The calculations proceed along the lines of subsection
\ref{subsec:two-body-aspects}.
The same ansatz (\ref{eq:mgsv2b}) for the two-body state vector
applies; the new term emerging from the action of $V_{\rm open}$ is
simply
\be
V_{\rm open} |\Psi\rangle = 
g_0\gammab \cdot \int\frac{d^3k}{(2\pi)^3} \chib(\mathbf{k})
a_{\mathbf{k}}^\dagger a_{-\mathbf{k}}^\dagger |0\rangle
\ee
where we have set 
\be
\gammab = \int\frac{d^3k}{(2\pi)^3} A(\mathbf{k}) \chib^*(\mathbf{k}).
\ee
From Schr\"odinger's equation at energy $E$,
we find the remarkable property
\be
(E-E_{\rm mol})\frac{\betab}{2\Lambda}+\gammab=\mathbf{0},
\ee
so that $\gammab$ can be expressed in terms of $\betab$ and 
the new reduced scattering amplitude has a form very similar to 
the previous one (\ref{eq:rsa}),
\be
f(k_0)=  \frac{-m k_0^2 e^{-k_0^2 b^2}/(4\pi\hbar^2)}
{\frac{3(E-E_{\mathrm{mol}})}{2\Lambda^2+g_0(E-E_{\mathrm{mol}})}
-\int \frac{d^3k}{(2\pi)^3}\, \frac{k^2 e^{-k^2 b^2}}
{E+i0^+ -\frac{\hbar^2 k^2}{m}}},
\label{eq:fnew}
\ee
where $E=\hbar^2 k_0^2/m$ is the energy of the two-body scattering state.
This leads to a modified expression for the scattering volume,
\be
\frac{1}{\mathcal{V}_s} = \frac{1}{2\pi^{1/2} b^3} - 
\frac{6\pi\hbar^2}{m\Lambda^2} \frac{E_{\rm mol}}
{1-g_0 E_{\rm mol}/(2\Lambda^2)}.
\label{eq:newvs}
\ee

We first analyze the result very far from the $p$-wave Feshbach
resonance.
Taking the limit $E_{\rm mol}\to \infty$ in
the above expression gives the background scattering volume
as a function of the open channel coupling constant,
\be
\frac{1}{\mathcal{V}_s^{\rm bg}} = \frac{1}{2\pi^{1/2} b^3} 
+\frac{12\pi\hbar^2}{m g_0}.
\ee
Since the van der Waals interaction is attractive, we 
take $g_0<0$ in all what follows. 
Then one sees that the background scattering
volume has a dependence with $|g_0|$ similar to the left part
of the Fig.\ref{fig:sw} calculated for a square well potential: 
For increasing values of
$m|g_0|/\hbar^2$ starting from zero, $\mathcal{V}_s^{\rm bg}$
 decreases
from zero to $-\infty$, it diverges on the
critical value
\be
\frac{m |g_0^{c}|}{\hbar^2} = 24 \pi^{3/2} b^3,
\ee
then it decreases from $+\infty$ down to $2\pi^{1/2} b^3$.
The divergence is due to the formation of a dimer 
in the open channel, and this dimer is deeply bound when
$g_0$ is away from the critical value $g_0^{\rm c}$.
The existence of a deeply bound dimer would deeply change
the physical nature of the three-body problem with respect
to our previous analysis:
E.g. it would open a decay channel to the trimer,
which would not exist as a true stationary state anymore but
at most as a resonance. We thus take from now on $|g_0|< |g_0^c|$
so that $\mathcal{V}_s^{\rm bg}<0$. We shall keep in mind that
\be
\mathcal{V}_s^{\rm bg} \approx -b^3
\ee
for a non-resonant interaction in the open channel.
To be complete, we have also calculated the value of the 
parameter $\alpha$ in the presence of open-channel interactions
only:
\be
\alpha_{\rm bg} = \frac{1}{\pi^{1/2} b} + \frac{b^2}{\mathcal{V}_s^{\rm bg}},
\label{eq:alpha_bg}
\ee
which can have any sign since the open-channel interaction is not
resonant.

We now come back to the vicinity of the Feshbach resonance,
where the closed channel coupling is no longer negligible.
First we can prove that, under the condition $\mathcal{V}_s^{\rm bg}<0$,
there is no bound state in the two-body problem
for $\mathcal{V}_s < 0$ and there is one for $\mathcal{V}_s>0$,
as expected; the proof was obtained
using the argument of a monotonic variation
of an appropriate function as done in the paragraph below 
(\ref{eq:mudelta}).

We restrict for simplicity
to the exactly resonant case $\mathcal{V}_s=\infty$. 
The corresponding values of the closed-channel
detuning $E_{\rm mol}$ and of $\alpha$ are then:
\bea
\label{eq:emolresnew}
\frac{g_0 E_{\rm mol}^{\rm res}}{2\Lambda^2} &=& 
\frac{\mathcal{V}_s^{\rm bg}}{2\pi^{1/2} b^3} \\
\alpha_{\rm res} &=&
\frac{1}{\pi^{1/2}b}+
\frac{6\pi\hbar^4}{m^2\Lambda^2
\left(1-\frac{\mathcal{V}_s^{\rm bg}}{2\pi^{1/2}b^3}\right)^2}.
\label{eq:alpharesnew}
\eea
We see that the range of variation of $\alpha_{\rm res}$
remains the same as in our previous model.
The scattering amplitude, analytically continuated to negative 
energies $E=-\hbar^2 q^2/m$, $q>0$, can be put in the simple
form
\be
\frac{e^{q^2 b^2}}{f(iq)} \stackrel{\mathcal{V}_s\to \infty}{=}
q e^{q^2b^2} \mbox{erfc}(qb) -\frac{1}{b \pi^{1/2}}
-\frac{\alpha_{\rm res}-1/(b\pi^{1/2})}{1+q^2/q_{\rm open}^2},
\ee
making it apparent that the main effect of the open channel interaction
is to introduce a new scale $q_{\rm open}$ for the wave vectors,
such that
\be
q_{\rm open}^2 = \frac{g_0 E_{\rm mol}^{\rm res}-2\Lambda^2}
{g_0\hbar^2/m}= 
\frac{-1/\mathcal{V}_s^{\rm bg}}{\alpha_{\rm res}-1/(\pi^{1/2}b)}.
\ee
This allows to reach first conclusions on the effect of the open channel
interaction on the properties of the original model (\ref{eq:hamil}):
\begin{itemize}
\item for a broad Feshbach resonance $\Lambda\gg\hbar^2b^{1/2}/m$,
we find that $q_{\rm open}b>1$, so the open channel interaction
should have a weak effect. In particular, this suggests
that the trimer states in the regime of rather broad resonances
should be weakly affected. Furthermore, if $\mathcal{V}_s^{\rm bg}$
is weak enough to have $q_{\rm open}b>1$ at the threshold
for the formation of the trimers in our previous model,
\be
|\mathcal{V}_s^{\rm bg}| < \frac{b^3}{\alpha_{\rm th}b-\frac{1}{\pi^{1/2}}}
\label{eq:empirique}
\ee
then the threshold itself 
should be weakly affected by the open channel interaction.
For a narrow Feshbach resonance, one has $q_{\rm open}b<1$ (except
for $|\mathcal{V}_s^{\rm bg}|\ll b^3$); in this case
the effect of the open channel coupling is more difficult to guess:
It may depend on the considered quantity and a more detailed
analysis is required.
\item if $q_{\rm open}$ was smaller than the estimate
$q_{\rm dim}\approx 1/(\alpha_{\rm res} \mathcal{V}_s)^{1/2}$ 
for the wave vector associated
to the dimer binding energy close to the resonance, then the open
channel interaction would have a dramatic effect. But one 
finds $q_{\rm dim}^2/q_{\rm open}^2 < |\mathcal{V}_s^{\rm bg}/\mathcal{V}_s|$ 
so that $q_{\rm dim}\ll q_{\rm open}$
in the resonant regime.
\end{itemize}
A last relevant quantity is the probability to find the dimer in the closed
channel. In the limit $\mathcal{V}_s\to +\infty$, after some calculation,
we find 
\bea
\label{eq:pclosednew}
p_{\rm closed}^{\rm res}  &=& \frac{6\pi\hbar^4}{m^2\Lambda^2}
\alpha_{\rm res}^{-1} \left(1-\frac{g_0E_{\rm mol}^{\rm res}}{2\Lambda^2}
\right)^{-2} \\
&=& 1 -\frac{1}{\pi^{1/2}\alpha_{\rm res} b}. 
\eea
The relation (\ref{eq:p_closed_infini}) is therefore affected by the
open channel interaction, whereas (\ref{eq:p_closed_infini_pas_univ})
is not.

\subsection{Modification of the three-body problem}

We now solve the three-body problem in presence
of both the closed channel coupling and the
open channel interaction.
The previous ansatz (\ref{eq:ansatz}) applies, but a new term
arises in Schr\"odinger's equation of eigenenergy $E$, 
\be
V_{\rm open} |\Psi\rangle =
g_0 \int \frac{d^3K d^3k}{(2\pi)^6}
\chib(\mathbf{k}) \cdot \gammab(\mathbf{K}) 
a^\dagger_{\frac{1}{2}\mathbf{K}+\mathbf{k}} 
a^\dagger_{\frac{1}{2}\mathbf{K}-\mathbf{k}}
a^\dagger_{-\mathbf{K}} |0\rangle
\ee
where we have set
\bea
\gammab(\mathbf{K}) &=&
 \int\frac{d^3k}{(2\pi)^3}
\left[A(\mathbf{K},\mathbf{k}) +\phantom{\frac{1}{1}}\right. \nonumber \\
&&\left. +2 A(-\frac{1}{2}\mathbf{K}+\mathbf{k},-\frac{3}{4}\mathbf{K}
-\frac{1}{2}\mathbf{k})\right]\chib^*(\mathbf{k}).
\label{eq:gamma3c}
\eea

Schr\"odinger's equation projected onto the molecular subspace
is unaffected by the open channel interaction, since there
is only one atom in that subspace. So Eq.(\ref{eq:mol_atom})
still holds exactly. Then one immediately sees 
that a simple relation relates $\gammab$ to $\betab$:
\be
\left[E-E_{\rm mol}-\frac{3\hbar^2 K^2}{4m}\right]
\frac{\betab(\mathbf{K})}{2\Lambda} + \gammab(\mathbf{K})
= \mathbf{0}.
\ee
One may then take as unknown any convenient combination
of $\gammab$ and $\betab$.

In Schr\"odinger's equation projected onto the atomic subspace,
a new term appears, but of the same structure as the term involving
$\betab$ in (\ref{eq:brute}). Thus the modification 
to (\ref{eq:A}) is minor,
\be
A(\mathbf{K},\mathbf{k}) =  A_0(\mathbf{K},\mathbf{k})
+\frac{[g_0\gammab(\mathbf{K})-\Lambda \betab(\mathbf{K})]
\cdot \chib (\mathbf{k})}
{E+i0^+-\frac{\hbar^2 }{m}\left(\frac{3}{4}K^2 + k^2\right)}.
\label{eq:newA}
\ee
This results immediately suggests which combination
of $\gammab$ and $\betab$ is convenient: We introduce
\be
\betab_{\rm eff}(\mathbf{K}) = \betab(\mathbf{K})
-\frac{g_0}{\Lambda} \gammab(\mathbf{K}),
\ee
the overall factor being such that $\betab_{\rm eff}$
reduces to $\betab$ in the absence of open channel interaction.
As a consequence, our unknown field is now
\be
\betab_{\rm eff}(\mathbf{K}) = 
\left[1+\frac{g_0}{2\Lambda^2}\left(E-E_{\rm mol} 
-\frac{3\hbar^2 K^2}{4m}\right)\right]
\betab(\mathbf{K}).
\label{eq:beta_eff}
\ee
Eliminating $\betab$ and $A$ in terms of this unknown field
in (\ref{eq:mol_atom}), we find after some calculations
and using (\ref{eq:fnew})
that $\betab_{\rm eff}(\mathbf{K})$ solves
{\sl the same equation}  (\ref{eq:integrale}) as $\betab$ in our previous model,
{\sl provided} that the modified scattering amplitude
(\ref{eq:fnew}) is used. As a consequence, all
the numerical and most of the analytical
techniques developed for the previous model may be reused for the
new model.

\begin{figure}[htb]
\includegraphics[width=8cm,clip]{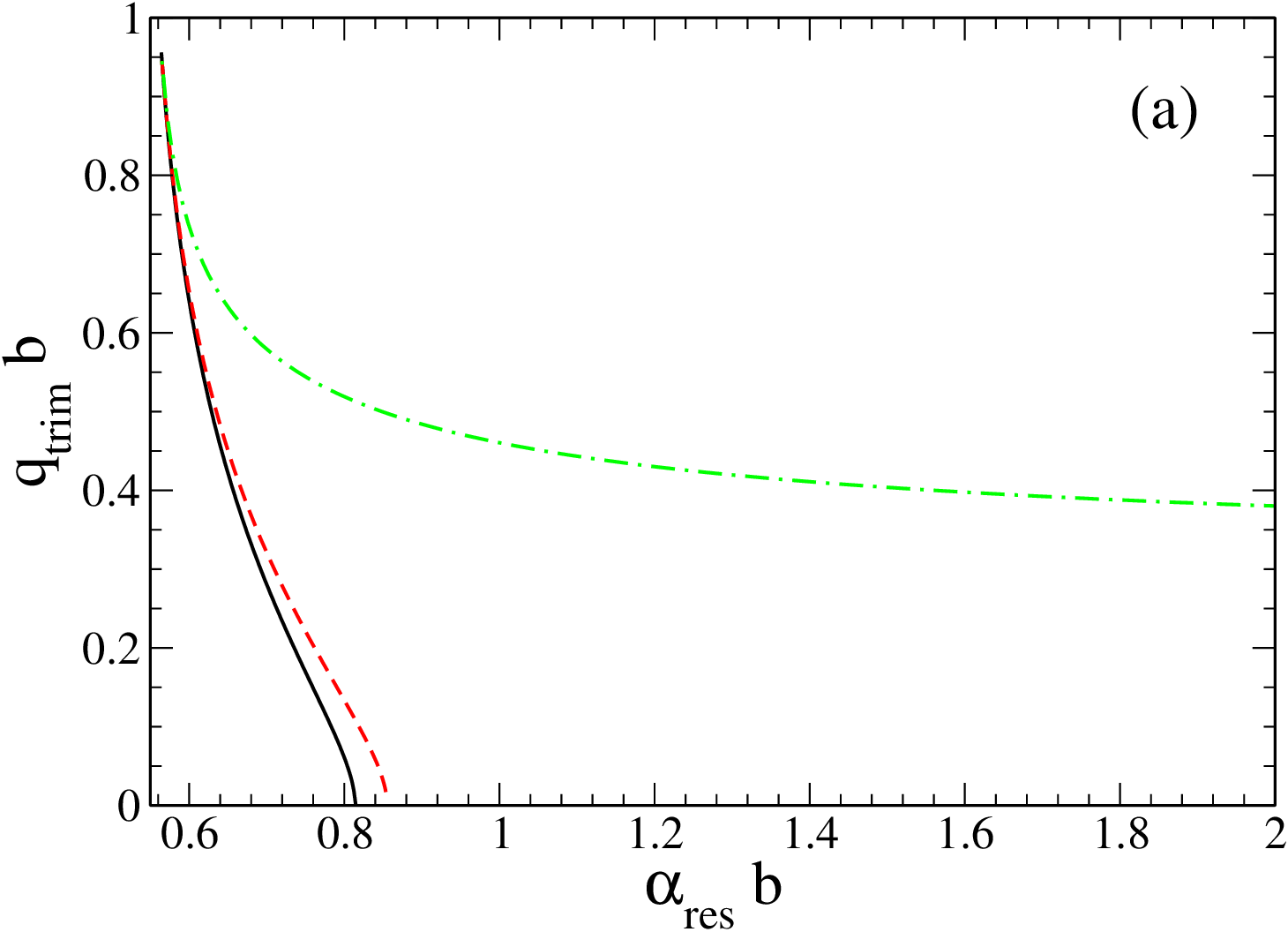}
\includegraphics[width=8cm,clip]{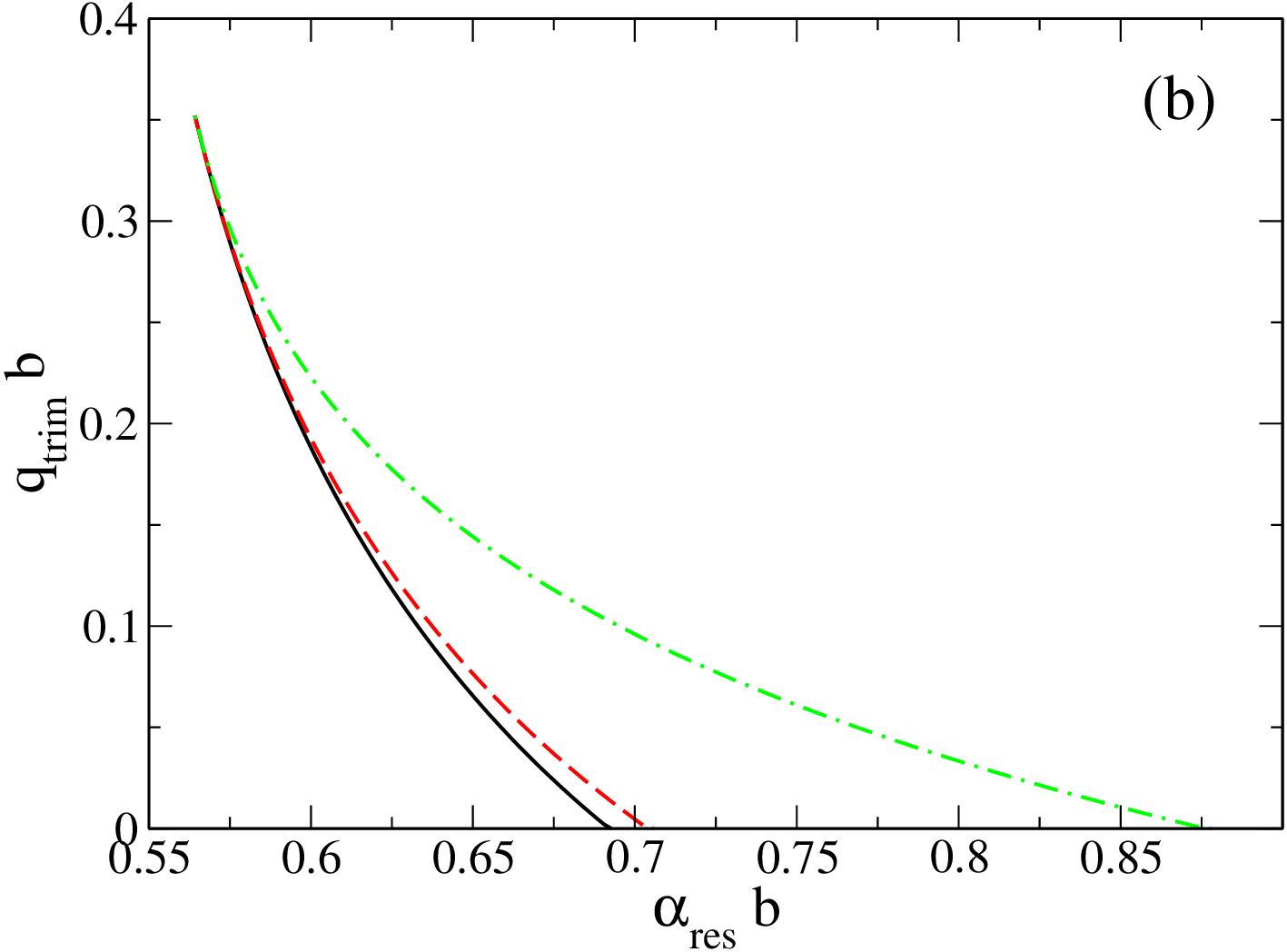}
\caption{(Color online) 
In presence of open channel interactions, characterized by a fixed
value of the background scattering volume $\mathcal{V}_s^{\rm bg}$,
parameter $q_{\rm trim}$ of the trimer (when it exists)
(a) in the even sector
and (b) in the odd sector,
as a function of $\alpha_{\rm res}b$. Here the scattering volume
$\mathcal{V}_s$ is infinite and $\alpha_{\rm res}$ is the corresponding
value of the parameter $\alpha$ for the modified scattering amplitude,
see Eq.(\ref{eq:alpharesnew}).
Solid line (black): $\mathcal{V}_s^{\rm bg}=0$. Dashed line (red):
$\mathcal{V}_s^{\rm bg}=-b^3$. Dashed-dotted line (green): 
$\mathcal{V}_s^{\rm bg}=-10 b^3$.}
\label{fig:trim_pair2}
\end{figure}

\noindent {\sl Existence of weakly bound trimers}:
We reproduce the numerical calculations of subsection \ref{subsec:ewbt}
with the scattering amplitude modified by the open channel
interaction. We restrict for simplicity to an infinite scattering
volume:
See Fig.\ref{fig:trim_pair2} giving $q_{\rm trim}b$ as a function
of $\alpha_{\rm res}b$, where the energy of the trimer is
$-\hbar^2 q_{\rm trim}^2/m$, for (a) the even sector and (b) the odd one.
Then we see that the trimer state still exists for low
$\alpha_{\rm res} b$. For a small background scattering
volume $\mathcal{V}_s^{\rm bg}=-b^3$,  
its energy dependence with $\alpha_{\rm res}b$
is only weakly affected by the open channel interaction, 
as expected from the qualitative condition
(\ref{eq:empirique}), and the threshold is only slightly shifted.
For a much more negative background scattering volume
$\mathcal{V}_s^{\rm bg}=-10 b^3$, there is simply a
larger shift in the odd sector, but the conclusion is radically changed
in the even sector, see Fig.\ref{fig:trim_pair2}a: 
the trimer seems to exist now for all values of $\alpha_{\rm res}b$.
This may be understood as follows: in the large $\alpha_{\rm res} b$
limit, the coupling $\Lambda$ to the closed channel tends to zero,
so does $E_{\rm mol}^{\rm res}$,  see (\ref{eq:emolresnew}),
and for a fixed non-zero $q$, the scattering amplitude
(\ref{eq:fnew}) converges to the one of a single channel model
with a scattering volume $\mathcal{V}_s^{\rm bg}$. If
$|\mathcal{V}_s^{\rm bg}|/b^3$ is large enough, then this single
channel model can indeed support a trimer \cite{pourquoi}.
On the other hand, the fact that the threshold for trimer formation
survives in the odd sector up to higher values of $|\mathcal{V}_s^{\rm bg}|
/b^3$ than in the even sector may be understood from the qualitative
argument (\ref{eq:empirique}).

\begin{figure}[htb]
\includegraphics[width=8cm,clip]{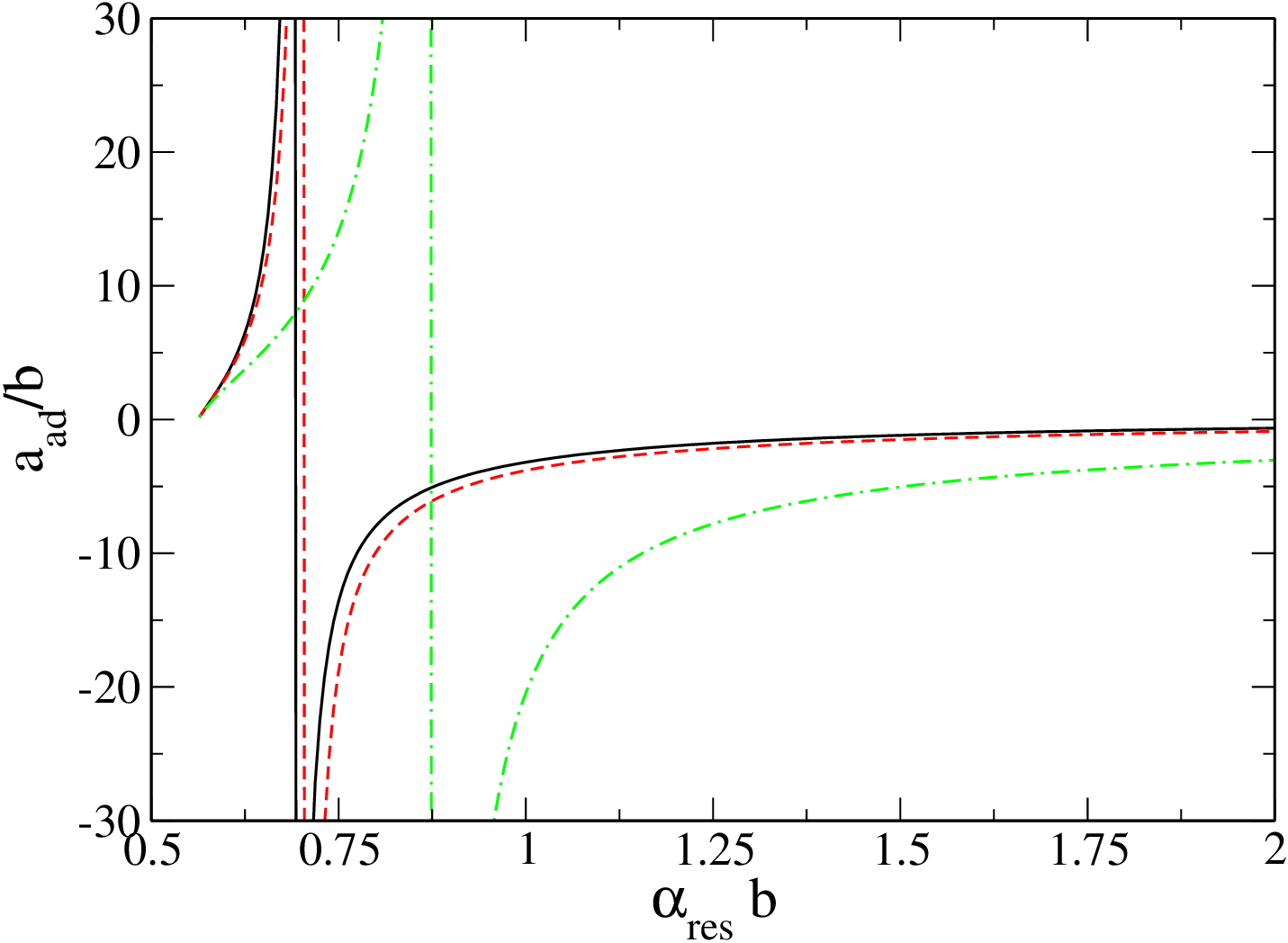}
\caption{(Color online) Atom-dimer scattering length
$a_{\rm ad}$ as a function of $\alpha_{\rm res}b$ for
the model including the open channel interaction,
for an infinite scattering volume $\mathcal{V}_s$.
Solid line (black): $\mathcal{V}_s^{\rm bg}=0$.
Dashed line (red): $\mathcal{V}_s^{\rm bg}=-b^3$.
Dashed-dotted line (green): $\mathcal{V}_s^{\rm bg}=-10 b^3$.}
\label{fig:a_ad2}
\end{figure}

\noindent{\sl $S$-wave atom-dimer scattering}:
We consider for $\mathcal{V}_s>0$ the scattering of an atom on a dimer 
in the limit of a vanishing relative kinetic energy, so that
$E=-E_{\rm dim}$, where $E_{\rm dim}$ is the dimer binding energy
for the new model, and the corresponding $s$-wave scattering is characterized
by the atom-dimer scattering length $a_{\rm ad}$.
Then the field $\betab(\mathbf{K})$ is given by
the ansatz (\ref{eq:with_delta}). As a consequence, the effective
field (\ref{eq:beta_eff}) will have the same structure; the delta distribution
$\delta(\mathbf{K})$ will simply be multiplied by the factor
$1-g_0 (E_{\rm dim} + E_{\rm mol})/(2\Lambda^2)$. Also the part of
$\betab_{\rm eff}$ diverging as $1/K^2$ will be related
to the part of $\betab(\mathbf{K})$ diverging as $1/K^2$ 
by exactly the same factor. So that one may read the value of
the atom-dimer scattering length directly from the effective field.
As a consequence, one has to solve the same integral equation
(\ref{eq:ieodd}), just changing the diagonal part $D(K)$
to account for the new scattering amplitude.
The corresponding numerical results for $a_{\rm ad}$ are presented
in Fig.\ref{fig:a_ad2}, as functions of $\alpha_{\rm res}$,
for an infinite scattering volume
$\mathcal{V}_s$. For the considered values
of $|\mathcal{V}_s^{\rm bg}/b^3|$, the open channel
interaction does not qualitatively change the result:
There exists a threshold for the
formation of a trimer in the odd sector, see Fig.\ref{fig:trim_pair2}b, and
we recover the divergence of $a_{\rm ad}$ at this threshold already
observed in our first model.

What happens to the atom-dimer scattering length 
in the large $\alpha_{\rm res}b$ limit~? 
Is the analytical prediction (\ref{eq:asympt_equiv_aad}) obtained
in our previous model still valid~? 
In presence of interactions in the open channel,
it seems surprising that $a_{\rm ad}$ can tend to zero 
at large $\alpha_{\rm res} b$, since the scattering amplitude 
for $\Lambda\to 0$ for a finite $q$ tends to the one of a single
channel model with a scattering volume $\mathcal{V}_s^{\rm bg}$.
This limit however is not reached uniformly in $q$: e.g.\ for
$q\simeq q_{\rm dim}$, $q\ll q_{\rm open}$ and
the scattering amplitude remains very close to the one of the
two channel model with no open channel interaction.
This non-uniformity of the $\Lambda\to 0$ limit 
is also revealed at $\mathcal{V}_s=\infty$
from the fact that the parameter $\alpha_{\rm res}$
is very different from the one (\ref{eq:alpha_bg}) that one
would have in the absence of coupling to the closed channel.
Mathematically, Eq.(\ref{eq:exact_relation}) still applies,
if one considers the effective field $\betab_{\rm eff}(\mathbf{K})$
rather than $\betab(\mathbf{K})$,
but in the infinite scattering volume limit analysis,
the function $h(K)$ in (\ref{eq:atsf}) is changed by the open-channel
interaction in
\be
h(K) =q e^{q^2b^2} \mbox{erfc}(qb) +\left(
\alpha_{\rm res}-\frac{1}{b \pi^{1/2}}\right)
\frac{q^2}{q^2+q_{\rm open}^2}
\ee
with $q=\sqrt{3}K/2$.
Contrarily to our previous model, the function $h(K)$ increases
from $0$ to $\alpha_{\rm res}$ when $K$ increases from $0$ to infinity,
in practice to values $\gg 1/b$.
We can no longer assume $1/(\alpha_{\rm res}-h(K))
\simeq 1/\alpha_{\rm res}$ for all $K$ in the large $\alpha_{\rm res}$ limit,
so we have only the weaker result that the right hand side
of (\ref{eq:exact_relation}) is $O(1)$ in this limit.
Since $h(K=0)=0$, $D(K)/K^2$ still converges to
$\alpha_{\rm res}/4$ in $K=0$ and we get
\be
a_{\rm ad} \stackrel{\alpha_{\rm res}b\gg 1}{=}
O\left(\frac{1}{\alpha_{\rm res}}\right).
\label{eq:new_equiv}
\ee
We have successfully compared this analytical prediction to the numerics.
For very large values of $\alpha_{\rm res} b$ (not shown),
we numerically find for $\mathcal{V}_s^{\rm bg} <0$ that $a_{\rm ad}$
tends to zero as $C/\alpha_{\rm res}$, where the constant $C$
depends on the background scattering volume, $C\simeq -0.69$ for
$\mathcal{V}_s^{\rm bg}=-b^3$, and $C\simeq -4.1$ for
$\mathcal{V}_s^{\rm bg}=-10b^3$. 

\begin{figure}
\includegraphics[width=8cm,clip]{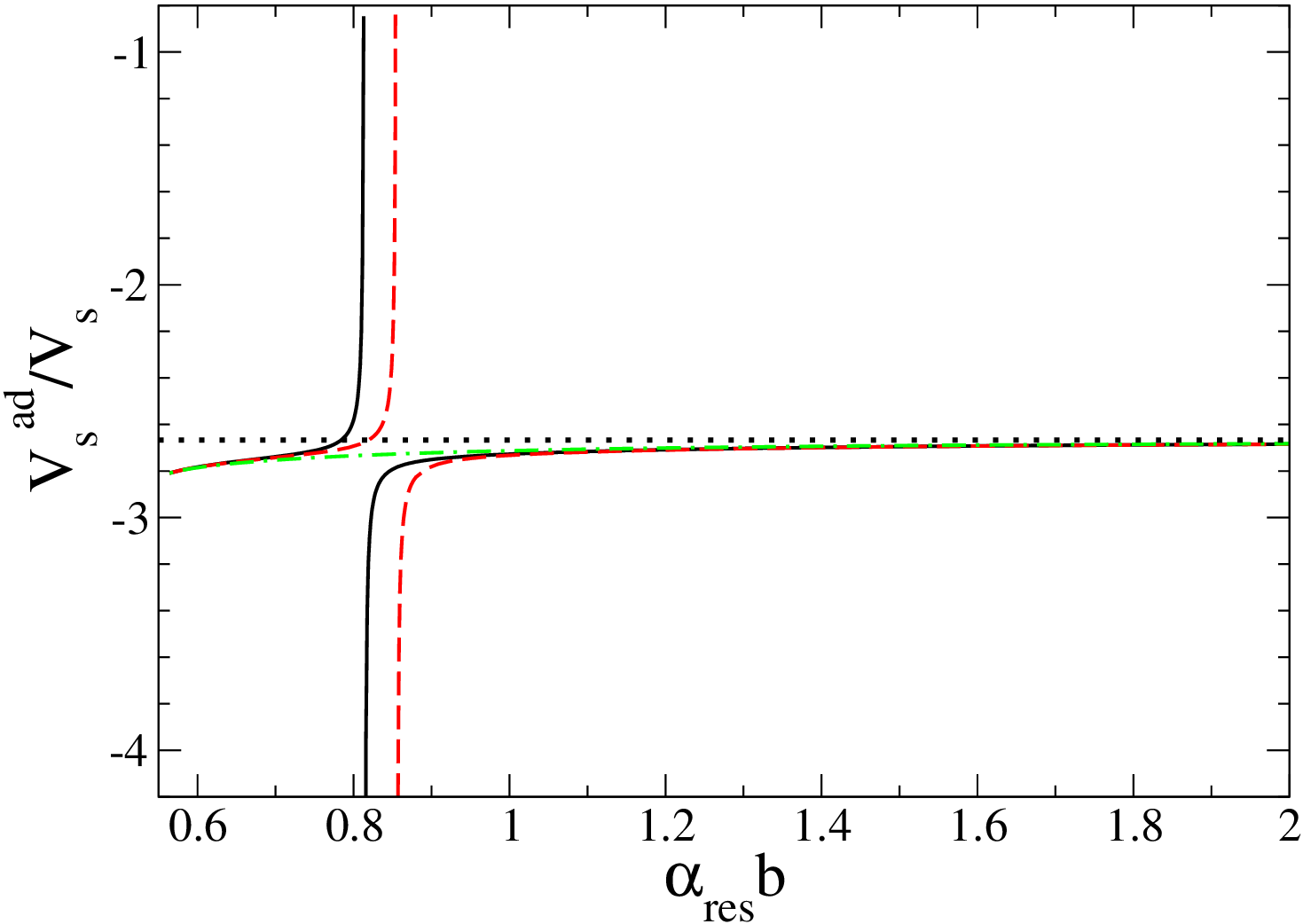}
\caption{(Color online) In the model including an open channel
interaction, atom-dimer scattering volume $\mathcal{V}_s^{\rm ad}$ for a total spin $J=1$ (as detailed in subsection \ref{subsec:ads})
as a function of $\alpha_{\rm res}b$, 
for a fixed value of the atom-atom scattering volume
$\mathcal{V}_s=10^4 b^3$ (solid black line), but for
various background scattering volumes in the open channel:
$\mathcal{V}_s^{\rm bg}=0$ (black solid line),
$\mathcal{V}_s^{\rm bg}=-b^3$ (dashed red line),
$\mathcal{V}_s^{\rm bg}=-10b^3$ (dashed-dotted green line).
$\mathcal{V}_s^{\rm ad}$ is expressed in units of $\mathcal{V}_s$. 
Dotted horizontal line: analytical prediction
(\ref{eq:asympt_volad})
in the limit $\mathcal{V}_s\to +\infty$.}
\label{fig:volad_oc}
\end{figure}

\noindent{\sl $P$-wave atom-dimer scattering:}
Now the incoming atom and dimer have a relative orbital momentum
$L=1$, so the vanishing kinetic energy limit of the scattering
$E\to -E_{\rm dim}$ is characterized by an atom-dimer scattering volume 
$\mathcal{V}_s^{\rm ad}$, with the ansatz (\ref{eq:ansatz_volad}) for 
$\betab(\mathbf{K})$ in the sector of total spin $J=1$
(see subsection \ref{subsec:ads}).
Since the factor linking $\betab_{\rm eff}(\mathbf{K})$ to
$\betab(\mathbf{K})$ varies only quadratically with $K$, it may be replaced
by its $K=0$ value in front of the gradient of $\delta(\mathbf{K})$,
so that $\betab_{\rm eff}$ and $\betab$ have the same low-$K$ 
behavior, from which $\mathcal{V}_s^{\rm ad}$ is readily extracted.
We numerically solve (\ref{eq:intadp}) updating the values of
the scattering amplitude $f(k)$ and the dimer binding energy
$E_{\rm dim}$ to include the open channel interaction.
The dependence of $\mathcal{V}_s^{\rm ad}$ with $\alpha_{\rm res}$ is
shown in Fig.\ref{fig:volad_oc}, for a fixed and large scattering
volume $\mathcal{V}_s$ and for various values of the background
scattering volume. It is apparent that $\mathcal{V}_s^{\rm ad}$
is weakly affected by the open channel interaction, apart in the vicinity
of the even trimer formation threshold (if it exists).
This can be understood analytically, realizing that the reasoning
leading to (\ref{eq:asympt_volad}) still applies in presence of the considered
open channel interaction. This confirms the `universality' of
the asymptotic behavior
\be
\mathcal{V}_s^{\rm ad} \sim -\frac{8}{3} \mathcal{V}_s
\ee
which was expected in subsection \ref{subsec:ads} from the fact
that it does not depend on the potential range $b$.

\noindent{\sl Recombination rate to weakly bound dimers:}
Finally, we consider the scattering problem of three atoms
in the zero total energy limit, for $\mathcal{V}_s>0$
and in presence of open channel interactions.
We take for the effective field $\betab_{\rm eff}(\mathbf{K})$
the same ansatz as in (\ref{eq:ansatz_merveilleux}), putting
the subscript ``eff" on the various functions of the ansatz.
The function $\mathcal{G}_{\rm eff}(K)$ is still given
by (\ref{eq:valG}) but the scattering amplitude is now changed.
As $\mathcal{G}_{\rm eff}$ is however multiplied by the gradient
of a delta, only its first derivatives in $\mathbf{K}=\mathbf{0}$
matter, so only the scattering volume comes out, and
$g_{\rm eff}$ obeys the integral equation (\ref{eq:intevensec}) 
with the same source term and the updated scattering amplitude.
In the large scattering volume limit, away from the even trimer formation
threshold if it exists, we can thus recycle (\ref{eq:Finf0}),
since it involves only the low-momentum behavior $K\leq K_{\rm dim}
\ll q_{\rm open}$ of the scattering amplitude and properties
of the function $C_0-C_2$; this leads to
\be
g_{\rm eff}(K_{\rm dim}) \sim - \frac{144 \pi \hbar^2}
{m\Lambda \alpha_{\rm res}} \frac{\mathcal{V}_s}{K_{\rm dim}}
K_0 k_0^\perp.
\ee
On the other hand, the reasoning leading to the recombination
constant $\mathcal{K}_{\rm rec}$ in (\ref{eq:Krec_e}) still holds,
with the function $g(K)$ (and not $g_{\rm eff}(K)$) and
the closed channel probability $p_{\rm closed}$ modified
by the open channel interaction.
The functions $g$ and $g_{\rm eff}$ differ, because $\betab$
and $\betab_{\rm eff}$ differ by a factor depending on $K$.
Since the factor in between square brackets in (\ref{eq:beta_eff})
cannot vanish for our choice $\mathcal{V}_s^{\rm bg}<0$,
$\betab(K)$ and $\betab_{\rm eff}$ both have a single pole
in $K=K_{\rm dim}$, with
\be
g(K_{\rm dim}) = 
\frac{g_{\rm eff}(K_{\rm dim})}{1-\frac{g_0}{2\Lambda^2}(E_{\rm mol}
+\hbar^2 q_{\rm dim}^2/m)}.
\ee
In the large $\mathcal{V}_s/b^3$ limit, one can neglect $q_{\rm dim}$
in the denominator of this expression, which amounts to neglecting 
$q_{\rm dim}$
with respect to $q_{\rm open}$.
Then we use the expression (\ref{eq:pclosednew}) and remarkably
we recover exactly the same asymptotic behavior 
for $\mathcal{K}_{\rm rec}$ as in the previous model, under
the assumption $\mathcal{V}_s^{\rm bg}<0$:
\be
\mathcal{K}_{\rm rec} 
\stackrel{\mathcal{V}_s\gg b^3}{\sim}
\frac{\hbar}{m} (48\pi)^2 
\left(\frac{\mathcal{V}_s^5}{3\alpha_{\rm res}}\right)^{1/2}.
\label{eq:Kasymptnew}
\ee
This indicates some `universality' of this result, which could be
hoped from the fact that it does not depend on the interaction
range $b$.
%toto

\section{Conclusion}
\label{sec:conclusion}

We have solved the free space
three-body problem for single spin state fermions resonantly
interacting in $p$-wave {\it via} a two-channel Feshbach resonance,
in the sector of total angular momentum one.

The central model that we used to describe the interaction 
depends on three parameters, the scattering volume $\mathcal{V}_s$
which diverges on resonance, an effective range parameter
$\alpha$ and the spatial range of the interaction $b$.
Whereas $b$ is of the order of the van der Waals length, the
parameter $\alpha$ on resonance
can range from a strictly positive minimal value of the order
of $1/b$ up to plus infinity, the minimal value
being model-dependent and equal to $1/(b\pi^{1/2})$ for our Gaussian
cut-off function.
In present
experiments one estimates $\alpha b\approx 3$ for ${}^{40}$K
\cite{Ticknor} and $\alpha b\approx 3$ for ${}^{6}$Li \cite{Chevy}
on resonance.
The two-body scattering
amplitude for $\mathcal{V}_s=\infty$ is $f_k\simeq -1/\alpha$
for low relative momenta $k \ll 1/b$, so that $|f_k|$ right on resonance
is at most of the order of $b$, which is extremely
small as compared to the usual $s$-wave unitary limit.
As a consequence, the resonant three-body problem has very different
properties from the $s$-wave one. 

First, it does not exhibit
the Efimov effect but it admits two trimers, one with
even parity and the other with odd parity, for low enough values
of $\alpha b$. Since the considered sector is of angular momentum one,
each trimer is three-fold degenerate. For 
$\mathcal{V}_s/b^3$ large and negative, our model Hamiltonian 
does not have a two-body bound state, so that these trimer states
are examples of Borromean states. However, we estimate that the spontaneous decay
rate $\Gamma_{\rm loss}$ of the trimers, due to the formation of deeply bound dimers
present in current experiments with real atoms, eventually becomes larger
than the binding energy of the trimers (over $\hbar$) if one gets very
close to their formation
threshold: in the limit of a vanishing trimer binding energy,
$\Gamma_{\rm loss}$ tends to a non-zero limit 
for the even trimer, 
and vanishes as the square root of the trimer binding energy for the odd trimer.

Second, the atom-dimer scattering length $a_{\rm ad}$, characterizing
the low-energy (that is $s$-wave) scattering of an atom on a dimer,
assumes small values, of the order
of $b$ (or even below in absolute value for ultra-narrow Feshbach
resonances), except close to the odd dimer
formation threshold where it diverges.
The fact that $a_{\rm ad}$ depends on the interaction range $b$ shows
that it is not a `universal' quantity and it is sensitive to the
microscopic details of the interaction.
Furthermore, the loss constant in the inelastic atom-dimer
$s$-wave scattering (due to the formation of deeply bound dimers) is 
proportional to $\hbar b/m$, away from the trimer threshold, so that
the inelastic rate may dominate over the elastic one.
A similar conclusion was reached for the elastic vs inelastic scattering
of $p$-wave weakly bound dimers \cite{Gurarie3}. 
We have also studied the atom-dimer scattering when the incoming relative wave
is a $p$-wave: in the considered sector of total angular momentum one,
the corresponding atom-dimer scattering volume $\mathcal{V}_s^{\rm ad}$
is shown analytically to become proportional to the atom-atom 
scattering volume $\mathcal{V}_s$ away from the even trimer formation 
threshold, see (\ref{eq:asympt_volad}). This asymptotic result looks `universal',
since it does not involve the interaction range $b$.

The recombination rate of three atoms into weakly bound dimers,
calculated in this work in the limit of low relative atomic wavevectors
$k\ll (\alpha \mathcal{V}_s)^{-1/2}$,
has properties more similar to the $s$-wave case.
What remains specific to the $p$-wave case is that the rate is proportional
to the square of the mean kinetic energy per particle,
see (\ref{eq:la_recette}), as already pointed out in \cite{Greene}. 
Apart from that, it includes as a factor
the recombination constant $\mathcal{K}_{\rm rec}$, which is
large close to the resonance: 
it diverges as $\mathcal{V}_s^{5/2}/\alpha^{1/2}$ in the large 
scattering volume limit, 
see (\ref{eq:Kasympt}),
an asymptotic expression valid away from
the even trimer formation threshold and which is `universal', since
it does not involve the interaction range $b$ and it is not sensitive
to the choice of the cut-off function $\chib(\mathbf{k})$ 
of the two-channel
model.
In the large scattering volume limit,
the recombination constant towards weakly bound dimers
dominates over the recombination constant towards deeply bound dimers,
which scales as $\mathcal{V}_s^2$ only (still in the limit of low relative
atomic momenta with respect to $1/(\alpha |\mathcal{V}_s|)^{1/2}$).
If one applies this last result to a degenerate macroscopic
gas, with a Fermi momentum $k_F \approx 1/(\alpha|\mathcal{V}_s|)^{1/2}$,
one finds a number of recombination events to deeply bound dimers
per unit of time and volume scaling as
\be
\gamma_{\rm rec} = O\left(\frac{\hbar}{m} b^4 n^3\right)
\ee
where we used $1/\alpha=O(b)$ and 
$n$ is the gas density.

In the last part of the paper, we have made the modelization more
realistic by including a fourth parameter, a direct attractive interaction between atoms
in the open channel.
Physically, this interaction is supposed to be not
resonant so that it has a weak background
scattering volume $\mathcal{V}_s^{\rm bg}$, of the order of $b^3$
and much smaller than $\mathcal{V}_s$.
To stay in the regime where no deeply bound dimers exist in the
Hamiltonian on resonance, one further imposes $\mathcal{V}_s^{\rm bg}<0$.
We then find that the existence of the trimers is preserved.
They remain weakly bound in the vicinity of some threshold values of
$\alpha$, provided that $|\mathcal{V}_s^{\rm bg}|$ does not
exceed a few $b^3$; these threshold values for $\alpha$ however depend
on $b$ and are not `universal'.
The atom-dimer scattering length $a_{\rm ad}$ is significantly
changed by the open channel interaction in the limit of ultra-narrow
Feshbach resonances, where it now tends to zero for large $\alpha b$
as $1/\alpha$, rather than as $1/(\alpha^2b)$ in our 3-parameter
model; this confirms the non-universal character of $a_{\rm ad}$.
On the contrary, in presence of open channel interactions,
the {\sl same} asymptotic expressions (\ref{eq:asympt_volad}) for
the atom-dimer scattering volume $\mathcal{V}_s^{\rm ad}$ 
and (\ref{eq:Kasympt}) for the recombination
constant $\mathcal{K}_{\rm rec}$ to weakly bound dimers are obtained as in our
3-parameter model, in terms of $\mathcal{V}_s$, 
and in terms of $\mathcal{V}_s$ and $\alpha$ respectively,
which confirms the `universal' character of these results.

\begin{acknowledgments}
We thank Fr\'ed\'eric Chevy,
Victor Gurarie, 
Christophe Mora,
Leticia Tarruell,
Christophe Salomon
and 
Servaas Kokkelmans 
for useful discussions. 
One of us (M.\ J.-L.) thanks Fondazione Angelo Della Riccia
and IFRAF for financial support. The cold atom group at Laboratoire Kastler Brossel
is a member of IFRAF.
\end{acknowledgments}

\appendix

\section{Integral equations for the $K$-dependent part of the Ansatz for $\betab$}
\label{appen:ad}

When injecting the ansatz Eq.(\ref{eq:odd}) or Eq.(\ref{eq:even}) in the homogeneous
part of the equation Eq.(\ref{eq:integrale}) for $\betab$, one faces
the calculation of the following angular averages over the direction of $\mathbf{k}$,
\bea
\mathbf{I}_0(\mathbf{K},k) 
&=& \int \frac{d\Omega_{\hat {\mathbf{k}}}}{4\pi}\,
\left(\frac{1}{2}\mathbf{K}+\mathbf{k}\right) \nonumber \\
&\times&
\frac{(\mathbf{K}\cdot \mathbf{e}_z) + \frac{1}{2} (\mathbf{k} \cdot \mathbf{e}_z)}
{q^2 + K^2 + k^2 + \mathbf{K}\cdot \mathbf{k}} e^{-\mathbf{K} \cdot \mathbf{k}\,b^2} \\
\mathbf{I}_1(\mathbf{K},k) &=& \int \frac{d\Omega_{\hat {\mathbf{k}}}}{4\pi}\,
\left(\frac{1}{2}\mathbf{K}+\mathbf{k}\right) \nonumber \\
&\times& \frac{\mathbf{K}\cdot
\left[\left(\frac{\mathbf{k}\cdot\mathbf{e}_z}{k}\right)\mathbf{e}_y-
\left(\frac{\mathbf{k}\cdot\mathbf{e}_y}{k}\right)\mathbf{e}_z\right]}
{q^2 + K^2 + k^2 + \mathbf{K}\cdot \mathbf{k}}  e^{-\mathbf{K} \cdot \mathbf{k}\,b^2} \\
\mathbf{I}_2(\mathbf{K},k) &=& \int \frac{d\Omega_{\hat {\mathbf{k}}}}{4\pi}\,
\left(\frac{1}{2}\mathbf{K}+\mathbf{k}\right) \nonumber \\
&\times&
(\mathbf{k}\cdot \mathbf{e}_z)\frac{\frac{1}{2}+\frac{\mathbf{K} \cdot \mathbf{k}}{k^2}}
{q^2 + K^2 + k^2 + \mathbf{K}\cdot \mathbf{k}} e^{-\mathbf{K} \cdot \mathbf{k}\,b^2}
\eea
Note that, for the calculations performed in this paper, one can restrict
to the case of a non-positive total energy $E$ so that we have
set $E=-\hbar^2 q^2/m$, $q\geq 0$, and we have omitted $i0^+$.

To perform this angular integration, we use spherical coordinates of polar axis
$\mathbf{K}/K$.
We need to evaluate first the integral over the azimuthal angle $\varphi$,
\bea
\mathbf{B}_1 = \int_0^{2\pi} \frac{d\varphi}{2\pi}\, \mathbf{k} = 
k \cos\theta\, \frac{\mathbf{K}}{K}\\
\mathbf{B}_2 = \int_0^{2\pi} \frac{d\varphi}{2\pi}\, (\mathbf{k}\cdot \mathbf{e}_\alpha)
\mathbf{k} =
\sum_{i,j} k^2 B_{ij}(\mathbf{E}_i\cdot\mathbf{e}_\alpha)\mathbf{E}_j
\eea
where $\alpha$ stands for $y$ or $z$,
$\{\mathbf{E}_{i=1,2,3}\}$ is an orthonormal basis with $\mathbf{E}_3=\mathbf{K}/K$
and 
\bea
B_{ij} &=& \int_0^{2\pi} \frac{d\varphi}{2\pi}\, (\mathbf{k}\cdot \mathbf{E}_i)
(\mathbf{k}\cdot \mathbf{E}_j) \\
&=& \frac{1}{2}\left[(1-\cos^2\theta)\delta_{ij}
+(3\cos^2\theta-1)\delta_{i3}\delta_{j3}\right].
\eea
This leads to
\be
\mathbf{B}_2 = \frac{k^2}{2}\left[(1-\cos^2\theta)\mathbf{e}_\alpha
+(3\cos^2\theta-1)\frac{(\mathbf{K}\cdot \mathbf{e}_\alpha)\mathbf{K}}{K^2}\right]
\ee

The integration over the polar angle $\theta$ involves basically the integral
\be
C_n = \int_{-1}^1 \frac{dx}{2}\, \frac{x^n e^{-t x}}{v+x}
\label{eq:Cn}
\ee
with $v>1$, $t>0$ and integer $n$. The result is
\bea
\label{eq:C0}
C_0 &=& \frac{e^{v t}}{2}\left[E_1(vt-t)-E_1(vt+t)\right] \\
\label{eq:C1}
C_1 &=& -v C_0 + j_0(i t) \\
\label{eq:C2}
C_2 &=& -v C_1 + i j_1(i t) \\
\label{eq:C3}
C_3 &=& -v C_2 - j_2(i t)- \frac{i}{t} j_1(i t)
\eea
where $E_1(z)$ is the exponential integral
\be
E_1(z)=\int_1^{+\infty} ds\,\frac{e^{-s z}}{s}
\ee
and $j_n(z)$ are the usual spherical Bessel functions. A straightforward integration over
the $\theta$ angle leads to
\bea
\mathbf{I}_0(\mathbf{K},k) &=& \frac{u}{4}(C_0-C_2)\mathbf{e}_z \nonumber \\
&+& \left[
\frac{u}{4}(3 C_2-C_0) + \frac{5}{4} C_1 + \frac{1}{2u} C_0
\right] \frac{(\mathbf{K}\cdot \mathbf{e}_z)\mathbf{K}}{K^2} \\
\mathbf{I}_1(\mathbf{K},k) &=&  \frac{1}{2}(C_2-C_0) 
\left[\frac{\mathbf{K}\cdot\mathbf{e}_z}{K}\mathbf{e}_y-
\frac{\mathbf{K}\cdot\mathbf{e}_y}{K}\mathbf{e}_z\right] \\
\mathbf{I}_2(\mathbf{K},k) &=& 
\left[\frac{u}{4}(C_0-C_2)+\frac{1}{2}(C_1-C_3)\right]\mathbf{e}_z \nonumber \\
&+& 
\left[\frac{u}{4}(3 C_2-C_0)+\frac{1}{2}(3 C_3-C_1) + \frac{1}{4} C_1 + \frac{1}{2u} C_2\right]
\nonumber \\
&\times& \frac{(\mathbf{K}\cdot \mathbf{e}_z)\mathbf{K}}{K^2}
\eea
with
\be
\label{eq:uv}
u=\frac{k}{K} \qquad v=\frac{q^2+K^2+k^2}{Kk} \qquad t=b^2 K k
\ee

For the odd sector, $E<0$ so the source term $A_0$ vanishes.
Projecting Eq.(\ref{eq:integrale})
onto the two components $\mathbf{e}_z$
and $(\mathbf{K}\cdot \mathbf{e}_z)\mathbf{K}/K^2$,
we obtain an integral system of coupled equations for $B_{L=0}(K)$ and
$B_{L=2}(K)$, given in the main text, see Eq.(\ref{eq:ieodd}),
where we have introduced the two by two matrix
\begin{widetext}
\be
M(K,k) = \left(
\begin{array}{cc}
\frac{k}{4K}(C_0-C_2) & -\frac{k}{4K}(C_0-C_2) - \frac{1}{2} (C_1-C_3) \\
-\frac{k}{4K}(3C_2-C_0) -\frac{5}{4}C_1-\frac{K}{2k} C_0   &
\frac{k}{4K}(3C_2-C_0)+\frac{3}{2} C_3 -\frac{1}{4} C_1 +\frac{K}{2k} C_2
\end{array}
\right).
\ee
\end{widetext}
The resulting integral operator can be made hermitian
by the change of variables
\be
\left(\begin{array}{c} B_{L=0}(K) \\ B_{L=2}(K) \end{array}\right)=
P^{-1} 
\left(\begin{array}{c} b_0(K)/K \\ b_2(K)/K \end{array}\right)
\ee
with the two by two transformation matrix
\be
P= \left(\begin{array}{cc} 2^{1/4} & 0 \\ -2^{-1/4} & 2^{-1/4} \end{array}\right).
\ee
This results in the hermitian integral equation
\bea
0&=&D(K) \left(\begin{array}{c} b_0(K) \\ b_2(K) \end{array}\right)
+\frac{4}{\pi} \int_0^{+\infty} dk\, \nonumber \\
& & Kke^{-5(K^2+k^2)b^2/8}
N(K,k) \left(\begin{array}{c} b_0(k) \\ b_2(k) \end{array}\right),
\eea
with the two by two matrix $N(K,k)=P M(K,k) P^{-1}$ satisfying
$N^\dagger(K,k)=N(k,K)$.
For the even sector, the integral equation for $B_{L=1}$ is given directly
in the main text, see Eq.(\ref{eq:intevensec}).

To conclude this appendix, we briefly explain how to normalize the state vector
of the trimer (when it exists). The normalization can be done directly in momentum
space by integration over internal variables, that is after having singled
out the total momentum variables $\mathbf{Q}$. In the sector of (\ref{eq:ansatz}) with
one molecule, using the fact that the parameterization of the molecular and atomic
momenta $\mathbf{k}_{\rm mol}=\mathbf{Q}/2+\mathbf{K}$,
$\mathbf{k}_{\rm at}=\mathbf{Q}/2-\mathbf{K}$, has a unit Jacobian, we find
\be
\langle \Psi_{\rm mol}| \Psi_{\rm mol}\rangle = \int \frac{d^3\mathbf{K}}{(2\pi)^3} 
|\betab(\mathbf{K})|^2.
\ee
In the purely atomic sector of (\ref{eq:ansatz}), using the fact that the
parameterization of the atomic momenta $\mathbf{k}_1=\mathbf{Q}/3+\mathbf{K}/2+\mathbf{k}$,
$\mathbf{k}_2=\mathbf{Q}/3+\mathbf{K}/2-\mathbf{k}$,
$\mathbf{k}_3=\mathbf{Q}/3-\mathbf{K}$, has a unit Jacobian, using Wick's theorem and
the fact that $A(\mathbf{K},\mathbf{k})$ is an odd function of $\mathbf{k}$,
we obtain
\bea
\langle \Psi_{\rm at}| \Psi_{\rm at}\rangle &=&  2\, 
\int \frac{d^3\mathbf{K}d^3\mathbf{k}}{(2\pi)^6}
A(\mathbf{K},\mathbf{k})\left[A^*(\mathbf{K},\mathbf{k})\phantom{\frac{1}{1}}\right. \nonumber \\
&-& \left. 2A^*\left(\mathbf{k}-\frac{1}{2}\mathbf{K},\frac{3}{4}\mathbf{K}+
\frac{1}{2}\mathbf{k}\right)\right].
\eea
Using the specific form of the ansatz (\ref{eq:even}) and (\ref{eq:odd})
for $\betab$, with $B_{L=0}(K)$, $B_{L=1}(K)$ and
$B_{L=2}(K)$ real, and the link (\ref{eq:A}) 
between $A$ and $\betab$ written here for $A_0\equiv 0$, 
one can get integrals of lower dimensions:
\begin{widetext}
\bea
\langle \Psi_{\rm mol}^{\rm even}| \Psi_{\rm mol}^{\rm even}\rangle & = &
\int_0^{+\infty} \frac{dK}{3\pi^2} 
\mathcal{B}^2(K) 
\label{eq:norm_mol_even}
\\
\langle \Psi_{\rm at}^{\rm even}| \Psi_{\rm at}^{\rm even}\rangle & =& 
\frac{m^2 \Lambda^2}{6\pi^4\hbar^4}
\int_0^{+\infty} \! \! \! \! dK \int_0^{+\infty}\! \!  \! \! dk 
\left[
\frac{\frac{2}{3}k^4 \mathcal{B}^2(K) e^{-k^2b^2}}
{\left(q_{\rm trim}^2+k^2+\frac{3}{4} K^2\right)^2}
+
2\mathcal{B}(K)\mathcal{B}(k)
(D_0-D_2)(K,k) e^{-\frac{5}{8}(K^2+k^2)b^2}
\right]
\label{eq:norm_at_even}
\\
\langle \Psi_{\rm mol}^{\rm odd}| \Psi_{\rm mol}^{\rm odd}\rangle &=& 
\int_0^{+\infty} \frac{dK}{3\sqrt{2}\pi^2}
\mathbf{b}^2(K)
\label{eq:norm_mol_odd}
 \\
\langle \Psi_{\rm at}^{\rm odd}| \Psi_{\rm at}^{\rm odd}\rangle & = &
\frac{m^2 \Lambda^2}{6\sqrt{2}\pi^4\hbar^4}
\int_0^{+\infty} \! \!  \! \! dK \int_0^{+\infty}\! \!  \! \! dk
\left[
\frac{\frac{2}{3}k^4 \mathbf{b}^2(K) e^{-k^2b^2}}
{\left(q_{\rm trim}^2+k^2+\frac{3}{4} K^2\right)^2}
-4 \mathbf{b}(K)\cdot N_D(K,k) \mathbf{b}(k) e^{-\frac{5}{8}(K^2+k^2)b^2}
\right].
\label{eq:norm_at_odd}
\eea
\end{widetext}
We have set $E=-\hbar^2 q_{\rm trim}^2/m$,
$\mathcal{B}(K)= K B_{L=1}(K)$ and
\be
D_n = \int_{-1}^1 \frac{dx}{2}\, \frac{x^n e^{-t x}}{(v+x)^2},
\ee
which is minus the derivative of $C_n$ with respect to $v$ for fixed $t$ and
obeys the recursive relation $D_{n+1}=C_n-vD_n$.
Also, the vector $\mathbf{b}(K)$ has components $b_0(K),b_2(K)$,
and the two by two matrix $N_D(K,k)$ is obtained in replacing each $C_n$ by $D_n$
in the matrix $N(K,k)$.

\section{Prescription for the recombination rate of fermions}
\label{appen:dim_rate}

We wish to derive the formula (\ref{eq:la_recette}) giving the rate of
dimer formation in a gas of fermions at low kinetic energy, in terms of
a recombination constant depending on the interaction and the expectation
value $\langle\ldots\rangle_0$ of some operator in the unperturbed state of the gas.
We start with the intuitive idea that a dimer formation can take place
by three-body collision when the mutual distances of the atoms
are at most of the order of the dimer radius $\sigma$,
hence the heuristic formula
\bea
&&\frac{d}{dt}N_{\rm dim} \propto \int d^3r_1\, d^3r_2\,
d^3r_3\, e^{-\rho^2/2\sigma^2}  \nonumber \\
&&\times \langle \hat{\psi}^\dagger(\mathbf{r}_1)\hat{\psi}^\dagger(\mathbf{r}_2)
\hat{\psi}^\dagger(\mathbf{r}_3) \hat{\psi}(\mathbf{r}_3) 
\hat{\psi}(\mathbf{r}_2) \hat{\psi}(\mathbf{r}_1)\rangle_0.
\eea
We have taken for simplicity a Gaussian cut-off function,
in terms of the hyperradius $\rho$ defined as
\begin{equation}
\rho^2=\sum_{i=1}^{3} (\mathbf{r}_i-\mathbf{R})^2 =
\frac{1}{2}(\mathbf{r}_1-\mathbf{r}_2)^2 + \frac{2}{3}\left[\mathbf{r}_3
-(\mathbf{r}_1+\mathbf{r}_2)/2\right]^2
\label{eq:hyperradius}
\end{equation}
where $\mathbf{R}=\sum_{i=1}^3 \mathbf{r}_i/3$ is the center of mass position.
Under the assumption that the typical wavevectors populated in the
uncorrelated state of the gas are much smaller than $1/\sigma$,
which is the case here since we assume relative momenta  
$<q_{\rm dim}$, we can expand each field operator $\hat\psi(\mathbf{r}_i)$
around the center of mass position $\mathbf{R}$ of the three $\mathbf{r}_i$
in powers of $\mathbf{r}_i-\mathbf{R}$.
Since $\hat{\psi}^2=0$ one has to go to second order in $\delta r$:
\bea
\hat{\psi}(\mathbf{r}_3)\hat{\psi}(\mathbf{r}_2)\hat{\psi}(\mathbf{r}_1)&=&
\sum_{\alpha\neq\beta} \mathcal{A}_{\alpha\beta}\, \hat{\psi}(\mathbf{R})
\left[\partial_{R_\alpha}\hat{\psi}(\mathbf{R}) \right]
\left[\partial_{R_\beta}\hat{\psi}(\mathbf{R})\right] \nonumber \\
& +& O(\delta r^3)
\eea
where $\alpha$ and $\beta$ run over the three directions of space
$x$, $y$ and $z$, and the matrix $\mathcal{A}$ is given by
\begin{equation}
\mathcal{A}_{\alpha\beta} = \delta r_{2,\alpha}\delta r_{1,\beta}
-\delta r_{3,\alpha}\delta r_{1,\beta}
+\delta r_{3,\alpha}\delta r_{2,\beta}.
\end{equation}
We restrict to this leading order in $\delta r$.  It remains
to integrate the Gaussian weighted
products $\mathcal{A}_{\alpha\beta} \mathcal{A}_{\gamma\delta}$ 
over the internal variables for fixed center of mass position 
$\mathbf{R}$, by using the Jacobi coordinates. Since $\mathcal{A}_{\alpha\beta}$
for $\alpha\neq\beta$
is odd with respect to the reflection along direction $\alpha$ or along
direction $\beta$,
this integral vanishes if $\{\alpha,\beta\}\neq \{\gamma,\delta\}$.
The invariance of the integral by permutation of the $x$, $y$
and $z$ axis leads to the final prescription (\ref{eq:la_recette}).

\section{Recombination constant close to the trimer formation threshold}
\label{appen:singular}

The goal is to derive the approximate formula (\ref{eq:fano}) giving the recombination
constant $\mathcal{K}_{\rm rec}$ for large scattering volumes and for a value
of $\alpha_{\rm res}$ close to the threshold for the even trimer formation.

To this end, we rewrite the integral equation (\ref{eq:intevensec}) 
as a sum of its $\mathcal{V}_s=\infty$ value and a remainder, then we treat the
remainder perturbatively. Taking as unknown $F(K)= K\,g(K)/\mathcal{V}_s$, using 
Eq.(\ref{eq:f_expl}) and the identity
$1/(X-i0^+) = \mathcal{P}\frac{1}{X}+i\pi\delta(X)$, where
$\mathcal{P}$ is the Cauchy principal value, we obtain the rewriting
\be
\frac{\alpha_{\rm res}}{4} F(K) -I_0[F](K) - I_1[F](K)
-i A(K) F(K) = S(K).
\label{eq:compact}
\ee
We have introduced the two functions
\bea
A(K) &=& K(C_0-C_2)(K,K_{\rm dim})\, e^{-\frac{5}{8}b^2(K^2+K_{\rm dim}^2)} \\
S(K) &=& -\frac{36\pi\hbar^2}{m\Lambda} K_0 k_0^\perp e^{-5 b^2 K^2/8},
\eea
and the two operators
\begin{widetext}
\bea
I_0[F](K) &=& \frac{h(K)}{4} F(K)
+ \frac{2}{\pi} \int_0^{+\infty} \!\!\! dK'\, \frac{K}{K'} (C_0-C_2)(K,K') 
e^{-\frac{5}{8}b^2 (K^2+K'^2)} F(K'), \\
I_1[F](K) &=& \frac{K_{\rm dim}^2}{4} \frac{h(K)-h(K_{\rm dim})}
{K^2-K_{\rm dim}^2}\, F(K)
+\frac{2 K_{\rm dim}^2}{\pi} \int_0^{+\infty} \!\!\! dK'\, \frac{K}{K'} (C_0-C_2)(K,K')
e^{-\frac{5}{8}b^2 (K^2+K'^2)} \mathcal{P}\frac{F(K')}{K'^2-K_{\rm dim}^2}
\eea
\end{widetext}
where we have set $h(K)=q\exp(q^2b^2)\,\mathrm{erfc}(qb)$ with $q=\sqrt{3}K/2$.

The operator $J$ such that $J[G](K)= I_0[F](K)/K$ where $F(K)=K\, G(K)$,
is self-adjoint and thus has real eigenvalues; we find numerically that it has
one and only one discrete eigenvalue, that we called 
$\alpha_{\rm th}^{\rm even}/4$, with the 
corresponding normalized eigenvector $K\to g_0(K)$. 
As a consequence, the operator $I_0$ admits, with the eigenvalue $\alpha_{\rm th}^{\rm even}/4$,
a discrete eigenvector $u_0(K)= K g_0(K)$, with the corresponding adjoint (left-eigenvector)
$v_0(K) = g_0(K)/K$. We note that the kernel in 
$I_0^\dagger$ behaves as $K'^2/(K^2+K'^2)$
at low momenta, which is bounded, so that $v_0(K)$ is bounded, $g_0(K)$ vanishes
linearly with $K$ in $K=0$ and $u_0(K)$ vanishes quadratically \cite{justif_meilleure}.
Apart for this discrete eigenvalue, we numerically find that $I_0$ has a continuous spectrum
extending from 0 to $1/(4\sqrt{\pi}b)$.

For $\alpha_{\rm res}$ close to $\alpha_{\rm th}^{\rm even}$, in the large scattering volume
limit, a small denominator (of the order of $\delta\equiv \alpha_{\rm res}-\alpha_{\rm th}^{\rm even}$) appears
in the direction of $u_0$ 
when one solves (\ref{eq:compact}). This small denominator is weakly perturbed
by $I_1$ (which shifts the value of $\alpha_{\rm res}$ for which the denominator
has a minimal modulus) and by the imaginary part involving the function $A(K)$
(which prevents the denominator from exactly vanishing). These effects can be
included systematically by using the ansatz
\be
F(K) = F_{\rm bg}(K) + c_0 u_0(K).
\label{eq:ansatzbg}
\ee
The background part of the solution, $F_{\rm bg}(K)$, and its derivative $F_{\rm bg}'(K)$,
are supposed to be uniformly bounded
in $\mathcal{V}_s$ and $\alpha_{\rm res}$,
even in the vicinity of $\alpha_{\rm res}=\alpha_{\rm th}^{\rm even}$. 
From the low $K$ behavior of $C_0(K,K')-C_2(K,K')$,
one finds that the solution $F(K)$ 
satisfies $F(0)=4S(0)/\alpha_{\rm res}$, this value being reached quadratically in $K$.
Since $|F'|$ is not uniformly bounded
in $\mathcal{V}_s$ and $\alpha_{\rm res}$, this does not give information on
the value $F(K_{\rm dim})$. 
One has also $F_{\rm bg}(0)=4S(0)/\alpha_{\rm res}$, but for a $|F'_{\rm bg}|$ bounded
by $c_{\rm bg}$, where the constant $c_{\rm bg}$ does not depend on $\mathcal{V}_s$
or $\alpha_{\rm res}$, one now has
\be
|F_{\rm bg}(K_{\rm dim}) -\frac{4S(0)}{\alpha_{\rm res}}| \leq c_{\rm bg} K_{\rm dim}.
\ee
This is the property of $F_{\rm bg}$ that we shall need.

Injecting the ansatz (\ref{eq:ansatzbg}) into (\ref{eq:compact}) and projecting
onto $u_0$ by using the adjoint vector $v_0$, we obtain the exact expression
\bea
[\delta/4-\langle v_0|I_1|u_0\rangle -i u_0(K_{\rm dim}) \langle v_0|A\rangle]c_0 =
\langle v_0|S\rangle \nonumber \\
+ (\delta/4) \langle v_0|F_{\rm bg}\rangle 
+\langle v_0| I_1| F_{\rm bg}\rangle +i F_{\rm bg}(K_{\rm dim}) \langle v_0|A\rangle,
\eea
where we used $\langle v_0|I_0|F_{\rm bg}\rangle = (\alpha_{\rm th}^{\rm even}/4) 
\langle v_0 | F_{\rm bg}\rangle$ (after justification).
Here Dirac's notation means $\langle f|g\rangle = \int_0^{+\infty} dK\, f^*(K) g(K)$.
Replacing $c_0$ by its expression in (\ref{eq:ansatzbg}), setting $K=K_{\rm dim}$
and putting all terms on a common denominator, we see that the imaginary contribution
$iu_0(K_{\rm dim}) \langle v_0|A\rangle$ exactly cancels in the numerator.
We finally obtain the still exact expression
\bea
\frac{F(K_{\rm dim})}{F_{\rm bg}(K_{\rm dim})}  &=& \mathcal{F}
\left[\delta/4-\langle v_0|I_1|u_0\rangle -i u_0(K_{\rm dim}) \langle v_0|A\rangle\right]^{-1}
\nonumber
\\
&\times& \{\delta/4-\langle v_0|I_1|u_0\rangle/\mathcal{F} + u_0(K_{\rm dim})[\langle v_0|S\rangle 
\nonumber \\
&+& \langle v_0| I_1| F_{\rm bg}\rangle]/[F_{\rm bg}(K_{\rm dim}) \mathcal{F}]\}
,
\label{eq:compli}
\eea
where $\mathcal{F} \equiv 1-\langle v_0|F_{\rm bg}\rangle u_0(K_{\rm dim})/F_{\rm bg}(K_{\rm dim})$.

The last step is to expand the various terms to leading order in $K_{\rm dim}$.
Expanding $I_1$ to leading order in $K_{\rm dim}$ and using the fact that $u_0$
is an eigenvector of $I_0$ to simplify the integral expression appearing in
this leading order form of $I_1$, we obtain
\be
\lim_{K_{\rm dim}\to 0} \frac{\langle v_0| I_1| u_0\rangle}{K_{\rm dim}^2} 
=\frac{\alpha_{\rm th}^{\rm even}}{4} \langle v_0 | v_0\rangle >0,
\ee
and the quantity $\langle v_0|v_0\rangle$
is readily evaluated numerically. This gives a position of the peak 
in $\mathcal{K}_{\rm rec}$ shifted to a value of $\alpha_{\rm res}$ larger
than $\alpha_{\rm th}^{\rm even}$ by a $O(K_{\rm dim}^2)$, see (\ref{eq:alpha1}).
Since $u_0(K)$ vanishes quadratically in $K$, one has $u_0(K_{\rm dim})$ of
the order of $K_{\rm dim}^2$; since $A(K)$ vanishes linearly with 
$K_{\rm dim}$ for a fixed $K$,
one has $\langle v_0| A\rangle$ of the order of $K_{\rm dim}$. A numerical calculation
of the corresponding coefficients leads to (\ref{eq:Da}).
Amusingly, using the low-$K$ expansion of $C_0(K,K')-C_2(K,K')$, we find the mathematical
equivalence in the zero $K_{\rm dim}$ limit,
\be
\langle v_0| A \rangle \sim \frac{2}{3} K_{\rm dim} \frac{\langle v_0|S\rangle}{S(0)}
\ee
which leads to (\ref{eq:amus}).
Another result is
\be
\langle v_0 | I_1 | F_{\rm bg}\rangle = O(K_{\rm dim}^2\ln K_{\rm dim})
\ee
so that this contribution, being multiplied by $u_0(K_{\rm dim})$ in (\ref{eq:compli}),
may be neglected at this order.
Finally, we note that $F_{\rm bg}(K_{\rm dim})$ and $\mathcal{F}$ differ
from $4S(0)/\alpha_{\rm res}$ and $1$ respectively 
by terms of order $K_{\rm dim}$ and $K_{\rm dim}^2$, that we neglect to obtain (\ref{eq:fano}).


\begin{thebibliography}{99}
\bibitem{wdam}
One can equivalently consider a dimer wavefunction of
a well defined angular moment component $m=0,\pm 1$ along the quantization axis $z$,
by replacing $(3/4\pi)^{1/2} r_\gamma/r$ in Eq.(\ref{eq:phidim}) by $Y_1^m(\theta,\varphi)$.
\bibitem{apz}
If one assumes $\alpha_{\rm res}=0$, taking
$u(k)=1/(k^2\mathcal{V}_s)$ for low $k$, as done in \cite{Macek},
one gets a dimer for $\mathcal{V}_s<0$
(contrarily to the realistic case) and one reaches a contradiction in
Eq.(\ref{eq:norm}), $|\mathcal{N}|^2<0$.
\bibitem{quasimodo} In the resonant limit $|\mathcal{V}_s|\to +\infty$,
on the side $\alpha_{\rm res} \mathcal{V}_s<0$ of the resonance,
there is a so-called quasi-bound state in the two-body problem, 
which was studied experimentally
in \cite{Gaebler}: From (\ref{eq:lkeu}) and using the fact
 that $\alpha_{\rm res}>0$, one finds that
the scattering amplitude $f(k)$ has a complex pole with a real part
$k_0^0\simeq 1/(\alpha_{\rm res} |\mathcal{V}_s|)^{1/2}$ much larger
than its imaginary part $\simeq -(k_0^0)^2/(2\alpha_{\rm res})$.
Physically this is related to the existence of a centrifugal barrier
when the two particles approach with non-zero relative momentum, here $l=1$.
This quasi-bound state does not play an important role in the present work:
in the exceptional case where we consider a negative rather than positive scattering
volume (see end of \ref{subsec:loss_recomb}), the total energy $E$ in the
three-body problem is still assumed to be much smaller than $\hbar^2 (k_0^0)^2/m$
so that the relative momentum $k_{\rm rel}$ appearing in the three-body problem,
as defined in (\ref{eq:Erel}), cannot reach the value $k_0^0$.
\bibitem{general}
This reasoning extends to an arbitrary, everywhere finite envelope function $f(k)$
such that $\chib(\mathbf{k})=\mathbf{k}\, f(k)$.
\bibitem{check_ok}
Checking the consistency of these two ways of calculating $\beta$
is actually useful: contour integration in the space of complex $k$
was used to derive Eq.(\ref{eq:norm}) from Eq.(\ref{eq:clos}), and it is not straightforward
that the contribution of the contour integral on an infinite radius half-circle
is zero, since $|f(i q)|$ actually rapidly diverges as $q\to\infty$ for our two-channel
model.
This check reduces to proving that $I(\mathbf{r}) = \int \frac{d^3k}{(2\pi)^3}
e^{-k^2 b^2/2} e^{i\mathbf{k}\cdot\mathbf{r}}/(k^2+q_{\rm dim}^2)
\sim e^{-q_{\rm dim}r} e^{q_{\rm dim}^2 b^2/2}/(4\pi r)$
for $r\to\infty$. This can be done by using the analytic expression for $I(\mathbf{r})$
or by noting that $I(\mathbf{r})$ is the convolution of a Gaussian with a Yukawa
function.
\bibitem{technical}
In the second term, an integral of the form 
$I=\int_0^{K_{\rm max}} dK'\, v(K') \mathcal{P}[1/(K'^2-K_{\rm dim}^2)]$
appears, with $K_{\rm max}$ being the numerical momentum cut-off,
which may lead to numerical difficulties since the integrand diverges in $K=K_{\rm dim}$.
We rewrite it using $v(K')=[v(K')-v(K_{\rm dim})]+v(K_{\rm dim})$. Using
$\int_0^{K_{\rm max}} dK'\, \mathcal{P}[1/(K'^2-K_{\rm dim}^2)]=
-\ln[(K_{\rm max}+K_{\rm dim})/(K_{\rm max}-K_{\rm dim})]/(2K_{\rm dim})$, we are left
with the numerical evaluation of 
$\int_0^{K_{\rm max}} dK'\, [v(K')-v(K_{\rm dim})]/(K'^2-K_{\rm dim}^2)$.
\bibitem{aide_wick}
We get $\langle\partial_{R_\alpha}\hat{\psi}^\dagger \partial_{R_\beta}\hat{\psi}\rangle_0
=M_{\alpha\beta}/L^3$, with $M_{\alpha\beta}=\sum_{i=1}^3 k_{i,\alpha}^0 k_{i,\beta}^0$.
From Wick's theorem the expectation value $\langle\ldots\rangle_0$
in (\ref{eq:la_recette})
is equal to $3L^{-9}(M_{\alpha\alpha} M_{\beta \beta}-M_{\alpha\beta}^2)$.
One may then check that $\sum_{(\alpha,\beta)} M_{\alpha\alpha} M_{\beta\beta} -M_{\alpha\beta}^2=
3(\mathbf{K}_0\times\mathbf{k}_0)^2$, where the sum is taken over the the three
values $(x,y),(x,z)$ and $(y,z)$ of $(\alpha,\beta)$.
\bibitem{pdd}
One sees that the quantity $t$ appearing in the definition of $C_0$ 
and $C_2$ in
the appendix \ref{appen:ad} is small and can be neglected; 
explicit expressions
without relying on special functions can be obtained; 
they can be further simplified
by performing a high $v$ expansion, where the parameter $v$ appears
in appendix \ref{appen:ad}, and by keeping the leading term 
$2/(3v)$.
Since the next term in the large $v$ expansion is $O(1/v^3)$,
this further approximation of keeping the leading term $\propto 1/v$ only
is already good at the minimal possible value of $v$, $v=2$, where it 
introduces an extra error of $\sim 5\%$.
\bibitem{theoreme} 
One can show the following result, for a fixed value of $\alpha_{\rm res}$: 
if there exist constants $A$ and $B$ such that,
for all $\mathcal{V}_s$ above some value, one has $|F(K)|< A$ and
$|F'(K)|< B$ at all $K$, then (\ref{eq:Kasympt}) is exact.
\bibitem{tres_technique}
We shall also assume that the first order derivative
of $\mathcal{E}(K)$ with respect to $K$ is uniformly bounded
as a function of $K$ and $K_{\rm trim}$.
\bibitem{justif}
From the low-$K$
expansion of the matrix elements in the second line of $M(K,k)$,
we get $B_{L=2}(K)=O(1/\sqrt{K^2+K_{\rm trim}^2})$, so that $K |B_{L=2}(K)|$
is uniformly bounded in $K$ and $K_{\rm trim}$, similarly to what happens
in the even sector. The crossed terms in $B_{L=0} B_{L=2}$ give at most
a logarithmic divergence in the normalization integrals.
\bibitem{pc}
We use here the convention that, in the two-body calculation of the dimer wavefunction
$\phi(\mathbf{r})$, $\betab=p_{\rm closed}^{1/2} \mathbf{e}_z$.
\bibitem{pourquoi}
The infinite $\mathcal{V}_s^{\rm bg}$ pure single channel model
is equivalent to an infinite $\mathcal{V}_s$ two channel model 
with no open channel interaction and
with $\Lambda\to +\infty$, since both models have
the same scattering amplitude in these limits.
\bibitem{justif_meilleure}
More precisely, if one assumes that $v_0(K)$ diverges as $1/K^\gamma$ for low $K$, 
$\gamma < 3$ to ensure convergence in the integral, one sees that the integral in the limit
$K\to 0$ either diverges
as $1/K^{\gamma-1}$, for $\gamma>1$, or does not diverge, for $\gamma < 1$. 
Since $\alpha_{\rm th}^{\rm even}>0$, this leads to a contradiction 
in the integral equation $\alpha_{\rm th}^{\rm even} v_0(K)/4 = I_0^\dagger[v_0](K)$,
the left-hand side diverging more rapidly than the right-hand side.
\end{thebibliography}
\end{document}